%% file: main.tex
\documentclass[manuscript]{acmart}

\usepackage{amsmath}
\usepackage[ruled,vlined]{algorithm2e}
\usepackage{graphicx}
\usepackage{geometry}
\usepackage{booktabs}
\usepackage{multirow}

\usepackage{framed}
\usepackage{subfigure}
\usepackage{listings}

\usepackage{todonotes}

\AtBeginDocument{%
  }

\newcommand{\new}[1]{#1} 
\newcommand{\rev}[1]{#1} 

\begin{document}

\title{How Many Shots Are Enough for a Quantum Circuit?}

\author{Giuseppe Bisicchia}
\email{giuseppe.bisicchia@phd.unipi.it}
\orcid{0000-0002-1187-8391}
\affiliation{%
  \institution{University of Pisa}
  \city{Pisa}
  \country{Italy}
}

\author{Alessandro Bocci}
\email{alessandro.bocci@unipi.it}
\orcid{https://orcid.org/0000-0002-7000-2103}
\affiliation{%
  \institution{University of Pisa}
  \city{Pisa}
  \country{Italy}
}

\author{Ernesto Pimentel}
\email{epimentel@uma.es}
\affiliation{%
  \institution{University of Malaga}
  \city{Malaga}
  \country{Spain}
}

\author{Antonio Brogi}
\email{antonio.brogi@unipi.it}
\affiliation{%
  \institution{University of Pisa}
  \city{Pisa}
  \country{Italy}
}

\renewcommand{\shortauthors}{Bisicchia et al.}

\begin{abstract}
Quantum algorithms require repeated circuit executions -- known as \textit{shots} -- to estimate output distributions accurately. Determining the minimal number of shots needed to meet a target accuracy is crucial to reduce costs and resource usage, especially on today’s noisy and expensive quantum hardware. 
In this paper, we address the shot optimisation problem in a \textit{black-box setting}, where \textit{no assumptions} are made about the structure of the quantum circuit or the noise model of the backend.
We introduce \textsc{IncrementalExecution}, a novel online framework that dynamically determines when to stop executing shots based on the principle of \textit{point of diminishing returns}: the point at which additional shots no longer significantly alter the empirical distribution \new{of a fixed circuit}.
The framework supports customisable policies for shot management, enabling flexible trade-offs between execution cost and result fidelity \new{within static execution scenarios}.
We assess our proposal through an extensive experimental evaluation spanning \new{$33{,}750$} framework configurations across \new{180} unique \new{static} quantum circuit--backend combinations, for a total of \new{7.3M} independent experiments.
Unlike prior work that relies on problem-specific knowledge or algorithm-dependent assumptions \new{(e.g., variational or adaptive workflows)}, our approach is \new{applicable to a large set of static circuits} and immediately deployable on current quantum cloud platforms.
\end{abstract}




\received{20 February 2007}
\received[revised]{12 March 2009}
\received[accepted]{5 June 2009}

\maketitle

\input{src/intro}
\input{src/related}
\input{src/problem}
\input{src/framework}
\input{src/policies}
\input{src/exp}
\input{src/applicability}
\input{src/threats}
\input{src/conclusions}

\bibliographystyle{ACM-Reference-Format}
\bibliography{biblio}

\newpage
\input{src/thAnalysis}

\end{document}

%% file: src/intro.tex
\section{Introduction}
\label{sec:intro}

Quantum computing is inherently probabilistic~\cite{dirac1981principles}. When executing a quantum circuit, the result is not deterministic but rather sampled from an unknown probability distribution defined by the quantum state measured~\cite{von2018mathematical}. Consequently, obtaining meaningful and accurate results from quantum computations requires executing the circuit multiple times -- a process known as \emph{performing} \emph{shots}. Each shot corresponds to a single execution of the circuit, and the empirical distribution obtained from repeated shots approximates the true underlying quantum distribution, or \textit{ground truth}~\cite{nielsen2010quantum}.

When executing a quantum circuit, quantum software developers choose a fixed number of shots based on the size of the circuit or prior knowledge.
The number of shots chosen to execute a quantum circuit significantly impacts both the \emph{accuracy} of the resulting measurement statistics and the \emph{cost} of computation, in terms of both execution time and resource consumption~\cite{awspricing,azurepricing}. This becomes especially significant in current \textit{Noisy Intermediate-Scale Quantum} (NISQ) devices, which are resource-constrained and prone to fluctuations in noise characteristics~\cite{nisq,miranskyy2025feasibility}. Thus, accurately determining the number of shots to execute is increasingly recognised as an essential practical challenge in quantum computing~\cite{bisicchia2024quantum,zhao2020quantum,zhang2025empirical,garcia-alonso2026soq}.

\new{Shot allocation has been extensively studied in the context of \emph{variational} and other \emph{adaptive} quantum algorithms~\cite{zhu2024optimizing,gu2021adaptive,kahani2023novel,liang2024artificial}. However, these workloads introduce intrinsic non-stationarity, as the circuit structure or parameters evolve across optimisation iterations, causing the underlying output distribution to change over time.
Instead, we focus on static, non-variational quantum circuits\footnote{\new{
A \emph{static quantum circuit execution} refers to the repeated execution of a quantum circuit whose gate sequence, topology, and parameters remain fixed across all shots. Under this assumption, the ground-truth output probability distribution is stationary throughout the execution. In contrast, variational, adaptive, or feedback-driven executions induce non-stationary output distributions due to circuit updates across iterations and require different statistical guarantees and stopping criteria.}}, whose structure and parameters remain fixed across repeated executions. This setting captures a broad and practically relevant class of workloads, including benchmarking, calibration, device characterisation, and the execution of fixed quantum algorithms.}

In this work, therefore, we investigate the following
\new{main}
research question \new{(MRQ)}: 

\begin{center}
\emph{\textbf{MRQ:} Given a \new{static, non-variational} quantum circuit and a noisy quantum processing unit (QPU), is it possible to determine how many shots are sufficient to reach a required target accuracy?}
\end{center}

\rev{It is worth noting that the statistical structure of this problem is not intrinsically quantum. At its core, it is an online sample-optimisation problem for a stationary black-box stochastic process with finite outcomes: after each batch of samples, one must decide whether additional samples are still changing the empirical distribution sufficiently to justify their cost. Classical instances include Monte Carlo estimation of categorical distributions, repeated benchmarking of randomized algorithms, stochastic simulators with discrete outputs, probabilistic software testing, and workload or failure-mode characterisation in distributed systems. Quantum circuit execution provides a particularly compelling instantiation because samples correspond to hardware shots, which incur monetary cost, queueing latency, and noise-dependent variability.}

\noindent
Specifically, we aim to design a methodology capable of approximating the \emph{optimal number of shots}, defined as the \textit{point of diminishing returns} where additional shots no longer meaningfully alter the empirical output distribution\footnote{\new{While many quantum applications ultimately aim to estimate expectation values of specific observables, in this work we deliberately focus on stabilisation of the output distribution. This choice is motivated by generality: a sufficiently accurate empirical distribution enables the evaluation of a wide class of observables, whereas the converse does not hold~\cite{nielsen2010quantum}.}} \new{for a fixed circuit instance}.

We tackle this question under a general \textbf{black-box setting}, where \textit{no assumptions} are made about \new{the internal} structure of the circuit or the noise model of the quantum hardware\footnote{The same problem can also be formulated in a \emph{grey-box} setting, where partial information is available -- such as aggregated properties of the circuit structure, output distribution, or noise model -- and in a \emph{white-box} setting, where full access to the circuit and noise model is assumed. In this work, we focus on the \emph{black-box} formulation as the most \new{assumption-free} scenario \new{within the scope of static circuit execution}. \new{Extending the methodology to \emph{variational} or other \emph{adaptive} workloads requires handling non-stationary distributions induced by circuit updates, and we leave this to future work.}}.

Prior work on this problem typically assumes knowledge about the quantum system, such as known noise models or the use of specific algorithmic families (e.g., Variational Quantum Algorithms (VQAs)~\cite{cerezo2021variational} or Quantum Machine Learning (QML)~\cite{biamonte2017quantum}). These assumptions restrict the scope of applicability and compromise robustness in realistic scenarios, where hardware noise is inherently stochastic, time-dependent, and hard to model accurately~\cite{arute2019quantum}. \new{Moreover, approaches designed for variational settings do not directly transfer to static circuits, where the primary challenge lies in efficiently estimating a fixed but unknown output distribution.}

Our approach fundamentally differs from prior work by reframing the problem into one of detecting the \textit{point of diminishing returns}, a state where additional shots no longer result in statistically significant updates to the empirical output distribution \new{of a static circuit}. Once this point is reached, further execution becomes unnecessary, as it no longer improves the estimation of the underlying distribution or the accuracy of the results. \new{This reframing is particularly well suited to static circuits, where the target distribution remains stationary across executions.}

Building on this principle, we introduce a general, iterative framework -- named \texttt{IncrementalExecution} -- which iteratively performs a limited number of shots until the point of diminishing returns is reached. The framework supports a variety of \emph{execution policies} designed to detect and exploit the point of diminishing returns efficiently while enabling customisation and adaptation to different use cases and optimisation goals \new{within static execution scenarios}. These policies govern the shot allocation process in an online, dynamic fashion, enabling the framework to adaptively decide when to stop execution.

In summary, the main contributions of our work are:
\begin{enumerate}
    \item We formally define the \textbf{black-box shot optimisation problem} \new{for static, non-variational quantum circuits}, where the goal is to determine the minimal number of quantum circuit executions (shots) required to meet a target accuracy, without making assumptions about the circuit structure or QPU noise model.
    
    \item We introduce a notion of \textbf{\textit{a posteriori} optimality}, which determines the optimal number of shots \textit{after} execution by identifying the minimal number required to achieve the point of diminishing returns for a specific static execution, assuming full access to the output samples.
    
    \item We design and implement the \textsc{IncrementalExecution} framework, an online, iterative approach that leverages the point of diminishing returns to stop execution as soon as sufficient statistical confidence is reached \new{for static circuits}. The framework directly addresses the black-box shot optimisation problem within this scope.
    
    \item We define a suite of flexible \textbf{execution policies} that operationalise the framework and allow adaptation to various practical requirements and optimisation goals, enabling a controlled trade-off between cost and accuracy \new{in static execution contexts}.
    
    \item We propose and release a new comprehensive \textbf{benchmark dataset}~\cite{incexcdataset,qsimbench} that includes the iterative, \textit{shot-by-shot} execution of various \new{static, non-variational} quantum circuits on different quantum hardware backends, enabling reproducible evaluation of shot-optimisation techniques.
\end{enumerate}

Our experimental evaluation indicates that the \textsc{IncrementalExecution} Framework effectively and efficiently approximates the optimal number of shots in black-box settings \new{for static circuits}. Furthermore, we show how suitable execution policies allow users to explicitly control and balance execution costs and statistical accuracy.

To the best of our knowledge, this work is the first to address the problem of shot-count optimisation using an online strategy in a \new{black-box setting for static, non-variational quantum circuits}, without relying on assumptions about circuit structure or noise models.

The remainder of this paper is organised as follows. Section~\ref{sec:related} reviews related literature. Section~\ref{sec:problem} formally defines the shot optimisation problem. Section~\ref{sec:framework} introduces the \textsc{IncrementalExecution} Framework. Section~\ref{sec:policies} describes execution policies. Section~\ref{sec:exp} presents an empirical evaluation. Section~\ref{sec:applicability} discusses applicability. Section~\ref{sec:threats} addresses threats to validity, and Section~\ref{sec:conclusions} concludes.

%% file: src/related.tex
\section{Related Work}
\label{sec:related}

Optimising the number of shots in quantum computations is a critical research challenge, particularly for NISQ devices, where execution cost, latency, and hardware noise significantly affect result quality. Existing approaches to shot optimisation can be broadly categorised into four main classes: \new{(1) analytical noise- or concentration-based estimation methods, (2) adaptive shot allocation techniques for variational and other adaptive quantum algorithms, (3) learning-based strategies, and (4) online black-box execution control.} In the following, we review these categories and position our work within this landscape.

\subsection{Analytical Noise- and Concentration-Based Estimation}

Analytical approaches estimate the required number of shots by deriving error bounds under explicit statistical or noise-related assumptions.

Seksaria and Prabhakar~\cite{seksaria2025shots} propose a systematic framework that incorporates detailed hardware noise models, including state preparation and measurement (SPAM) errors, amplitude damping ($T_1$), phase damping ($T_2$), and gate errors. Their approach leverages the Central Limit Theorem to derive closed-form expressions for measurement variance. While effective, its accuracy critically depends on precise noise characterisation and has primarily been validated on VQE workloads~\cite{tilly2022variational}.

\new{A complementary line of work relies on concentration inequalities to estimate the number of shots required to bound statistical error with high probability~\cite{mayer2023estimating,veltheim2025optimizing,gresch2025reducing}. Hoeffding-style inequalities~\cite{hoeffding1963probability} are commonly used to derive worst-case bounds on the estimation error of expectation values or probabilities under the assumption of independent and identically distributed samples. Weissman-type inequalities~\cite{weissman2003inequalities,mardia2018concentration} further extend this analysis by providing finite-sample guarantees on the $\ell_1$ distance between the empirical and true output distributions. While these bounds offer strong theoretical guarantees, they typically lead to conservative shot estimates and require users to predefine confidence levels and error thresholds \emph{a priori}, limiting their practical applicability in quantum execution settings~\cite{conc_hoef_ineq,miranskyy2025cost}.}

\subsection{Adaptive Shot Allocation in Variational and Adaptive Quantum Algorithms}

The most extensively explored class of shot optimisation techniques targets variational and other adaptive quantum algorithms, where shot allocation is dynamically adjusted across optimisation iterations. In these settings, the circuit structure or parameters evolve over time, inducing non-stationary output distributions.

\paragraph{Variance-Preserved Shot Reduction (VPSR)}
Zhu \emph{et al.}~\cite{zhu2024optimizing} propose the VPSR approach, which dynamically reallocates measurement shots based on real-time variance estimates of Hamiltonian terms grouped into commuting cliques. Their method achieves substantial shot reductions in molecular ground-state simulations (e.g., $H_2$ and $LiH$), but relies on explicit Hamiltonian structure and is inherently tied to VQE-style workflows.

\paragraph{Coupled Adaptive Shot Strategies}
Gu \emph{et al.}~\cite{gu2021adaptive} introduce global Coupled Adaptive Number of Shots (gCANS), optimising shot allocation by maximising expected utility derived from gradient magnitudes at each iteration. Kübler \emph{et al.}~\cite{kubler2020adaptive} refine this approach with individual CANS (iCANS), allocating shots at the level of individual gradient components. Both strategies assume access to gradient information and are designed for iterative parameter optimisation.

\paragraph{Distribution-Adaptive Dynamic Shot Allocation (DDS)}
Kim \emph{et al.}~\cite{kim2024distribution} propose DDS, which links the required number of shots to the entropy of the circuit output distribution and dynamically adapts shot counts across optimisation iterations. While effective in reducing average shot counts, DDS assumes an adaptive execution model where circuit parameters change over time.

\paragraph{Shot Optimisation in Quantum Machine Learning}
Phalak and Ghosh~\cite{phalak2023shot} explore adaptive shot scheduling in quantum machine learning by varying shot counts across training epochs using linear and step-based schedules. Their approach achieves large shot reductions with minimal accuracy degradation, but relies on supervised learning signals and task-specific performance metrics.

\paragraph{Estimator--Optimiser Framework}
Kahani and Nobakhti~\cite{kahani2023novel} propose a modular estimator–optimiser framework for VQAs, explicitly separating observable estimation from parameter optimisation. Their approach dynamically controls estimation error through sensitivity analysis, but assumes iterative circuit updates and estimator-specific tuning.

\subsection{Learning-Based Methods}

Learning-based strategies use machine learning techniques, most notably reinforcement learning (RL), to infer shot allocation policies from execution data. Liang \emph{et al.}~\cite{liang2024artificial} propose an RL-based framework that learns shot allocation decisions during VQE optimisation. While reducing reliance on handcrafted heuristics, this approach is tightly coupled to variational workflows and requires training over representative optimisation trajectories.

\subsection{\new{Positioning of This Work}}

Most existing shot optimisation strategies rely on explicit assumptions about circuit structure, noise characteristics, access to gradient information, or adaptive execution semantics. While effective within their intended domains, these approaches are not directly applicable to \new{static, non-variational quantum circuit executions}, where the primary challenge lies in efficiently estimating a fixed but unknown output distribution under hardware noise.

In contrast, our proposed \textsc{IncrementalExecution} framework adopts an \new{online, black-box execution control} perspective specifically tailored to static circuits. Rather than estimating accuracy \emph{a priori} or relying on algorithm-specific signals, the framework dynamically detects the \emph{point of diminishing returns}—the point at which additional shots no longer induce statistically meaningful changes in the empirical output distribution—and halts execution accordingly.

This enables principled and assumption-light shot optimisation for static workloads, complementing existing analytical, adaptive, and learning-based techniques, and addressing a distinct and practically relevant gap in the literature.

Table~\ref{tab:related-work} summarises the reviewed methods, highlighting their optimisation metrics, algorithmic contexts, and underlying assumptions.

\begin{table}[ht]
\centering
\caption{Comparison of existing shot optimisation strategies.}
\label{tab:related-work}
\begin{tabular}{p{3.4cm}p{3.1cm}p{3.0cm}p{3.3cm}}
\toprule
\textbf{Work} & \textbf{Optimisation Metric} & \textbf{Algorithmic Context} & \textbf{Underlying Assumptions} \\
\midrule
\multicolumn{4}{l}{\textbf{Analytical and Concentration-Based Methods}} \\
\midrule
Seksaria \& Prabhakar~\cite{seksaria2025shots} & Variance & Static / VQA & Explicit noise model (SPAM, $T_1$, $T_2$, gate errors) \\
\midrule
\new{Hoeffding \new{Inequality~\cite{hoeffding1963probability}}} & \new{Error bound} & \new{Static, non-variational} & \new{i.i.d.\ samples, predefined confidence level} \\
\midrule
\new{Weissman Inequality~\cite{weissman2003inequalities}} & \new{$\ell_1$ distance bound} & \new{Static, non-variational} & \new{Finite-sample concentration, conservative bounds} \\
\midrule
\multicolumn{4}{l}{\textbf{Adaptive Shot Allocation in Variational Algorithms}} \\
\midrule
Zhu et al. (VPSR)~\cite{zhu2024optimizing} & Variance & VQE & Hamiltonian structure, commuting observables \\
\midrule
Gu et al. (gCANS)~\cite{gu2021adaptive} & Utility per iteration & VQA & Access to cost gradients \\
\midrule
Kübler et al. (iCANS)~\cite{kubler2020adaptive} & Gradient variance & VQA & Partial derivatives available \\
\midrule
Kim et al. (DDS)~\cite{kim2024distribution} & Entropy & VQA & Iterative parameter updates \\
\midrule
Phalak \& Ghosh~\cite{phalak2023shot} & Accuracy & QML & Supervised learning signal \\
\midrule
Kahani \& Nobakhti~\cite{kahani2023novel} & Estimator sensitivity & VQA & Iterative optimisation loop \\
\midrule
\multicolumn{4}{l}{\textbf{Learning-Based Methods}} \\
\midrule
Liang et al.~\cite{liang2024artificial} & Convergence performance & VQE & Trained RL policy \\
\midrule
\multicolumn{4}{l}{\textbf{Online, Black-Box Execution Control}} \\
\midrule
\textit{\textbf{This Work}} & \textit{\textbf{Point of Diminishing Returns}} & \textit{\textbf{Static, Non-Variational}} & \textit{\textbf{Stationary distribution, black-box access only}} \\
\bottomrule
\end{tabular}
\end{table}

%% file: src/problem.tex
\section{The Black-box Sample Optimisation Problem}
\label{sec:problem}

\new{Quantum circuit execution is inherently stochastic: each execution produces a measurement outcome sampled from an unknown output distribution. Accurately estimating this distribution requires repeated sampling---or \emph{shots}---yet each additional shot incurs cost and latency, especially on real quantum hardware. In this paper, we study the fundamental question of how to determine, as efficiently as possible, how many shots are sufficient for a given execution, i.e., when additional shots yield diminishing returns.}

\new{Importantly, we restrict attention to static, non-variational circuits: the circuit structure and parameters remain fixed across all shots of the execution, so that the effective output distribution can be treated as \emph{stationary} during that execution. Variational and other adaptive workloads, in which circuits change across optimisation iterations and induce non-stationary output distributions, require different stopping criteria and statistical guarantees and are outside our scope.}

\new{This section formalises the problem of sample optimisation in a black-box setting and then instantiates it for static quantum circuit execution. Specifically:}
\begin{itemize}
    \item \new{We first define the general problem of estimating an unknown discrete distribution under a hard sampling budget, and we introduce an \emph{a posteriori} procedure to compute the minimal sufficient number of samples from a complete execution trace.}
    \item \new{We then define an online, observable proxy for convergence---the \emph{point of diminishing returns}---that can be detected without access to the true distribution.}
    \item \new{Finally, we instantiate these notions in the quantum setting for static circuits executed on noisy quantum backends.}
\end{itemize}

\subsection{General Formulation}
\label{sec:problem:general}

\new{Consider a black-box stochastic process that produces outcomes in a finite discrete space $\mathcal{X}$ according to an unknown target distribution $P$. Each draw yields one sample $x \in \mathcal{X}$; repeating the process $n$ times yields the empirical distribution $\hat{P}_n$. Throughout this section we assume that, for the duration of an execution, samples are generated from a stationary distribution (i.e., $P$ does not change with $n$).}

\rev{This formulation is intentionally domain-independent. The only assumptions are that outcomes lie in a finite discrete space, repeated samples are drawn from a stationary distribution during the execution, and each sample has non-negligible cost. Quantum circuit execution is introduced in Section 3.3 as one concrete instantiation, where the sampling oracle is a circuit–backend pair and samples are measurement shots.}

A distance or divergence metric $\mathcal{D}(\cdot,\cdot)$ is used to quantify the difference between two probability distributions over $\mathcal{X}$. Given a threshold $\delta > 0$ representing the maximum acceptable estimation error and a hard budget $\mathcal{B}$ on the maximum number of available samples, we define the following problem.

\begin{definition}[\textbf{\new{The Black-box Sample Optimisation Problem}}]
\label{def:problem}
Given an unknown distribution $P$ over a discrete space $\mathcal{X}$,
a divergence metric $\mathcal{D}$, a maximum estimation error threshold
$\delta$, and a budget $\mathcal{B} \in \mathbb{N}$, find the smallest
positive integer $n^* \leq \mathcal{B}$ such that
\[
n^* = \min \left\{ \left\{ n \leq \mathcal{B} \mid
\mathcal{D}(\hat{P}_m, P) \leq \delta,\ \forall m \ge n \right\}
\cup \left\{ \mathcal{B} \right\} \right\},
\]
where $\hat{P}_m$ denotes the empirical distribution based on the first
$m$ samples.
\end{definition}

\new{
To justify the previous definition (and the subsequent ones), we
observe that the divergence between the empirical distribution
$\hat{P}_n$ and the target distribution $P$ converges almost surely
to $0$. Let $\mathcal{X}$ be a finite discrete outcome space,
$\rho$ a fixed quantum state on a finite-dimensional Hilbert space
$\mathcal{H}$, and $\mathcal{E}=\{E_x\}_{x\in\mathcal{X}}$ a POVM.
By the Born rule, $P(x)=\mathrm{Tr}(\rho E_x)$ defines a probability
distribution on $\mathcal{X}$. Assume that repeated executions of
the experiment produce i.i.d.\ outcomes $(X_t)_{t\ge1}$ with
distribution $P$, and then
\[
\hat P_n(x)=\frac{1}{n}\sum_{t=1}^n \mathbf{1}\{X_t=x\}
\]
as the empirical distribution after $n$ shots. By the Strong Law of
Large Numbers, $\hat P_n(x)\xrightarrow{\text{a.s.}} P(x)$ for each
$x\in\mathcal{X}$ as $n\to\infty$. Since $\mathcal{X}$ is finite,
this implies almost-sure convergence of the full vector $\hat P_n$
to $P$ in any equivalent norm on $\mathbb{R}^{|\mathcal{X}|}$, and
consequently,
\[
\mathcal{D}(\hat P_n,P)\xrightarrow{\text{a.s.}}0
\qquad \text{as } n\to\infty.
\]
The previous formulation captures the ideal objective: determine the smallest $n$ such that the empirical distribution is within $\delta$ of the (unknown) target distribution $P$. In many practical settings, including quantum circuit execution on real devices, $P$ cannot be accessed directly; consequently, this ideal definition cannot be evaluated online or even retrospectively without additional assumptions.}

\new{To obtain a computable notion of optimality under a fixed sampling budget, we adopt a standard practical reference: the empirical distribution obtained using the full budget, $\hat{P}_\mathcal{B}$. While $\hat{P}_\mathcal{B}$ remains an approximation of $P$, it is the best estimate available under the imposed budget and therefore provides a principled baseline for assessing whether earlier prefixes of the execution were already sufficient. We define the resulting \emph{a posteriori} optimality criterion as follows.}

\begin{definition}[\textbf{\new{The A Posteriori Black-box Sample Optimisation Problem}}]
\label{def:optimal-shots}
Given an execution history of subsequent $\mathcal{B}$ samples $\{x_1, \dots, x_\mathcal{B}\}$ drawn from an unknown distribution $P$, a divergence metric $\mathcal{D}$, and a target threshold $\delta > 0$, find the \emph{optimal number of samples} $n^* \leq \mathcal{B}$ such that
\[
n^* = \min \left\{ n \leq \mathcal{B} \mid \mathcal{D}(\hat{P}_m, \hat{P}_\mathcal{B}) \leq \delta, \forall m\geq n \right\},
\]
where $\hat{P}_m$ is the empirical distribution over the first $m$ samples and $\hat{P}_\mathcal{B}$ is the empirical distribution over all $\mathcal{B}$ samples.
\end{definition}

\new{This notion of optimality is \emph{a posteriori} because it requires access to the complete execution trace in order to compute $\hat{P}_\mathcal{B}$. It yields a computable and execution-specific ``grounded'' convergence point: the earliest prefix whose empirical distribution is within $\delta$ of the best estimate obtainable under the budget. In this sense, $n^*$ captures the \emph{point of diminishing returns} for that execution. Since $\mathcal{D}(\hat{P}_\mathcal{B}, \hat{P}_\mathcal{B})=0$, the worst case is $n^*=\mathcal{B}$.} \new{It is important to emphasise that this a posteriori notion of optimality is not intended to be used during execution. Rather, it provides a principled retrospective baseline that can only be computed after observing the full execution history. Its purpose is to quantify how early convergence could have been achieved and to serve as a reference for evaluating online strategies that do not have access to the full budget distribution.}

\paragraph{\new{A Posteriori Optimal Sample Count Algorithm.}}
\new{We next provide a simple procedure for computing $n^*$ from a complete execution history. While it is not usable online, it serves as a rigorous retrospective baseline in our evaluation.}

\begin{algorithm}[H]
\caption{A Posteriori Optimal Sample Count Computation (Suffix-Stable)}
\label{alg:aposteriori}
\KwIn{Measurement outcomes $\{x_1, x_2, \dots, x_\mathcal{B}\}$, threshold $\delta$, divergence metric $\mathcal{D}$}
\KwOut{Optimal number of samples $n^* \leq \mathcal{B}$}

// Step 1: Build the reference distribution $\hat{P}_\mathcal{B}$
\textbf{initialize} $\texttt{count\_B[$k$]} \gets 0$ for all outcomes $k$\;

\For{$i \gets 1$ \KwTo $\mathcal{B}$}{
    \texttt{count\_B[$x_i$]} $\gets$ \texttt{count\_B[$x_i$]} $+ 1$\;
}
\ForEach{key $k$ in \texttt{count\_B}}{
    $\hat{P}_\mathcal{B}[k] \gets$ \texttt{count\_B[$k$]} $/ \mathcal{B}$\;
}

// Step 2: Compute divergences $d[m] = \mathcal{D}(\hat{P}_m,\hat{P}_\mathcal{B})$ for all $m$
\textbf{initialize} $\texttt{count\_m[$k$]} \gets 0$ for all outcomes $k$\;
\texttt{total} $\gets 0$\;

\For{$m \gets 1$ \KwTo $\mathcal{B}$}{
    \texttt{count\_m[$x_m$]} $\gets$ \texttt{count\_m[$x_m$]} $+ 1$\;
    \texttt{total} $\gets$ \texttt{total} $+ 1$\;

    \ForEach{key $k$ in \texttt{count\_B}}{
        $f \gets \texttt{count\_m.get($k$, 0)}$\;
        $\hat{P}_m[k] \gets f / \texttt{total}$\;
    }

    $d[m] \gets \mathcal{D}(\hat{P}_m, \hat{P}_\mathcal{B})$\;
}

// Step 3: Find the earliest $n$ such that $\max_{m \ge n} d[m] \le \delta$
\texttt{suffix\_max} $\gets 0$\;
$n^* \gets \mathcal{B}$\;

\For{$m \gets \mathcal{B}$ \KwTo $1$}{
    \texttt{suffix\_max} $\gets \max(\texttt{suffix\_max}, d[m])$\;
    \If{\texttt{suffix\_max} $\le \delta$}{
        $n^* \gets m$\;
    }
}

\Return $n^*$\;
\end{algorithm}

\new{Algorithm~\ref{alg:aposteriori} returns the smallest prefix length $n^*$ such that \new{$\hat{P}_m$ is within $\delta$ of the full-budget empirical distribution $\hat{P}_\mathcal{B}$ for all $m\geq n^*$}. This procedure is not usable online, since $\hat{P}_\mathcal{B}$ is only available after consuming the full budget. Nevertheless, it provides (i) a concrete, computable baseline for evaluation and (ii) an operational interpretation of ``diminishing returns'' under a hard sampling budget.}

\subsection{The Point of Diminishing Returns as a Practical Approximation}
\label{sec:info-conv}

\new{To enable \emph{online} decision making without access to $P$ (and without requiring the full trace needed for $\hat{P}_\mathcal{B}$), we adopt an observable convergence proxy: the \emph{point of diminishing returns}. The core idea is to monitor how much the empirical distribution changes as additional samples are collected. When these changes become sufficiently small, the empirical distribution can be treated as stable for practical purposes, and further sampling is unlikely to yield meaningful improvements.}

\begin{definition}[\textbf{\new{Approximate Point of Diminishing Returns}}]
\label{def:info-conv}
Let $\hat{P}_n$ be the cumulative empirical distribution after $n \leq \mathcal{B}$ samples, and let $ \mathcal{D}(\cdot,\cdot)$ be a divergence metric over distributions. Given a look-back window $\tau$, the \textit{approximate} point of diminishing returns is said to occur at step $n$ if $n$ is the smallest value such that:
\[
\mathcal{D}\left( \hat{P}_n, \hat{P}_{n - \tau} \right) \leq \varepsilon
\]
where $\varepsilon > 0$ is a tolerance parameter controlling sensitivity to change.
\end{definition}

\new{Unlike $\delta$ in Definition~\ref{def:optimal-shots}, which is defined with respect to the full-budget reference $\hat{P}_\mathcal{B}$, $\varepsilon$ is an online tolerance that controls sensitivity to distributional changes observed during execution. In our \textsc{IncrementalExecution} framework, $\varepsilon$ is a configurable parameter.}

\new{While this criterion does not guarantee closeness to the unknown true distribution \(P\), it yields a practical, model-agnostic stopping signal. Moreover, the criterion is theoretically well-motivated in the sense that, under mild regularity assumptions, the divergence between two empirical distributions separated by a fixed lag \(\tau\) must vanish asymptotically. In particular, if the samples are i.i.d. (\textit{independent and identically distributed}) \ from \(P\) and \(\mathcal{D}\) is a divergence (or metric) that is continuous with respect to the convergence mode induced by \(\hat P_n\) (e.g., total variation, \(\ell_1\), \(\ell_2\), or standard \(f\)-divergences on finite support), then}
\[
\mathcal{D}(\hat{P}_{n}, \hat{P}_{n-\tau})
\xrightarrow{\text{a.s.}} 0
\qquad \text{as } n \to \infty.
\]
\new{A direct way to see this is to note that by the Strong Law of Large Numbers (SLLN), as we already justified, for any fixed \(x\in\mathcal{X}\),}
\[
\hat{P}_n(x) \xrightarrow{\text{a.s.}} P(x).
\]
\new{Hence, for each \(x\), both sequences \(\hat{P}_n(x)\) and \(\hat{P}_{n-\tau}(x)\) converge almost surely to the same limit \(P(x)\), implying \(\hat{P}_n(x)-\hat{P}_{n-\tau}(x)\to 0\) a.s. Coordinate-wise convergence, together with continuity of \(\mathcal{D}\), yields \(\mathcal{D}(\hat{P}_{n}, \hat{P}_{n-\tau})\to 0\) almost surely. Operationally, this means that for any fixed tolerance \(\varepsilon>0\), there exists (almost surely) a finite step \(n_\varepsilon\) after which the lag-\(\tau\) empirical divergence remains below \(\varepsilon\), providing a principled basis for using Definition~\ref{def:info-conv} as an online proxy for stabilization.}

\new{Importantly, the asymptotic guarantee above should be interpreted only as a consistency property of the proxy, not as a certificate of small error with respect to \(P\) at finite \(n\). In finite-budget regimes, the choice of \(\tau\) and \(\varepsilon\) controls a bias--variance trade-off: larger \(\tau\) tends to suppress spurious fluctuations but may delay detection, whereas smaller \(\varepsilon\) enforces stricter stabilization at the cost of longer sampling.}
\new{In Section~\ref{sec:exp}, we empirically show that this approximate point of diminishing returns provides an effective online proxy for the a posteriori optimality criterion in Definition~\ref{def:optimal-shots}.}



\subsection{The Quantum Black-box Shot Optimisation Problem}
\label{sec:quantum-problem}

\new{The Black-box Sample Optimisation Problem naturally instantiates in quantum computing~\cite{zhang2025empirical}. For a fixed (static) quantum circuit $C$ executed on a noisy backend $Q$, each shot yields a single measurement outcome and can be viewed as a sample from an unknown, effective output distribution $P(C,Q)$ induced by the circuit-backend pair. Repeating the execution $n$ times yields an empirical distribution $\hat{P}_n(C,Q)$,
and the shot budget $\mathcal{B}$ bounds the maximum number of executable shots.}

\new{Within our scope, we assume static circuit execution: the circuit structure and parameters do not change across shots, and the effective output distribution is treated as stationary over the duration of the execution. Under these assumptions, the goal is to determine the smallest $n$ such that the resulting empirical distribution is sufficiently stable (or sufficiently close to the best available reference under the budget).}

\new{Consequently, we define the quantum instance of the a posteriori black-box problem as follows.}

\begin{definition}[\textbf{The Quantum Black-box Shot Optimisation Problem Circuits}]
\label{def:quantum-problem}
Given a \new{static} quantum circuit $C$ and a noisy quantum processing unit $Q$, and a maximum shot budget $\mathcal{B} \in \mathbb{N}$, determine the smallest natural number $n^* \leq \mathcal{B}$ such that:
\[
n^* = \min \left\{ n \leq \mathcal{B} \mid \mathcal{D}(\hat{P}_m(C,Q), \hat{P}_\mathcal{B}(C,Q)) \leq \delta, \forall m \geq n \right\},
\]
where $\hat{P}_m(C, Q)$ denotes the empirical distribution obtained by performing $m$ shots of $C$ on $Q$; $\hat{P}_\mathcal{B}(C, Q)$ denotes the empirical distribution obtained using the full budget $\mathcal{B}$; and $\delta > 0$ is the tolerance threshold below which $\hat{P}_n(C, Q)$ is considered a sufficiently accurate approximation of $\hat{P}_\mathcal{B}(C, Q)$.
\end{definition}

\new{Since $\hat{P}_\mathcal{B}(C, Q)$ is the best estimate attainable under the budget, identifying an optimal $n^* \leq \mathcal{B}$ can save $\mathcal{B}-n^*$ shots while preserving essentially the same empirical distributional estimate achievable at full budget.}

\new{From this point forward, we instantiate the divergence metric $\mathcal{D}$ using three standard distributional distances/divergences: Total Variation Distance (TVD), Hellinger distance, and Jensen--Shannon (JS) divergence. We primarily report results with TVD for its simplicity and direct interpretability, and we use Hellinger and JS to validate our conclusions.}

\paragraph{Total Variation Distance (TVD)}
Given two discrete probability distributions $P$ and $P'$ over the same outcome space $\mathcal{X}$, the Total Variation Distance is defined as:
\[
\mathrm{TVD}(P, P') = \frac{1}{2} \sum_{x \in \mathcal{X}} \left| P(x) - P'(x) \right|.
\]
\new{TVD measures the maximal discrepancy in assigned probability mass across events and ranges from 0 (identical distributions) to 1 (disjoint support).}

\paragraph{\new{Hellinger distance.}}
\new{The (squared) Hellinger distance between $P$ and $P'$ is:}
\[
H^2(P,P') = \frac{1}{2}\sum_{x\in\mathcal{X}}\left(\sqrt{P(x)}-\sqrt{P'(x)}\right)^2,
\]
\new{and $H(P,P')=\sqrt{H^2(P,P')}$.}
\new{Hellinger distance is symmetric, bounded in $[0,1]$, and behaves smoothly for small probabilities, which is useful for sparse empirical distributions.}

\paragraph{\new{Jensen--Shannon (JS) divergence.}}
\new{Let $M = \tfrac{1}{2}(P + P')$. The Jensen--Shannon divergence is defined as:}
\[
\mathrm{JS}(P \parallel P') = \frac{1}{2}\mathrm{KL}(P \parallel M) + \frac{1}{2}\mathrm{KL}(P' \parallel M),
\]
\new{where $\mathrm{KL}(\cdot\parallel\cdot)$ denotes the Kullback--Leibler divergence.}
\new{JS is symmetric and always finite; when using $\log_2$, it is bounded in $[0,1]$~\cite{EndresSchindelin2003}.}

\new{While TVD, Hellinger distance, and JS divergence quantify distributional discrepancy in different ways, they are tightly connected and often interchangeable up to constants for the purpose of detecting stabilisation.
First, Hellinger controls TVD (and vice versa up to constants): under standard normalisations,}
\begin{equation}
H^2(P,Q) \;\le\; \mathrm{TVD}(P,Q) \;\le\; \sqrt{2}\,H(P,Q),
\label{eq:hellinger-tvd}
\end{equation}
\new{so convergence in Hellinger is equivalent to convergence in TVD (with rescaled thresholds).
Second, JS is a symmetric, bounded information divergence, and its square root is a metric (often called the Jensen--Shannon distance).
Moreover, JS can be bounded above and below by (squared) variation distance for categorical distributions; in particular, for the equal-weight case and common support, Corander \emph{et al.}~\cite{CoranderRemesKoski2021} show that JS admits two-sided bounds in terms of the squared (total) variation distance, implying that for small discrepancies,}
\begin{equation}
\mathrm{JS}(P \parallel Q) = \Theta\!\left(\mathrm{TVD}(P,Q)^2\right),
\label{eq:js-tvd-quadratic}
\end{equation}
\new{and providing explicit constants (and a sharper logarithmic lower bound) under mild conditions.
Practically, this means TVD reacts \emph{linearly} to probability-mass shifts, while Hellinger and JS are typically \emph{smoother} and often behave quadratically near convergence, which can reduce sensitivity to small residual fluctuations when used as online stopping signals.}

\begin{figure*}[t]
\centering
\subfigure[TVD to ideal distribution as a function of shot count.]{\includegraphics[width=0.49\textwidth]{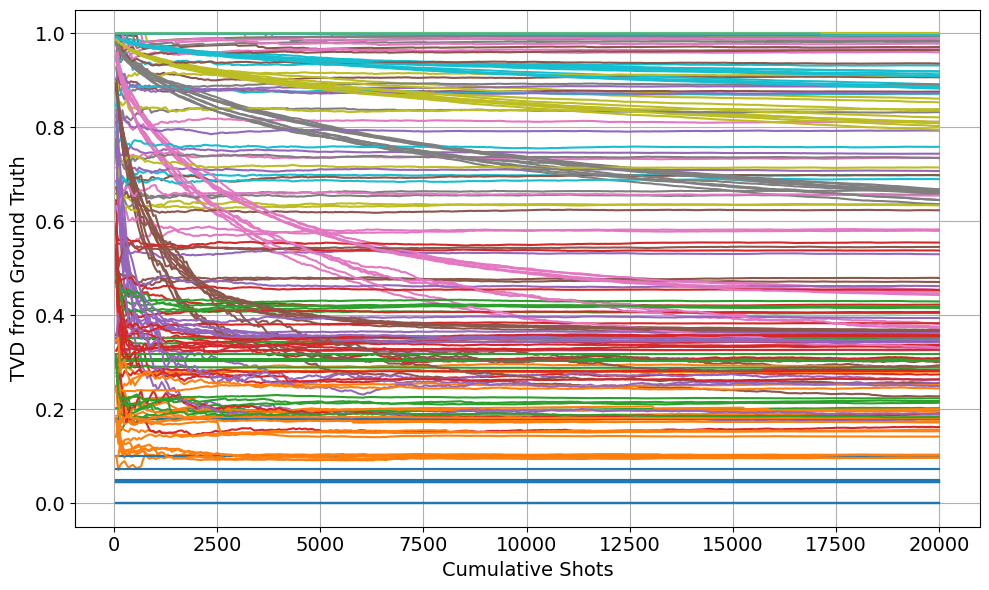}}
\hfill
\subfigure[Change in empirical distribution between successive windows ($\tau=50$).]{\includegraphics[width=0.49\textwidth]{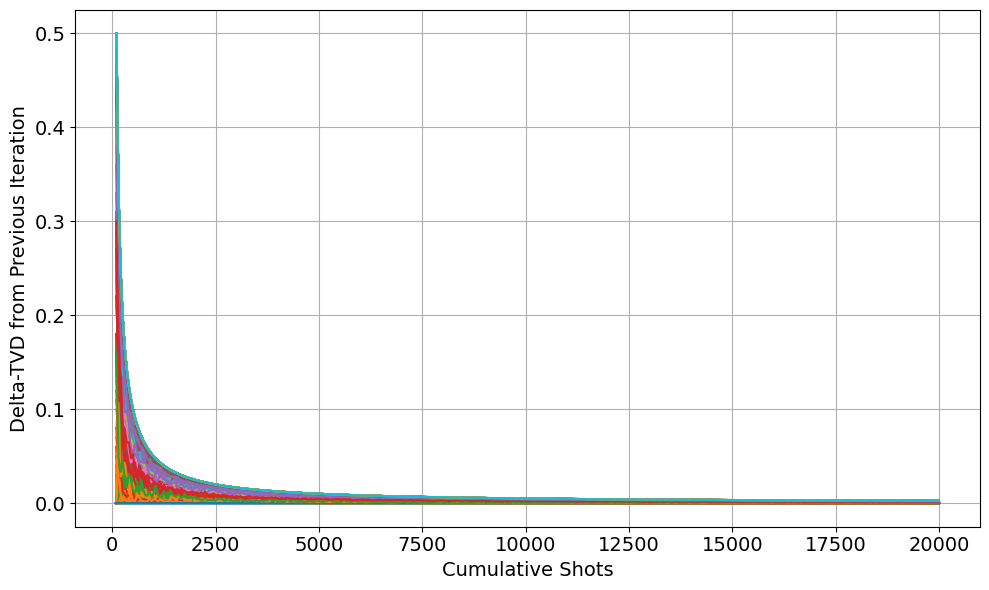}}
\caption{\new{Empirical behaviour of static quantum circuit output distributions as the shot count increases. Each line corresponds to a circuit--backend pair.}}
\Description{Two line plots. The first shows the Total Variation Distance (TVD) between the empirical and ideal distributions as a function of the number of shots for multiple circuit-backend pairs; TVD decreases and stabilises over time. The second shows the divergence between consecutive empirical distributions (Delta-TVD) over a sliding window of size 50; all lines show a rapid decay and plateau, indicating convergence.}
\label{fig:tvd}
\end{figure*}

\subsection*{\new{Preliminary Study}}
\new{To assess the feasibility of our approach in the static setting, we conducted a preliminary study to evaluate whether empirical output distributions exhibit stabilisation as the number of shots increases, and whether this behaviour can be captured by Definition~\ref{def:info-conv}.}

Figure~\ref{fig:tvd} presents empirical evidence of these dynamics. Each line corresponds to a unique pair consisting of a quantum circuit and a noisy quantum backend. We considered 28 different circuits with qubit counts ranging from 4 to 16 and executed them on five IBM noisy simulators. For each circuit-backend pair, we ran the circuit in batches of $\tau = 50$ shots, up to a total of $\mathcal{B}=20{,}000$ samples, recording the empirical distribution $\hat{P}_n$ after each batch.

\new{In subfigure~(a), we compute the TVD between each empirical distribution $\hat{P}_n(C,Q)$ and the corresponding ideal distribution $P_n(C,Q)$, 
obtained by executing the same circuit $C$ on a noiseless simulator $Q$. With few shots, the TVD is unstable due to high sampling variance. As $n$ increases, the empirical distribution concentrates and TVD decreases, reflecting improved estimation. Eventually, the curves plateau: additional shots provide only marginal gains, as residual discrepancies are increasingly dominated by systematic effects (e.g., device noise) rather than sampling noise.}

\new{Subfigure~(b) shows \emph{Delta-TVD}, i.e., the divergence between empirical distributions separated by a fixed look-back window $\tau=50$. The rapid decay followed by a plateau indicates that the empirical distribution stabilises, consistent with the emergence of a point of diminishing returns as defined in Section~\ref{sec:info-conv}.}

\paragraph{\new{Advantages in Quantum Contexts.}}
\new{The approximate point of diminishing returns provides a practical and flexible convergence signal for static quantum circuit execution:}
\begin{itemize}
\item \new{It does not require access to circuit internals, an explicit noise model, or the ideal distribution $P(C,Q)$,
making it applicable across hardware backends and circuit families within the static setting.}
\item \new{It can be computed online from observed outcomes, without requiring the full trace needed to compute $\hat{P}_\mathcal{B}(C,Q)$.}
\item \new{It naturally adapts to circuit- and backend-specific behaviour, enabling shot allocation that reflects empirical statistical complexity.}
\item \new{It is metric-agnostic and, in our evaluation, we instantiate it with TVD, Hellinger distance, and Jensen-Shannon divergence to assess robustness to the choice of $\mathcal{D}$.}
\end{itemize}

\new{In summary, static quantum circuit execution provides a concrete and practically relevant instantiation of the Black-box Sample Optimisation Problem. The point of diminishing returns yields an online stopping signal that underpins our \textsc{IncrementalExecution} framework, introduced next as a fully online procedure for black-box shot optimisation in static quantum experiments.}

%% file: src/framework.tex
\section{The \textsc{Incremental Execution} Framework}
\label{sec:framework}

Building on the problem formulation and a posteriori analysis of Section~\ref{sec:problem}, we now introduce the \textsc{IncrementalExecution} Framework: a practical and online method for dynamically approximating the optimal number of shots required to estimate the output distribution of a quantum circuit with sufficient \textit{accuracy}\footnote{Throughout this work, unless explicitly stated otherwise, we define accuracy in terms of closeness to the best empirically attainable distribution under a fixed budget of quantum circuit executions. Specifically, unless stated otherwise, we evaluate accuracy with respect to the empirical distribution $\hat{P}_\mathcal{B}$ obtained from all $\mathcal{B}$ available shots on a given noisy QPU, rather than the ideal (noiseless) distribution $P_C$. This choice reflects the fact that $\hat{P}_\mathcal{B}$ is the best estimate of the circuit’s true output distribution that can be practically obtained under budget constraints. In contrast, the ideal distribution $P_C$, derived from a noiseless simulation of the circuit, is typically inaccessible on real quantum hardware due to stochastic noise and uncharacterised error sources. As such, $\hat{P}_\mathcal{B}$ serves as a more realistic and actionable target for optimisation in real-world settings, where noise prevents convergence to the ideal output.}.

\subsection{From A Posteriori Optimality to Online Execution}
\label{sec:framework:idea}

As established in Definition~\ref{def:optimal-shots}, the optimal number of shots $n^*$ is the smallest $n \leq \mathcal{B}$ such that the empirical distribution $\hat{P}_n$ is within distance $\delta$ of the best possible approximation $\hat{P}_\mathcal{B}$ obtained using the full budget. While informative, this a posteriori approach requires all $\mathcal{B}$ samples to be collected upfront, making it unsuitable for online or cost-sensitive execution. \new{Nevertheless, despite its lack of operational applicability, a posteriori optimality serves as a crucial evaluation benchmark in this work. The objective of the \textsc{IncrementalExecution} framework is to approximate this ideal stopping point online, without ever executing the full shot budget.}

In this work, therefore, our goal is to approximate $n^*$ in an \emph{online} setting, where samples are acquired incrementally, and the decision to continue or stop is made dynamically based only on observed data. We reframe the problem as one of approximating the point of diminishing returns -- the point at which additional shots do not meaningfully change the cumulative output distribution.

This leads to the following guiding principle:
\begin{quote}
\textit{Continue performing shots only while they significantly improve the accuracy of estimated distribution. Stop as soon as further sampling yields negligible gain.}
\end{quote}

This philosophy transforms shot execution into a feedback-driven process that adapts online to circuit behaviour and hardware variability.

To operationalise this approach, our design goals are to:
\begin{itemize}
    \item \textbf{Operate under a black-box assumption}, requiring no prior knowledge of the circuit structure or the noise characteristics of the quantum backend.
    \item \textbf{Avoid over-execution}, reducing unnecessary cost while maintaining user-defined accuracy guarantees.
    \item \textbf{Dynamically determine} how many additional shots are required, based solely on the measurement outcomes observed so far.
\end{itemize}

To this end, we introduce a general-purpose, adaptive algorithm grounded in the notion of point of diminishing returns, which serves as a robust and efficient proxy for accuracy in real-world quantum execution environments.


\subsection{The Framework}
\label{sec:framework:alg}

The \textsc{IncrementalExecution} Framework (Algorithm~\ref{alg:incremental}) is structured as an adaptive, feedback-driven loop that incrementally refines the number of shots used to estimate a quantum circuit’s output distribution. Rather than deciding upfront how many shots are needed, the framework proceeds incrementally: it performs a small batch of shots, evaluates whether their outcomes meaningfully improve our knowledge of the circuit’s output distribution (i.e., if it reaches point of diminishing returns), and only continues performing shots if they do. 

\begin{algorithm}[ht]
\caption{\textsc{IncrementalExecution}}
\label{alg:incremental}
\SetKwInOut{Input}{Input}
\SetKwInOut{Output}{Output}

\KwIn{Quantum circuit $C$, backend $\mathcal{Q}$, initial batch size $b_0$}
\KwIn{Distance metric $\mathcal{D}$, convergence threshold $\varepsilon$, max shots $\mathcal{B}$}
\KwIn{Stopping criterion \textsc{StoppingCriterion}, stability criterion \textsc{StabilityCriterion}, allocation policy \textsc{AllocateShots}}
\KwOut{Estimated distribution $\hat{P}$, total shots used $N$}

$\hat{P} \leftarrow \emptyset$\;
$\textit{history} \leftarrow [\,]$, $\textit{stable} \leftarrow 0$\;
$n \leftarrow 0$, $b \leftarrow b_0$\;

\While{$n < \mathcal{B}$}{
    $R \leftarrow \textsc{Run}(C, \mathcal{B}, b)$ \tcp*{Execute next batch of $b$ shots}
    \textbf{append} $R$ to \textit{history}\;
    \textbf{update} $\hat{P}$ with outcomes in $R$ \tcp*{Incremental empirical distribution update}
    $n \leftarrow n + b$\;

    $(\textit{converged}, \textit{info}) \leftarrow \textsc{StoppingCriterion}(\textit{history}, \hat{P}, R)$\;
    \eIf{$\textit{converged}$}{
        $\textit{stable} \leftarrow \textit{stable} + 1$\;
    }{
        $\textit{stable} \leftarrow 0$\;
    }

    $k \leftarrow \textsc{StabilityCriterion}(\textit{history}, \hat{P}, R, \textit{info})$\;
    \If{$\textit{stable} \ge k$}{
        \textbf{break} \tcp*{Point of diminishing returns confirmed}
    }

    $b \leftarrow \textsc{AllocateShots}(\textit{history}, \hat{P}, R, \textit{info})$\;
}
\Return{$\hat{P}, N$}
\end{algorithm}

\paragraph{Initialisation.}  
The algorithm begins by setting up key variables:
\begin{itemize}
    \item $\hat{P} \leftarrow \emptyset$: the cumulative empirical distribution, updated incrementally.
    \item \textit{history}: a sequence storing all observed outcomes.
    \item $\textit{stable}$: a counter for the number of consecutive stable iterations.
    \item $n \leftarrow 0$: the total number of shots executed so far.
    \item $b \leftarrow b_0$: the initial size of the shot batch.
\end{itemize}
These initial values prepare the algorithm to start incremental execution and monitoring.

\paragraph{Main Execution Loop.}  
The algorithm proceeds iteratively until either the point of diminishing returns is detected or the maximum allowed shot budget $\mathcal{B}$ is reached. Each iteration includes the following steps:

\begin{itemize}
    \item \textbf{Batch Execution}:A batch of $b$ shots is executed on the quantum backend $Q$, producing a new set of measurement outcomes $R$.
    \item \textbf{Empirical Update:} The outcomes in $R$ are appended to the \textit{history} and used to update the cumulative distribution $\hat{P}$.
    \item \textbf{Convergence Check:} The function \textsc{StoppingCriterion} is invoked to determine whether the empirical distribution has changed significantly relative to the recent past (i.e., if it has reached the point of diminishing returns). It returns a boolean flag indicating local convergence and auxiliary information such as the computed divergence. If convergence is detected, the stability counter $\textit{stable}$ is incremented; otherwise, it is reset to zero.
    \item \textbf{Stability Enforcement}: To guard against transient fluctuations, the function \textsc{StabilityCriterion} computes how many consecutive stable iterations are required to declare global convergence. If the threshold $k$ is met or exceeded, execution terminates early.
    \item \textbf{Shot Allocation}: If convergence is not yet established, the function \textsc{AllocateShots} determines the size of the next shot batch based on the current history and convergence metadata.
\end{itemize}

\paragraph{Termination and Output.}  
The loop terminates either upon achieving stably the point of diminishing returns or reaching the maximum budget $\mathcal{B}$. The algorithm then returns:
\begin{itemize}
    \item $\hat{P}$: the final cumulative empirical distribution;
    \item $n$: the total number of shots executed.
\end{itemize}
The output reflects the estimated best possible estimation of the circuit's output distribution within the budget and convergence constraints.

\subsection*{Discussion.}  
This adaptive procedure approximates the a posteriori optimal shot count $n^*$ (Definition~\ref{def:optimal-shots}) in an online fashion. As access to the full budget reference distribution $\hat{P}_\mathcal{B}$ is unavailable during execution, the algorithm instead tracks changes in the empirical distribution over time. Convergence is inferred from distributional stability, under the assumption that small incremental changes imply diminishing returns.

A key strength of the framework lies in its modularity. The behaviour of the algorithm is governed by three pluggable policy components:

\begin{description}    
    \item[\textsc{StoppingCriterion} \texttt{(history, $\hat{P}$, R)} $\rightarrow$ \texttt{(bool, info)}:] Determines whether the most recent batch $R$ introduces a statistically significant change in the empirical distribution. It acts as an estimator of the point of diminishing returns. Typically compares the current empirical distribution $\hat{P}_i$ with that of the previous iteration $\hat{P}_{i - 1}$ using a divergence metric $\mathcal{D}$, and returns a boolean flag indicating convergence along with auxiliary diagnostic information.

    \item[\textsc{StabilityCriterion} \texttt{(history, $\hat{P}$, R, info)} $\rightarrow$ \texttt{int}:] Specifies how many consecutive stable iterations are required to confirm convergence. Its purpose is to reduce the possibility of false positives when signalling the point of diminishing returns. It can be constant or adaptively computed based on recent divergence values or system variability.

    \item[\textsc{AllocateShots} \texttt{(history, $\hat{P}$, R, info)} $\rightarrow$ \texttt{int}:] Determines the number of shots in the next batch. Can be fixed, randomised, or dynamically adjusted based on convergence trends or statistical uncertainty.
\end{description}

Each component can be configured independently, enabling diverse optimisation strategies:
\begin{itemize}
    \item \textbf{Dynamic thresholds} for $\varepsilon$ (the \textit{stopping criterion threshold}\footnote{See Definition~\ref{def:info-conv}.}) based on real-time calibration data considered application;
    \item \textbf{Multi-metric convergence}, where stability is defined over multiple divergence metrics;
    \item \textbf{Task-aware objectives}, combining convergence with downstream utility metrics (e.g., optimisation loss or classification accuracy).
\end{itemize}

\subsection{Implementation Overview}
\label{sec:framework:impl}

The \textsc{IncrementalExecution} Framework is implemented as an open-source\footnote{The full, open-source codebase is freely available at: \url{https://github.com/GBisi/quantum-incremental-execution}}, modular, extensible Python class that supports dynamic shot allocation in quantum circuit execution under black-box conditions. It is written in Python 3.12 and designed to be easily integrable with both simulation environments and real quantum backends.

\paragraph{Class Interface.}
The core class is defined as follows:

\begin{center}
\begin{minipage}{0.9\linewidth}
\begin{lstlisting}[language=Python]
class IncrementalExecution:
    def __init__(self,
                 stopping_criterion: Callable,
                 stability_criterion: Callable,
                 allocate_shots: Callable,
                 default_shots: int,
                 max_shots: Optional[int] = None,
                 runner: Optional[Callable] = None,
                 verbose: bool = True)

    def run(self,
            runner: Optional[Callable] = None,
            initial_guess: Optional[Dict[str, int]] = None,
            *args, **kwargs) -> Dict[str, int]

    def __call__(self, func: Callable) -> Callable
\end{lstlisting}
\end{minipage}%
\end{center}

This interface exposes the three key user-defined policy functions (i.e., \textsc{StoppingCriterion}, \textsc{StabilityCriterion}, and \textsc{AllocateShots}), while users may also provide:

\begin{itemize}
    \item \texttt{runner}: a callable that executes a quantum circuit (or synthetic simulation) and returns measurement outcome counts. Its only requirement is to feature a \textsc{shots} parameter.
    \item \texttt{default\_shots}: the number of shots to perform during the first iteration.
    \item \texttt{max\_shots}: maximum shots budget.
    \item \texttt{initial\_guess}: an optional empirical distribution to bootstrap early convergence; useful when are available some information on the "shape" of the target distribution.
\end{itemize}

\paragraph{Usage.}
The framework supports both functional and decorator-based usage. For example:

\begin{center}
\begin{minipage}{0.9\linewidth}
\begin{lstlisting}[language=Python]
# Functional interface
incremental_exec = IncrementalExecution(
    stopping_criterion=stop_crit,
    stability_criterion=stability_crit,
    allocate_shots=shot_policy,
    default_shots=50,
    max_shots=20000
)
result = incremental_exec.run(runner=my_runner)

# Decorator interface
@incremental_exec
def my_runner(*, shots): ...
result = my_runner()
\end{lstlisting}
\end{minipage}%
\end{center}


\subsection{Summary}

In summary, the \textsc{IncrementalExecution} framework provides a principled and flexible solution to the shot optimisation problem in quantum computing. It turns the fixed-budget question \emph{\textquotedblleft How many shots should I run?\textquotedblright} into a dynamic decision problem that adapts to empirical data. Grounded in the concept of \emph{the point of diminishing returns}, the framework provides a lightweight, accurate, and backend-agnostic solution for shot count optimisation.

\begin{enumerate}
  \item It avoids the need to access the true distribution or simulate noise models.
  \item It is robust to backend fluctuations and circuit-specific behaviours.
  \item It supports integration on quantum cloud platforms, such as IBM Quantum Sessions (see Section~\ref{sec:applicability}).
\end{enumerate}

In the following section, we instantiate this framework with concrete implementations of the three decision policies, each with its own way of interpreting the concept of the point of diminishing returns.

%% file: src/policies.tex
\section{Policy Space and Experimental Configurations}
\label{sec:policies}

The flexibility of the \textsc{IncrementalExecution} Framework lies in its policy modularity. This section presents the suite of execution policies implemented and tested in our experiments, encompassing stopping criteria, stability enforcement, and shot allocation strategies. Each policy is parameterised, allowing fine-grained control over the trade-off between convergence speed and estimation accuracy. We also describe the configuration grid used to generate thousands of framework instantiations. Such policies and parameters should be seen as hyperparameters of the framework and may need a fine-tuning phase to reach the framework's best performance on a specific application.

\subsection{Stopping Criterion Policies}
\label{sec:policies:stopping}

Stopping criteria determine whether the empirical distribution has reached the point of diminishing returns --- i.e., whether additional measurements yield negligible information gain. We implemented three families of stopping conditions, each relying on the comparison of the cumulative empirical distribution to past estimates \new{using the divergence metric $\mathcal{D}$ (viz., TVD, Hellinger or J-S):}

\begin{itemize}
    \item \textbf{Delta Distance Criterion (Delta):} Execution halts when the distance between the empirical distributions at the current and previous iterations $i-\tau$, $\hat{P}_i$ and $\hat{P}_{i-\tau}$, falls below a fixed threshold $\varepsilon$:
    \[
    \mathcal{D}(\hat{P}_i, \hat{P}_{i-\tau}) < \varepsilon
    \]
    \textbf{Parameters tested in the experiments:}
    \begin{itemize}
    \item \new{Thresholds $\varepsilon$: \{0.01, 0.025, 0.5, 0.1, 0.25\}}
    \item Look-back window $\tau$: \{1, 2, 3\}
    \end{itemize}

\item \textbf{Delta Moving Average Criterion (DMA):} \new{Instead of requiring individual divergence values to be below the threshold, this criterion compares the moving average of divergence values over the last $w$ iterations to the threshold. Let $d_j = \mathcal{D}(\hat{P}_{i-w+j}, \hat{P}_{i-w+j-\tau})$ for $j = 1, \dots, w$. Then:}
%
%
\[
\text{Avg}_w(d_1, \ldots, d_w) < \varepsilon
\]
where the average can be computed uniformly or as an exponentially weighted moving average (EWMA).

\textbf{Exponentially Weighted Moving Average (EWMA).} Given a sequence $d_1, \ldots, d_w$, the EWMA with smoothing factor $\alpha \in (0,1]$ is computed recursively as:
\[
\begin{cases}
s_1 = d_1 \\
s_t = \alpha d_t + (1 - \alpha) s_{t-1}, & \text{for } t = 2, \ldots, w
\end{cases}
\quad \Rightarrow \quad \text{EWMA}_{\alpha} = s_w
\]

\textbf{Parameters tested in the experiments:}
\begin{itemize}
    \item \new{Thresholds $\varepsilon$: \{0.01, 0.025, 0.5, 0.1, 0.25\}}
    \item Look-back window $\tau$: \{1, 2, 3\}
    \item Window sizes $w$: \{3, 5\}
    \item Smoothing factors $\alpha$: \{None, 0.5\}
\end{itemize}
\end{itemize}

\subsection{Stability Criterion Policy}
\label{sec:policies:stability}

To prevent early stopping due to transient fluctuations, we define a single stability criterion that requires a certain number of consecutive iterations to independently satisfy the stopping condition before termination is allowed. In this work, we implemented and tested a constant stability threshold:

\begin{itemize}
    \item \textbf{Constant Stability Criterion:} Convergence is accepted only if the stopping condition has held for $k$ consecutive iterations.

    \textbf{Parameters tested in the experiments:}
    \begin{itemize}
        \item $k \in \{1, 3, 5\}$
    \end{itemize}
\end{itemize}

\subsection{Shot Allocation Policies}
\label{sec:policies:allocation}

The shot allocation policy determines how many additional measurements are taken in each iteration. In this work, we implemented and tested both static and adaptive strategies for shot allocation.

\begin{itemize}
    \item \textbf{Constant Shot Allocation:} A fixed number of shots is collected in each iteration, regardless of convergence state:

    \textbf{Parameters tested in the experiments:} 
    \begin{itemize}
        \item \new{Shots: $\{50, 100\}$}
    \end{itemize}

    \item \textbf{Dynamic Shot Allocation:} The number of shots to be allocated at each iteration is adaptively estimated based on the evolution of the empirical distribution. Let $\varepsilon_i = \mathcal{D}(\hat{P}_i, \hat{P}_{i - 1})$ denote the observed divergence between the current empirical distribution $\hat{P}_i$ and the one from the previous iteration $\hat{P}_{i - 1}$. This sequence $\{\varepsilon_i\}_i$ represents the per-iteration change in the empirical distribution. Assuming smooth convergence (i.e., the distances $\varepsilon_i$ decrease over time), we can estimate the next divergence value by extrapolating the trend. Specifically, we compute the ratio:
$
\delta_i = \frac{\varepsilon_i}{\varepsilon_{i - 1}},
$
and use it to estimate the distance in the next iteration:
$
\varepsilon_{i + 1} \approx \delta_i \cdot \varepsilon_i = \frac{\varepsilon_i^2}{\varepsilon_{n - i}}.
$
To reach a fixed divergence threshold $\varepsilon$, the number of shots $S_{i+1}$ to allocate in the next iteration is scaled accordingly:
\[
S_{i+1} = \frac{\varepsilon_i^2}{\varepsilon_{i - 1}} \cdot S_i,
\]
where $S_i$ is the number of shots used in the current iteration.

The intuition is to project the diminishing returns of sampling, allocating more shots if convergence is slowing, or fewer if the distribution stabilises rapidly. 

To ensure stability in early iterations, where $\varepsilon_i$ and $\delta_i$ may fluctuate unpredictably, a fixed number of initial batches can be used before enabling dynamic scaling.

\textbf{Parameters tested in the experiments:}
\begin{itemize}
    \item \new{Minimum initial fixed iterations: $\{0, 1, 3\}$}
\end{itemize}
    
\end{itemize}

\subsection{Common Execution Parameters}
\label{sec:policies:budget}

\new{In all configurations, the maximum shot budget $\mathcal{B}$ was fixed to $20{,}000$ shots. The \texttt{default\_shots} parameter (i.e., the number of shots in the first iteration) was selected independently and tested with the same two values: \{50, 100\}. This ensured fair and consistent initialisation across all policies. We evaluated our framework considering five different $\delta$ thresholds: \{0.01, 0.025, 0.5, 0.1, 0.25\}. \new{Finally, all configurations were tested in combinations with three different divergence metrics: TVD, Hellinger and JS.}}

\subsection{Parameter Grid and Configurations}
\label{sec:policies:combos}
By combining all policies and their parameter values, we explored a total of $33,750$ unique configurations. These were computed as follows:

\begin{itemize}
    \item \textbf{Stopping criteria} (75 configurations):
    \begin{itemize}
        \item Delta: $5$ thresholds $\times$ $3$ look-back window sizes $= 15$
        \item DMA: $5$ thresholds $\times$ $3$ look-back window sizes $\times$ $2$ windows $\times$ $2$ alphas $= 60$
    \end{itemize}
    \item \textbf{Stability enforcement:} 3 values of $k$
    \item \textbf{Shot allocation:} (5 configurations):
        \begin{itemize}
        \item Constant: $2$ values of constant batch size
        \item Dynamic: $3$ minimum initial fixed iterations
    \end{itemize}
    \item \textbf{Default shot count for first batch:} 2 values
    \item \textbf{$\delta$ threshold}: 5 values
    \item \new{Divergence metrics}: 3 values
\end{itemize}

\new{Thus, the total configuration space is:
\[
75~\times~3~\times~5~\times 2~\times~5~\times~3~=~33,750~\text{total configurations}
\]
}

This thorough evaluation allows us to characterise the empirical behaviour of the \textsc{IncrementalExecution} Framework under a wide variety of convergence dynamics and cost-accuracy trade-offs. In Section~\ref{sec:exp}, we analyse these configurations and compare them against the optimal a posteriori policy defined in Alg.~\ref{alg:aposteriori}.

%% file: src/exp.tex
\section{Experimental Evaluation}
\label{sec:exp}

This experimental evaluation\footnote{Experimental scripts, raw data and analysis are freely available at: \url{https://github.com/GBisi/quantum-incremental-execution}.} section has a twofold objective: (i) to assess whether there exist configurations capable of achieving the target accuracy while reducing the total number of shots with respect to a fixed budget, and (ii) to provide practical insights and preliminary guidelines for policy fine-tuning as a function of quantum circuit characteristics.

\new{More in detail, we tackle the following research questions about our framework:
\begin{itemize}
    \item \textbf{RQ1: } How the parameters of \textsc{IncrementalExecution} impact on finding the target accuracy?
    \item \textbf{RQ2: } How the configuration of \textsc{IncrementalExecution} is affected by noise models and circuits?
    \item \textbf{RQ3: } Which are the best configurations of \textsc{IncrementalExecution}?
    \item \textbf{RQ4: } How the best configurations of \textsc{IncrementalExecution} behave in comparison with the state of the art?
\end{itemize}
}

\subsection{Experimental Setup}

\new{To evaluate the several configurations of our framework, we collected $180$ unique execution traces, each corresponding to a distinct quantum circuit and backend pair, in order to investigate how changing the quantum algorithm, circuit size, or noise model affects \textsc{IncrementalExecution}. In particular, we employed:
\begin{itemize}
    \item 5 IBM noisy backend simulators: \texttt{fake\_faze}, \texttt{fake\_kyiv}, \texttt{fake\_marrakesh}, \texttt{fake\_sherbrooke}, and \texttt{fake\_torino}.
    \item 6 quantum algorithms for static circuits: \texttt{dj}, \texttt{static qaoa}, \texttt{qnn}, \texttt{qft}, \texttt{random}, and \texttt{static vqe}.
    \item 6 qubit sizes \rev{(4, 6, 8, 10, 12, and 14)} for each circuit.
\end{itemize}}

\noindent
\new{We collected the output distribution for each trace by executing the circuits on the backends in batches of 50 shots, until reaching a total of $20,\!000$ shots per trace. From these samples, we compute the \textit{a posteriori} optimal value for each trace using the three divergence metrics introduced in Section~\ref{sec:quantum-problem}: TVD, Hellinger, and JS.}

\new{This data collection\footnote{The dataset is fully available on: \url{https://github.com/GBisi/qsimbench}~\cite{qsimbench}} also supported a preliminary study that shaped our experimental design in two ways: (i) it informed the choice of shot budgets used in the experiments, and (ii) it motivated a partition of circuits into \emph{small} and \emph{large} instances. We define these as size $<10$ and size $\geq 10$, respectively. Table~\ref{tab:shot-budget} (Table~2) reports, for each circuit size, an estimate of the \emph{a posteriori} optimal number of shots and the fraction of traces whose \emph{a posteriori} optimum exceeds $10,\!000$ shots, a budget that is half of that used to compute the a posterior optimal value for each trace. This percentage increases sharply with circuit size, from 8.15\% at 4 qubits to 78.33\% at 14 qubits, and it crosses 50\% at 10 qubits. Based on this evidence, we observed that using a shot budget below $20,\!000$ increases the number of traces that consume the entire budget before reaching the point of diminishing returns; therefore, we adopt $20,\!000$ shots as the budget $\mathcal{B}$ for all runs of \textsc{IncrementalExecution}. Moreover, since the fraction of traces with an optimal shot count above $10,\!000$ exceeds 50\% from 10 qubits onward, we use the same threshold to group circuits in the remainder of the analysis: \textit{small} circuits have 4, 6, or 8 qubits, while \textit{large} circuits have 10, 12, or 14 qubits.}

\begin{table}[ht!]
\centering
\caption{Shot-budget statistics from the preliminary study, grouped by circuit size.}
\label{tab:shot-budget}
\begin{tabular}{c c}
\hline
\textbf{Size} & \textbf{\% optimal $>10{,}000$} \\
\hline
4  & 8.15\%  \\
6  & 18.52\% \\
8  & 33.89\% \\
10 & 51.30\% \\
12 & 64.44\% \\
14 & 78.33\% \\
\hline
\end{tabular}
\end{table}


\new{We evaluated all \textsc{IncrementalExecution} configurations by exhaustively exploring the policy space described in Section~\ref{sec:policies}, for a total of $33,\!750$ unique configurations. Each configuration specifies (i) the divergence metric, (ii) the stopping criterion, (iii) the stability-enforcement policy, and (iv) the shot-allocation strategy.}

\new{We consider a configuration \emph{valid} if, for each of the 180 traces, it returns a number of shots $n$ such that the resulting distribution diverges from the full budget distribution for less than the user-defined tolerance $\delta$, more formally $\mathcal{D}(\hat{P}_n, \hat{P}_{20000}) \leq \delta$, which is based on Definition~\ref{def:quantum-problem}. In our experiments, we use values of $\delta \in \{0.01, 0.025, 0.05, 0.1, 0.25\}$.
}

\new{In all the following evaluations, we always compare the results by comparing the perspective of the three divergence metrics and the five values of $\delta$}

\subsection{Analysis of the Configuration Parameters}
\new{We analyse the configuration space of \textsc{IncrementalExecution} to understand how its parameters affect the identification of \emph{valid} configurations for the deltas and divergence metrics we consider. Specifically, we examine: (i) the threshold $\epsilon$, used by the stopping criterion to decide when to halt incremental iterations; (ii) the stopping-criterion policy, which specifies how to use the output distributions produced at each iteration to halt the incremental iterations; and (iii) the stability criterion, defined by a sliding window of size $k$, which requires the observed behavior to persist before a configuration is deemed stable.}

\new{By systematically varying $\epsilon$, the stopping policy, and $k$, we quantify how sensitive the search is to these parameters, and how this sensitivity changes with the delta values and the adopted divergence metric. This analysis allows us to characterise which parameter regimes facilitate the discovery of valid configurations, providing practical guidance for selecting robust settings across heterogeneous deltas and metrics.}

\subsubsection{Stopping Threshold $\epsilon$ analysis}

\begin{figure}[ht!]
    \includegraphics[width=\textwidth]{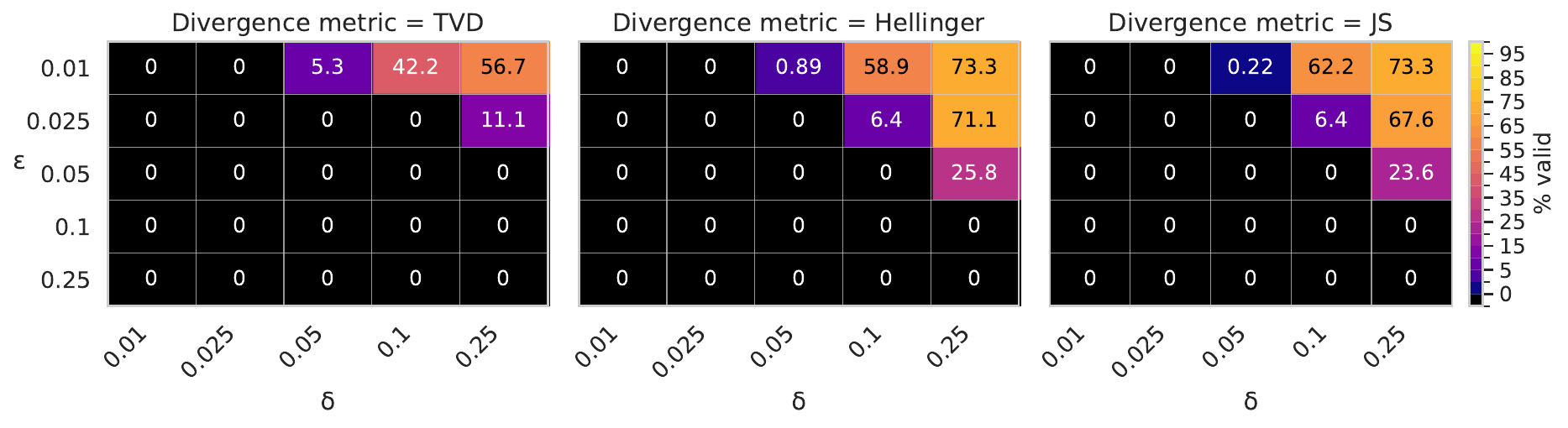}
    \caption{Heatmaps for $\epsilon$ 0.01, 0.025, 0.05, 0.1, and 0.25, for all the divergence metrics and $\delta$ values considered.}
    \label{fig:thr_heatmap}
\end{figure}

\new{Figure~\ref{fig:thr_heatmap} reports, for each divergence metric, a heatmap in which each cell indicates the percentage of valid configurations obtained for a given pair of values $(\epsilon,\delta)$. The plots show that choosing $\delta \in {0.01, 0.025}$ impedes for \textsc{IncrementalExecution} to find valid configurations: the divergence between the distribution estimated from the shots selected by \textsc{IncrementalExecution} and the reference distribution computed using the full budget rarely falls below these stringent $\delta$ thresholds, so the validity condition is rarely satisfied.}

\new{A clear pattern emerges when comparing divergence metrics and parameter regimes. First, validity is observed almost exclusively for the smallest threshold $\epsilon$ (0.01), and only when $\delta$ is sufficiently relaxed (typically $\delta \ge 0.1$). In this regime, \textsc{IncrementalExecution} achieves substantially higher validity rates under Hellinger and Jensen–Shannon (JS) than under TVD: at $\delta=0.1$, Hellinger and JS yield about $58.9\%$ and $62.2\%$ valid configurations, respectively, while TVD reaches $42.2\%$; at $\delta=0.25$, Hellinger and JS both reach $73.3\%$, compared to $56.7\%$ for TVD. This suggests that, for the same $(\epsilon,\delta)$, TVD behaves as the most demanding metric in terms of meeting the validity constraint, whereas Hellinger and JS are more permissive and provide very similar outcomes.}

\new{Second, increasing $\epsilon$ rapidly reduces the fraction of valid configurations across all metrics, indicating high sensitivity to this parameter. For $\epsilon=0.025$, TVD essentially collapses to near-zero validity (only $11.1\%$ at $\delta=0.25$), while Hellinger and JS remain comparatively robust at the loosest delta, still yielding $71.1\%$ and $67.6\%$ valid configurations at $\delta=0.25$, but dropping to $6.4\%$ at $\delta=0.1$. For larger thresholds ($\epsilon\ge 0.05$), validity is recovered only at $\delta=0.25$ and only for Hellinger/JS (about $25.8\%$ and $23.6\%$, respectively), whereas TVD yields no valid configurations.}

\new{Overall, the figure highlights a strong interaction between $\epsilon$, $\delta$ and the divergence metric: stringent deltas almost never allow validity, $\epsilon$ is always strictly lower than $\delta$ to find valid configurations, highlighting 
and when validity is attainable, Hellinger and JS consistently support a wider and more stable set of valid configurations than TVD.}

\subsubsection{Stopping Criterion analysis}

\begin{figure}[ht!]
    \includegraphics[width=\textwidth]{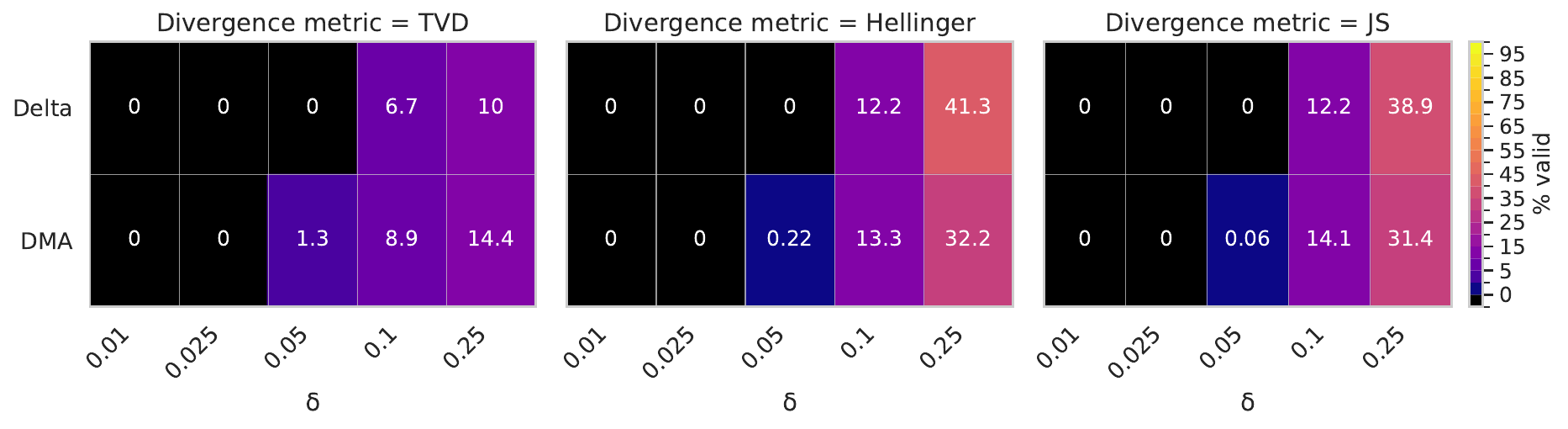}
    \caption{Heatmaps for Delta and DMA stopping criterion policies, for all the divergence metrics and $\delta$ values considered.}
    \label{fig:stopping_heatmap}
\end{figure}

\new{Figure~\ref{fig:stopping_heatmap} compares the two stopping-criterion policies, \textsc{Delta} and \textsc{DMA}, across the considered divergence metrics. Each heatmap cell reports the percentage of valid configurations for a given $\delta$, with rows corresponding to the stopping policy.}

\new{Across all metrics, validity is zero for the most stringent settings ($\delta\in{0.01,0.025}$), and it starts to appear only from $\delta=0.05$ onward. However, the two policies behave differently once validity becomes attainable. Under TVD, \textsc{Delta} yields no valid configurations at $\delta=0.05$ and only modest rates at larger deltas ($6.7\%$ at $\delta=0.1$ and $10\%$ at $\delta=0.25$). In contrast, \textsc{DMA} is consistently more effective for TVD, having valid configurations already at $\delta=0.05$ ($1.3\%$) and improving to $8.9\%$ at $\delta=0.1$ and $14.4\%$ at $\delta=0.25$.}

\new{The gap between the policies is even clearer for Hellinger and JS. For Hellinger, \textsc{Delta} reaches $12.2\%$ valid configurations at $\delta=0.1$ and $41.3\%$ at $\delta=0.25$, whereas \textsc{DMA} provides similar performance at $\delta=0.1$ ($13.3\%$) but is lower at $\delta=0.25$ ($32.2\%$), while still enabling a small fraction of valid configurations at $\delta=0.05$ ($0.22\%$) where \textsc{Delta} yields none. A comparable pattern holds for JS: \textsc{Delta} obtains $12.2\%$ at $\delta=0.1$ and $38.9\%$ at $\delta=0.25$, while \textsc{DMA} slightly improves at $\delta=0.1$ ($14.1\%$) but is lower at $\delta=0.25$ ($31.4\%$), again with a small advantage at $\delta=0.05$ ($0.06\%$) where \textsc{Delta} is zero.}

\new{Overall, \textsc{DMA} appears more conservative in the sense that it can occasionally obtain valid configurations under tighter deltas (notably at $\delta=0.05$) and it improves results for the most demanding metric (TVD). Conversely, when $\delta$ is sufficiently loose (especially $\delta=0.25$), and the divergence is measured with Hellinger or JS, \textsc{Delta} tends to achieve the highest validity rates. This highlights a trade-off between the two stopping policies: \textsc{DMA} offers slightly better robustness in stricter regimes, while \textsc{Delta} can be more effective in permissive regimes where many configurations can satisfy the validity condition.}

\subsubsection{Stability Windows $k$ analysis}

\begin{figure}[ht!]
    \includegraphics[width=\textwidth]{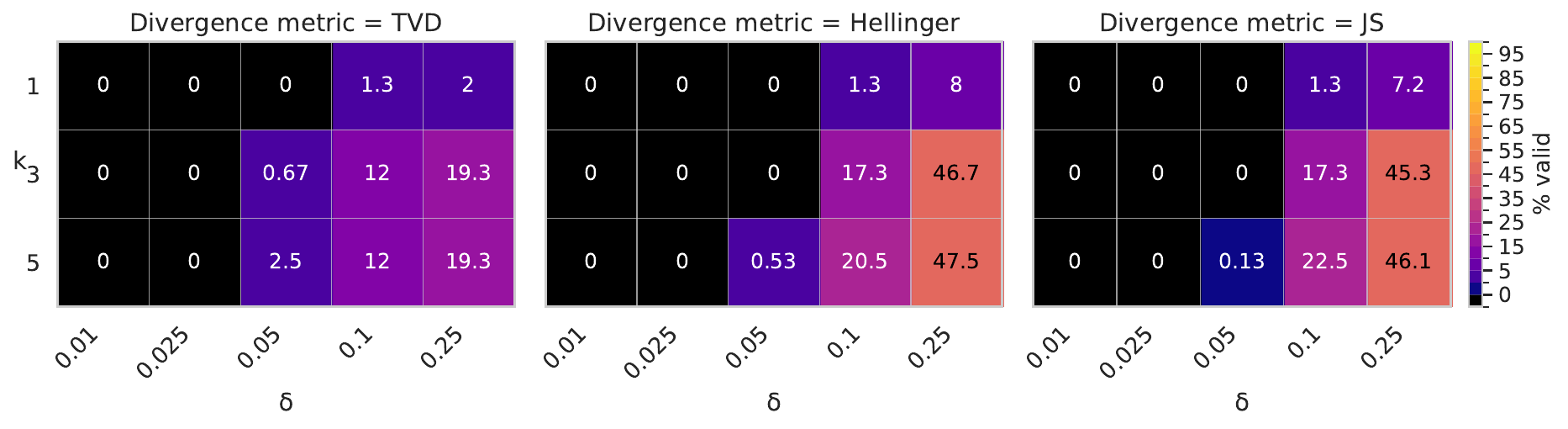}
    \caption{Heatmaps for stability window $k$ 1, 3 and 5, for all the divergence metrics and $\delta$ values considered.}
    \label{fig:stability_heatmap}
\end{figure}

\new{Figure~\ref{fig:stability_heatmap} analyses the effect of the stability window parameter (here shown as $k\in{1,3,5}$) on the percentage of valid configurations, across divergence metrics and $\delta$ values. As in the previous analyses, the strictest deltas ($\delta\in{0.01,0.025}$) lead to no valid configurations for any metric and any window size, confirming that these settings are too demanding for the observed divergences to consistently fall below the target threshold.}

\new{When validity becomes attainable ($\delta\ge 0.05$), increasing the window size generally improves the success rate, especially for Hellinger and JS. Under TVD, the effect is present but more limited: with $k=1$ no valid configurations are found at $\delta=0.05$, while $k=3$ and $k=5$ recover small but non-zero rates ($0.67\%$ and $2.5\%$, respectively). At $\delta=0.1$, TVD rises sharply from $1.3\%$ ($k=1$) to $12\%$ ($k=3$ and $k=5$), and at $\delta=0.25$ it increases from $2\%$ to $19.3\%$ for both $k=3$ and $k=5$. This suggests that, for TVD, a larger window is necessary to filter out transient fluctuations, but gains saturate quickly beyond $k=3$.}

\new{The benefit of larger windows is more pronounced for Hellinger and JS. For Hellinger, validity is zero at $\delta=0.05$ for $k\in{1,3}$ and becomes non-zero only with $k=5$ ($0.53\%$), while at $\delta=0.1$ it improves from $1.3\%$ ($k=1$) to $17.3\%$ ($k=3$) and $20.5\%$ ($k=5$). At $\delta=0.25$, the increase is substantial: from $8\%$ ($k=1$) to $46.7\%$ ($k=3$) and $47.5\%$ ($k=5$). JS shows the same trend: at $\delta=0.1$, validity rises from $1.3\%$ ($k=1$) to $17.3\%$ ($k=3$) and $22.5\%$ ($k=5$); at $\delta=0.25$, it increases from $7.2\%$ to $45.3\%$ and $46.1\%$, respectively, with a small non-zero value at $\delta=0.05$ only for $k=5$ ($0.13\%$).}

\new{Overall, these results indicate that requiring stability over a longer window markedly improves the likelihood of identifying valid configurations, particularly when using Hellinger or JS. Moreover, performance tends to saturate between $k=3$ and $k=5$, suggesting that moderate window sizes already provide most of the benefit, while TVD remains the most restrictive metric even under larger windows.}

\subsection{Impact of Circuit Size and Noise Models}
\new{We now analyse the experimental results from the perspective of (i) circuit size, by contrasting the \emph{small} and \emph{large} circuit groups introduced earlier, and (ii) the adopted noise model, instantiated through the five simulated backends used in our data collection. The goal of this section is to understand whether valid configurations of \textsc{IncrementalExecution} depend on the intrinsic complexity of the circuit and on the characteristics of the underlying noise profile.}

\new{Following the same methodology as in the previous section, we characterise validity by varying $\delta$ and the considered divergence metrics, and we measure how the percentage of valid configurations changes across circuit-size regimes and backends. This comparison highlights which settings remain robust when moving from smaller to larger circuits, and how different noise models alter the feasibility of satisfying the validity constraints under the same $(\delta,\text{metric})$ combinations.}

\subsubsection{Circuit size analysis.} 

\begin{figure}[ht!]
    \includegraphics[width=\textwidth]{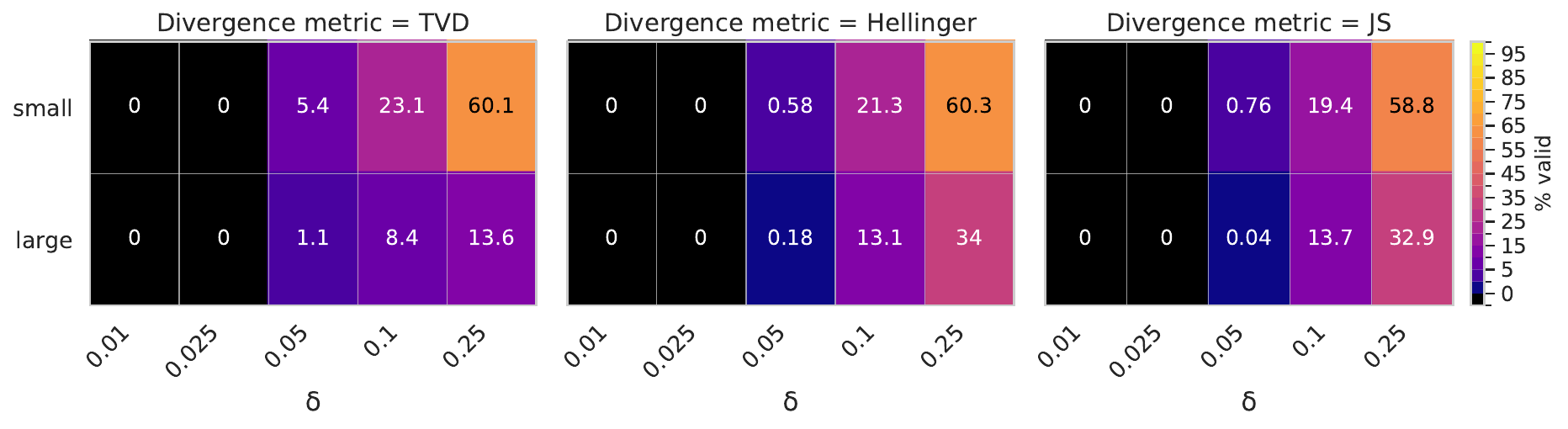}
    \caption{Heatmaps for small and large circuits, for all the divergence metrics and $\delta$ values considered.}
    \label{fig:size_heatmap}
\end{figure}

\new{Figure~\ref{fig:size_heatmap} compares the percentage of valid configurations obtained for smaller and larger circuits as $\delta$ varies, across the three divergence metrics. As expected, the strictest settings ($\delta\in{0.01,0.025}$) yield no valid configurations in any case, confirming that these targets are too stringent irrespective of circuit size and metric.}

\new{When $\delta$ is relaxed, circuit size becomes a key discriminant. For small circuits, validity increases sharply as $\delta$ grows: under TVD it rises from $5.4\%$ at $\delta=0.05$ to $23.1\%$ at $\delta=0.1$ and reaches $60.1\%$ at $\delta=0.25$. A very similar trend is observed for Hellinger and JS, which are slightly more favourable at the loosest setting: Hellinger goes from $0.58\%$ ($\delta=0.05$) to $21.3\%$ ($\delta=0.1$) and $60.3\%$ ($\delta=0.25$), while JS goes from $0.76\%$ to $19.4\%$ and $58.8\%$. Overall, for small circuits, $\delta=0.25$ enables around $60\%$ valid configurations across all metrics, indicating that \textsc{IncrementalExecution} can reliably meet the validity constraint when the delta tolerance is sufficiently permissive.}

\new{In contrast, large circuits exhibit systematically lower validity rates, even at relaxed deltas. Under TVD, validity peaks at only $13.6\%$ for $\delta=0.25$ (with $11\%$ at $\delta=0.05$ and $8.4\%$ at $\delta=0.1$). Hellinger and JS show the same qualitative behaviour, but yield higher rates than TVD at the loosest delta: for $\delta=0.25$, Hellinger reaches $34\%$ and JS $32.9\%$, while remaining low at $\delta=0.05$ ($0.18\%$ and $0.04\%$) and moderate at $\delta=0.1$ ($13.1\%$ and $13.7\%$).}

\new{Taken together, the figure highlights that increasing circuit size makes the validity condition harder to satisfy under all divergence metrics, consistent with larger circuits producing output distributions that are more strongly affected by noise and sampling variability. Moreover, while the three metrics are broadly consistent for small circuits, Hellinger and JS appear more inclined to have valid configurations than TVD for large circuits, especially at $\delta=0.25$, suggesting that metric choice becomes more impactful as circuit complexity grows.}

\subsubsection{Noise models analysis.}

\begin{figure}[ht!]
    \includegraphics[width=\textwidth]{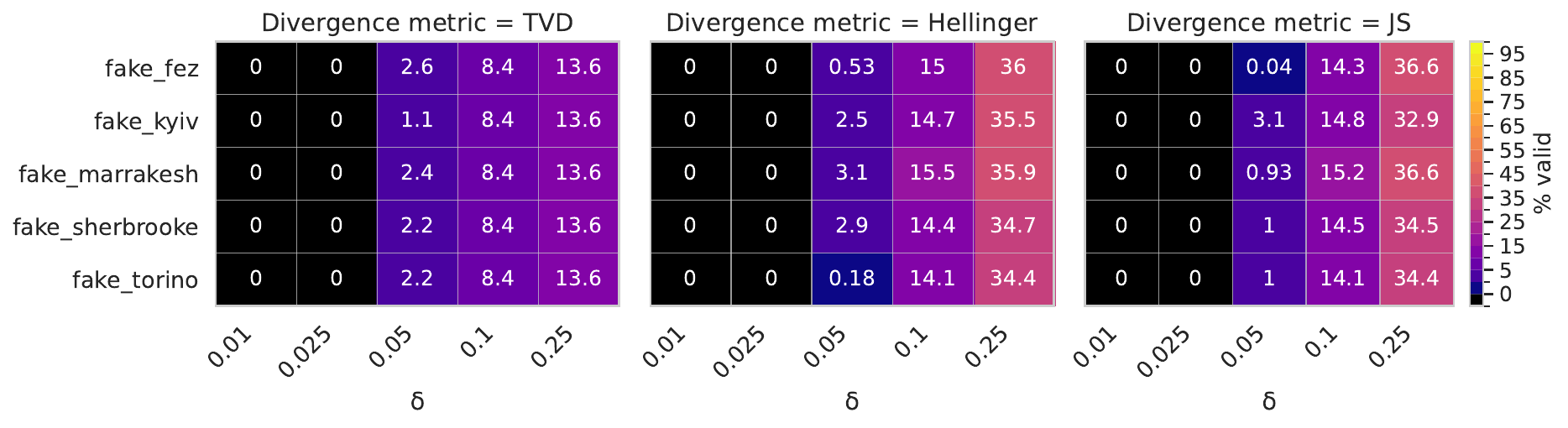}
    \caption{Heatmaps for the five considered backends, for all the divergence metrics and $\delta$ values considered.}
    \label{fig:backend_heatmap}
\end{figure}

\new{Figure~\ref{fig:backend_heatmap} breaks down the percentage of valid configurations by noise model, comparing the five simulated backends (\texttt{fake\_fez}, \texttt{fake\_kyiv}, \texttt{fake\_marrakesh}, \texttt{fake\_sherbrooke}, and \texttt{fake\_torino}) across $\delta$ values and divergence metrics. As in the previous plots, the strictest deltas ($\delta\in{0.01,0.025}$) produce no valid configurations for any backend and any metric, indicating that these targets are unattainable regardless of the noise profile.}

\new{For $\delta\ge 0.05$, the overall behaviour is remarkably consistent across backends, with only modest variations. Under TVD, validity remains low and largely backend-independent: for most backends it is around $2$--$3\%$ at $\delta=0.05$, $8.4\%$ at $\delta=0.1$, and $13.6\%$ at $\delta=0.25$. The only noticeable deviation occurs at $\delta=0.05$ for \texttt{fake\_kyiv}, which achieves a higher rate ($11\%$) than the other backends (roughly $2.2$--$2.6\%$), while still aligning with them at $\delta\ge 0.1$.}

\new{The picture changes when moving to Hellinger and JS: validity increases substantially at the loosest delta and remains broadly stable across noise models. With Hellinger, the percentage of valid configurations at $\delta=0.25$ lies in a narrow range around the mid-30s (from $34.4\%$ for \texttt{fake\_torino} up to $36\%$ for \texttt{fake\_faze}), and around $14$--$15.5\%$ at $\delta=0.1$. At $\delta=0.05$ the rates are low but non-zero and slightly more backend-dependent (e.g., from $0.18\%$ for \texttt{fake\_torino} up to $3.1\%$ for \texttt{fake\_marrakesh}, with \texttt{fake\_kyiv} at $2.5\%$). JS shows an almost identical trend: at $\delta=0.25$ all backends fall between $32.9\%$ and $36.6\%$, and at $\delta=0.1$ they cluster tightly around $14.1$--$15.2\%$. Differences are again most visible at $\delta=0.05$, ranging from near-zero for \texttt{fake\_faze} ($0.04\%$) to about $3.1\%$ for \texttt{fake\_kyiv}.}

\new{Overall, these results suggest that, within the considered parameter ranges, the choice of divergence metric and $\delta$ has a much stronger impact on validity than the specific backend noise model. Backend-specific effects are limited and mainly appear in the stricter regime ($\delta=0.05$), while for $\delta\in{0.1,0.25}$ all noise models exhibit very similar validity rates, especially under Hellinger and JS, indicating that the observed trends are robust across the considered noise profiles.}

\subsection{Selecting the Best Configurations}
\new{So far, we have analysed how parameters, noise-model backends, and circuit size individually affect the validity of configurations. We now turn to identifying the best configurations jointly across these factors, and to evaluating how well they generalise to unseen data. Specifically, we search for the best configuration for each triple $(\textit{size},\delta,\textit{divergence metric})$, where \textit{size} denotes the circuit family (small vs.\ large), $\delta$ is the target tolerance used to define validity, and the divergence metric is the distance adopted to compare output distributions. Since our previous results show that no valid configurations are found for $\delta=0.01$ and $\delta=0.025$, we exclude these values from the following selection analysis.}

\subsubsection{Scoring metrics.}
\new{To rank configurations, we define two complementary metrics that quantify how close a configuration comes to the optimal shot budget. Let $s_{c,t}$ be the number of shots selected by configuration $c$ on trace $t$, and let $o_t$ be the corresponding optimal number of shots (i.e., the minimum number of shots that satisfies the validity constraint for that trace).
First, the \emph{average difference} measures the typical absolute deviation from the optimum:
\[
\mathrm{AvgDiff}(c)=\frac{1}{|\mathcal{T}|}\sum_{t\in\mathcal{T}}\bigl|s_{c,t}-o_t\bigr|.
\]
This metric is expressed in shots and captures the absolute inefficiency of a configuration: lower values indicate that, on average, the configuration selects a number of shots close to the optimum.}

\new{Second, the \emph{median ratio} captures the typical multiplicative overhead:
\[
\mathrm{MedRatio}(c)=\mathrm{median}_{t\in\mathcal{T}}\left(\frac{s_{c,t}}{o_t}\right).
\]
A value close to $1$ indicates that the configuration usually matches the optimal budget; values larger than $1$ indicate overshooting (using more shots than necessary). Values below $1$ indicate undershooting, which should never happen for valid configurations, as the optimal value is defined as the lowest number of shots to satisfy the target $\delta$.}

\subsubsection{Train/test protocol.}
\new{Our dataset comprises $180$ traces, each corresponding to a specific combination of algorithm, circuit instance, and backend noise model. We first partition the traces by circuit size into \emph{small} and \emph{large} groups. Within each size group, we select the best configuration using an $80/20$ split: configurations are ranked on the validation set (80\% of the traces) and then evaluated on the held-out test set (20\%).}

\new{On the training split, we compute $\mathrm{AvgDiff}$ and $\mathrm{MedRatio}$ for every candidate configuration, normalise the two metrics to make them comparable, and combine them with equal weight into a single score. The configuration with the lowest combined score is selected as the best one for the corresponding $(\textit{size},\delta,\textit{divergence metric})$ setting. In the following, we report the performance of the selected configuration on the test split, providing an unbiased estimate of how well the chosen configuration generalises to unseen traces within the same size regime.}

\subsubsection{Best configurations and validity on the test set}
\begin{table}[ht!]
\centering
\caption{Best configuration (selected on the validation split) for each $(\textit{size},\delta,\textit{distance})$ setting, and its validity percentage on the test split.}
\label{tab:best_configs_validity}
\begin{tabular}{lll lr}
\toprule
\textbf{Size} & \textbf{Distance} & \boldmath$\delta$ & \textbf{Best conf. (val)} & \textbf{Validity on test (\%)} \\
\midrule
small & tvd       & 0.05 & conf\_7    & 100.00 \\
small & hellinger & 0.05 & conf\_3567 & 100.00 \\
small & js        & 0.05 & conf\_3650 & 100.00 \\
small & tvd       & 0.10 & conf\_358  & 94.44  \\
small & hellinger & 0.10 & conf\_465  & 100.00 \\
small & js        & 0.10 & conf\_3682 & 100.00 \\
small & tvd       & 0.25 & conf\_525  & 100.00 \\
small & hellinger & 0.25 & conf\_561  & 100.00 \\
small & js        & 0.25 & conf\_1320 & 100.00 \\
\midrule
large & tvd       & 0.05 & conf\_105  & 94.44  \\
large & hellinger & 0.05 & conf\_3559 & 83.33  \\
large & js        & 0.05 & conf\_275  & 100.00 \\
large & tvd       & 0.10 & conf\_84   & 100.00 \\
large & hellinger & 0.10 & conf\_303  & 94.44  \\
large & js        & 0.10 & conf\_3921 & 100.00 \\
large & tvd       & 0.25 & conf\_3740 & 100.00 \\
large & hellinger & 0.25 & conf\_3945 & 94.44  \\
large & js        & 0.25 & conf\_327  & 94.44  \\
\bottomrule
\end{tabular}
\end{table}

\new{Table~\ref{tab:best_configs_validity} summarises, for each combination of circuit size (small vs.\ large), $\delta$, and divergence metric, the identifier of the configuration selected as best on the validation split (80\% of the traces) and its corresponding \emph{validity percentage} when evaluated on the held-out test split (20\%). Recall that the validity percentage represents the fraction of test traces for which the selected configuration satisfies the validity constraint under the given $(\delta,\text{distance})$ setting.}

\new{Overall, the selected configurations generalise well, but circuit size clearly influences robustness. For \emph{small} circuits, the best configurations achieve $100\%$ validity in almost all settings, with the only exception being TVD at $\delta=0.10$ ($94.44\%$). For \emph{large} circuits, instead, the validity of the selected configurations is more variable: while several settings still reach $100\%$, a few cases show noticeable drops (e.g., Hellinger at $\delta=0.05$ reaches $83.33\%$, and both Hellinger and JS show $94.44\%$ in some of the $\delta\in\{0.10,0.25\}$ settings). This confirms the trend observed in the previous analyses: larger circuits make the validity condition harder to satisfy consistently, and configuration choices that appear optimal on the validation set may generalise slightly less reliably when circuit complexity increases.}

\subsubsection{Incremental vs. optimal shots on the test set}
\begin{figure*}[ht!]
  \centering
  \includegraphics[width=0.85\textwidth]{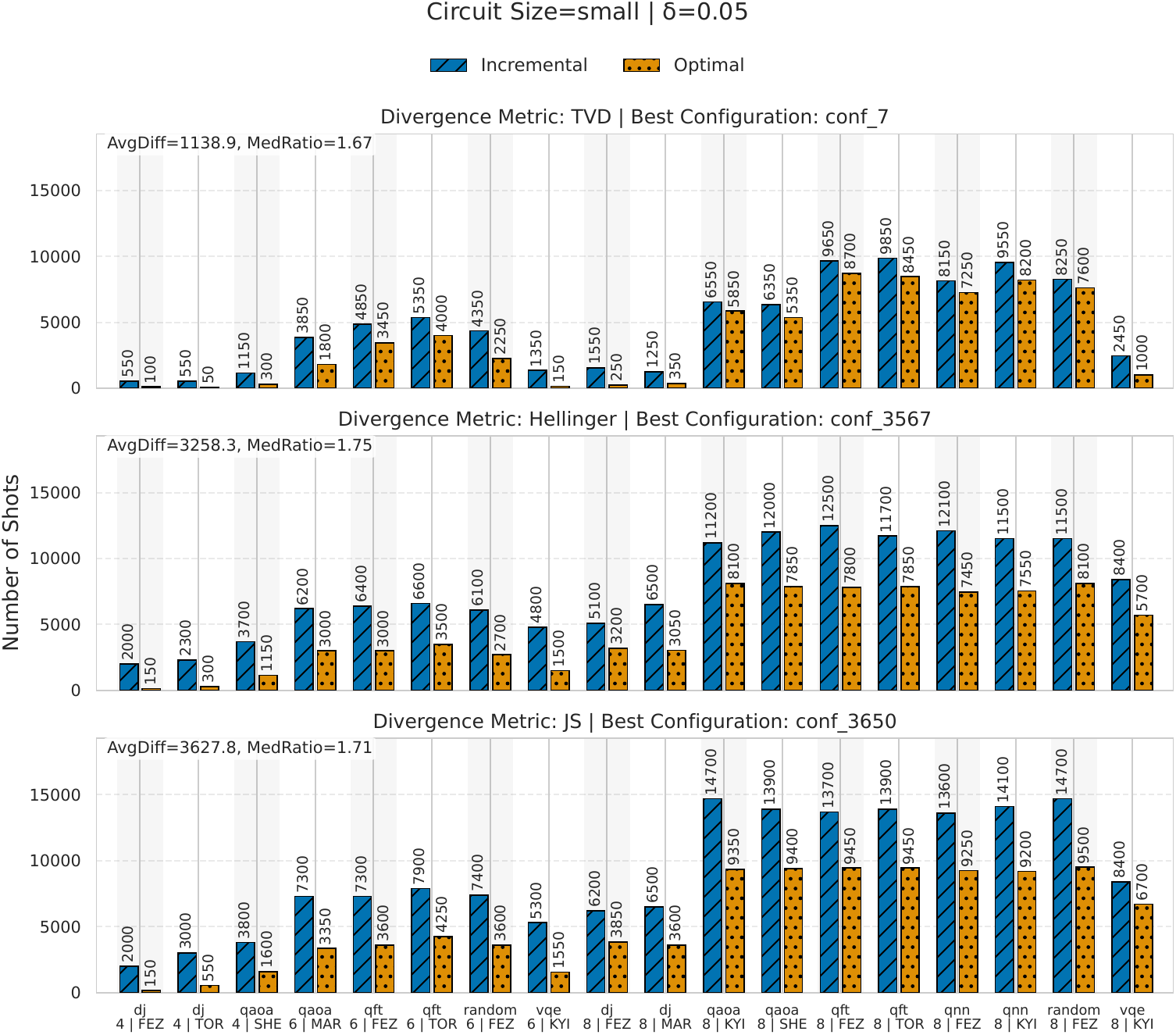}
  \caption{Small circuits and $\delta=0.05$: incremental shots (best configuration) vs.\ optimal shots on the test set.}
  \label{fig:best_small_d005}
\end{figure*}

\begin{figure*}[ht!]
  \centering
  \includegraphics[width=0.85\textwidth]{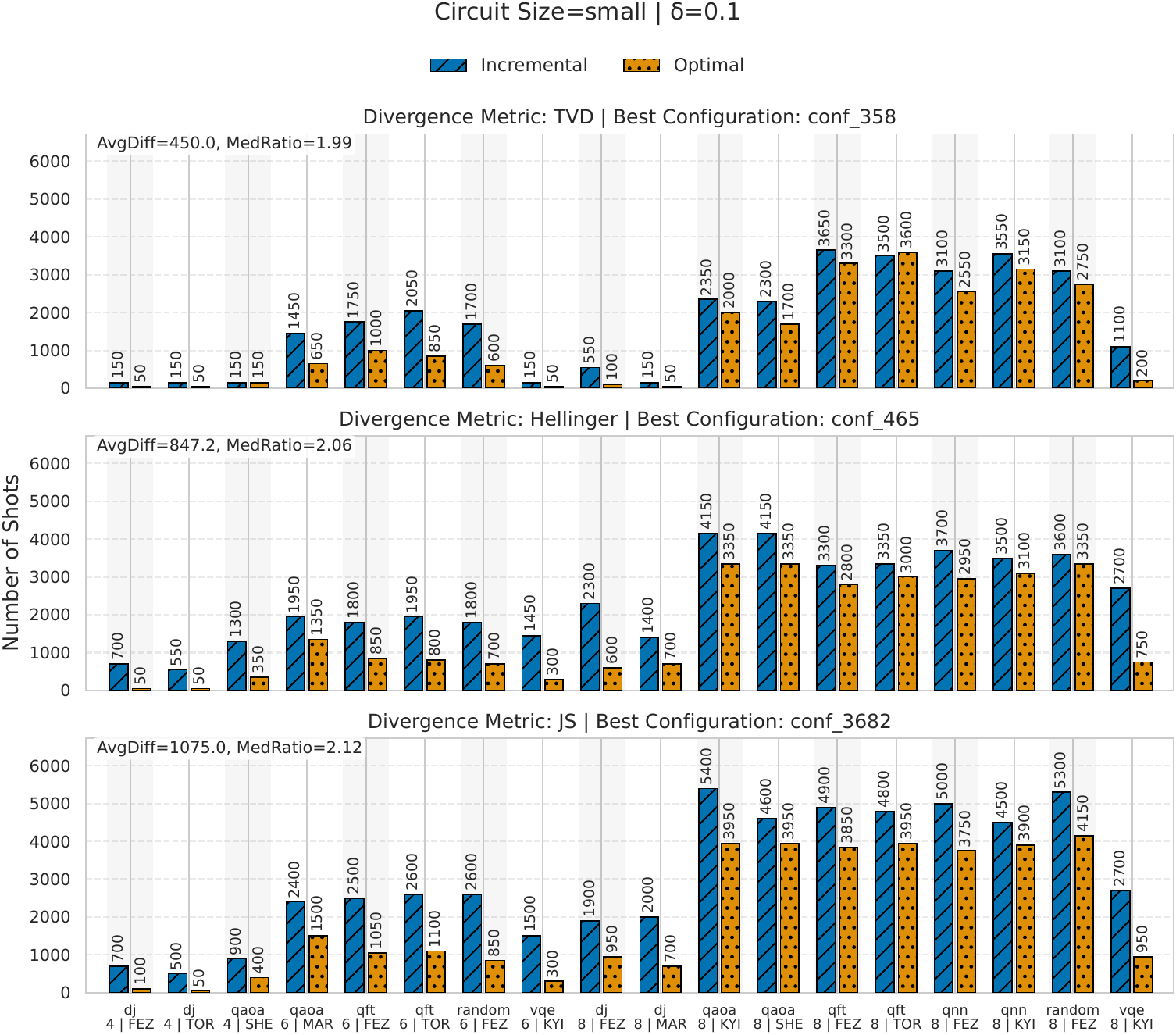}
  \caption{Small circuits and $\delta=0.10$: incremental shots (best configuration) vs.\ optimal shots on the test set.}
  \label{fig:best_small_d01}
\end{figure*}

\begin{figure*}[ht!]
  \centering
  \includegraphics[width=0.85\textwidth]{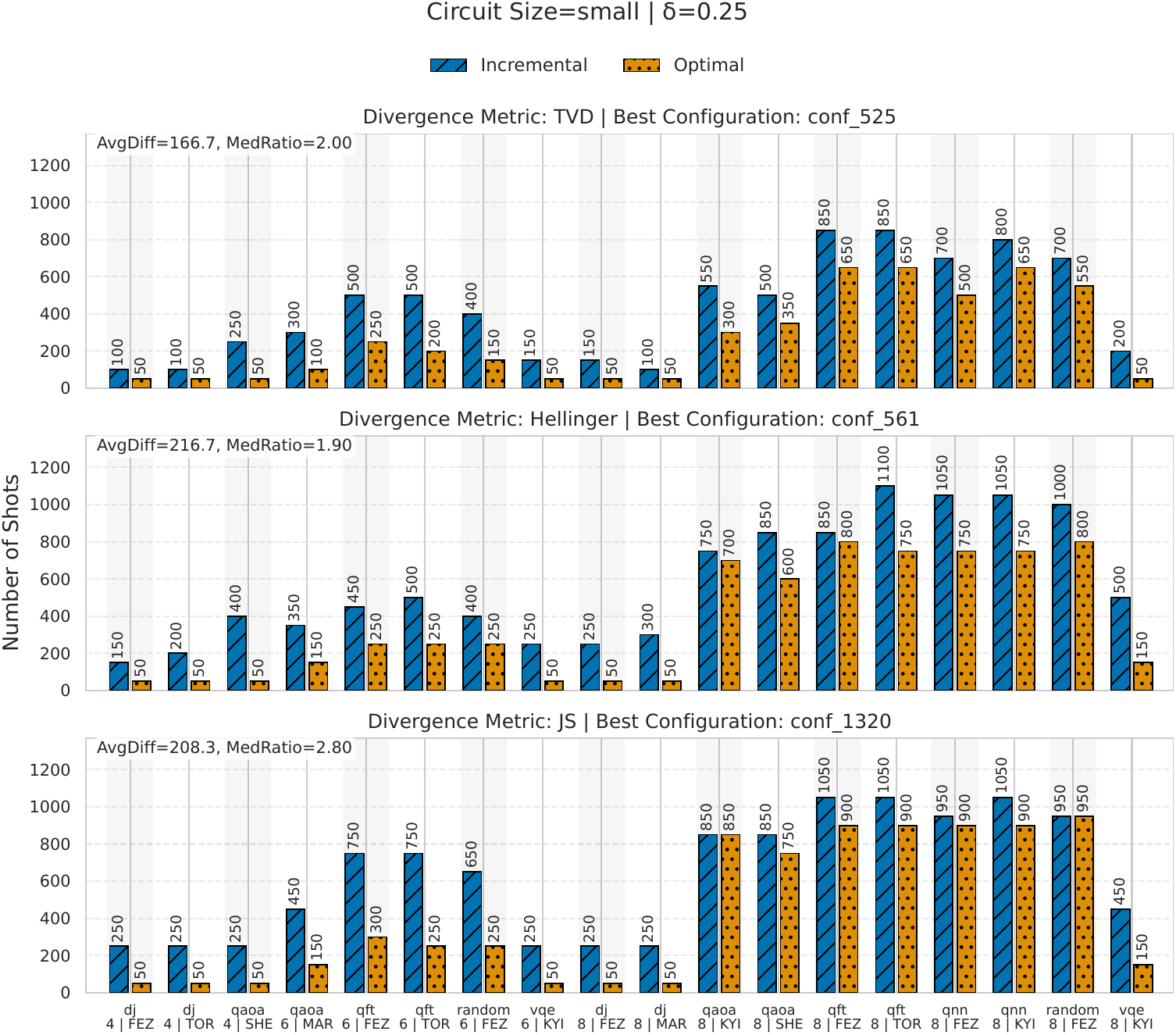}
  \caption{Small circuits and $\delta=0.25$: incremental shots (best configuration) vs.\ optimal shots on the test set.}
  \label{fig:best_small_d025}
\end{figure*}

\begin{figure*}[ht!]
  \centering
  \includegraphics[width=0.85\textwidth]{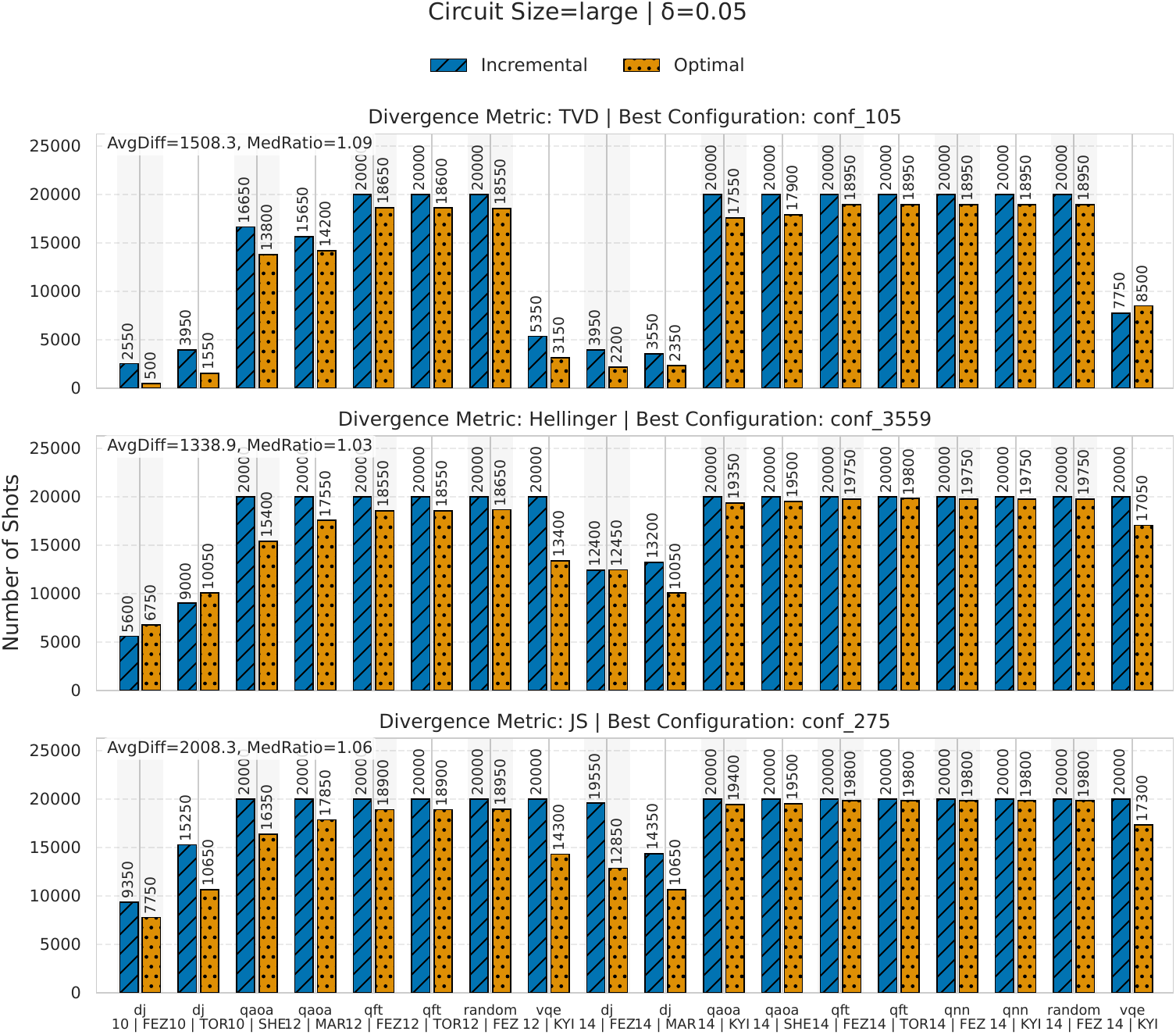}
  \caption{Large circuits and $\delta=0.05$: incremental shots (best configuration) vs.\ optimal shots on the test set.}
  \label{fig:best_large_d005}
\end{figure*}

\begin{figure*}[ht!]
  \centering
  \includegraphics[width=0.85\textwidth]{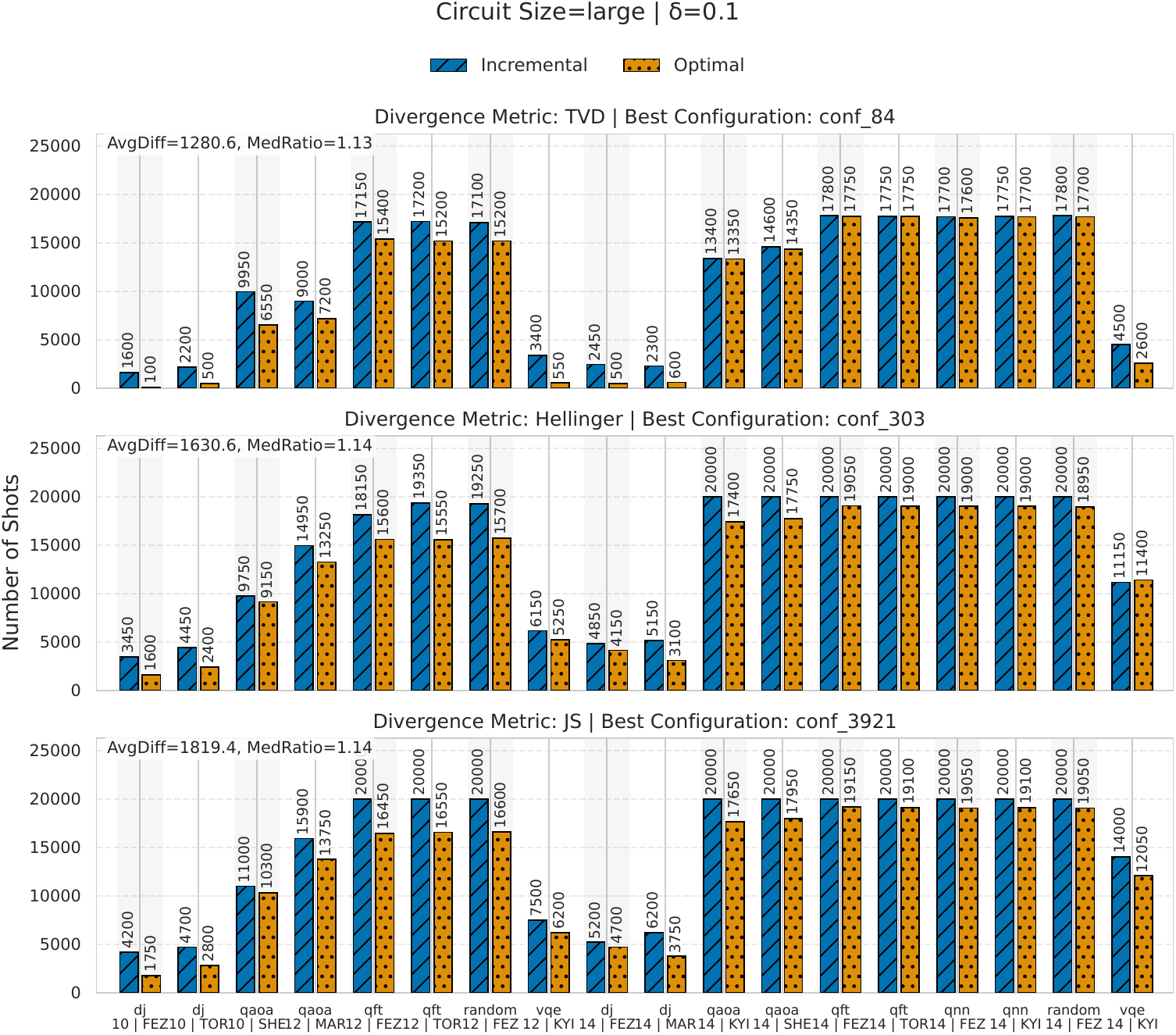}
  \caption{Large circuits and $\delta=0.10$: incremental shots (best configuration) vs.\ optimal shots on the test set.}
  \label{fig:best_large_d01}
\end{figure*}

\begin{figure*}[ht!]
  \centering
  \includegraphics[width=0.85\textwidth]{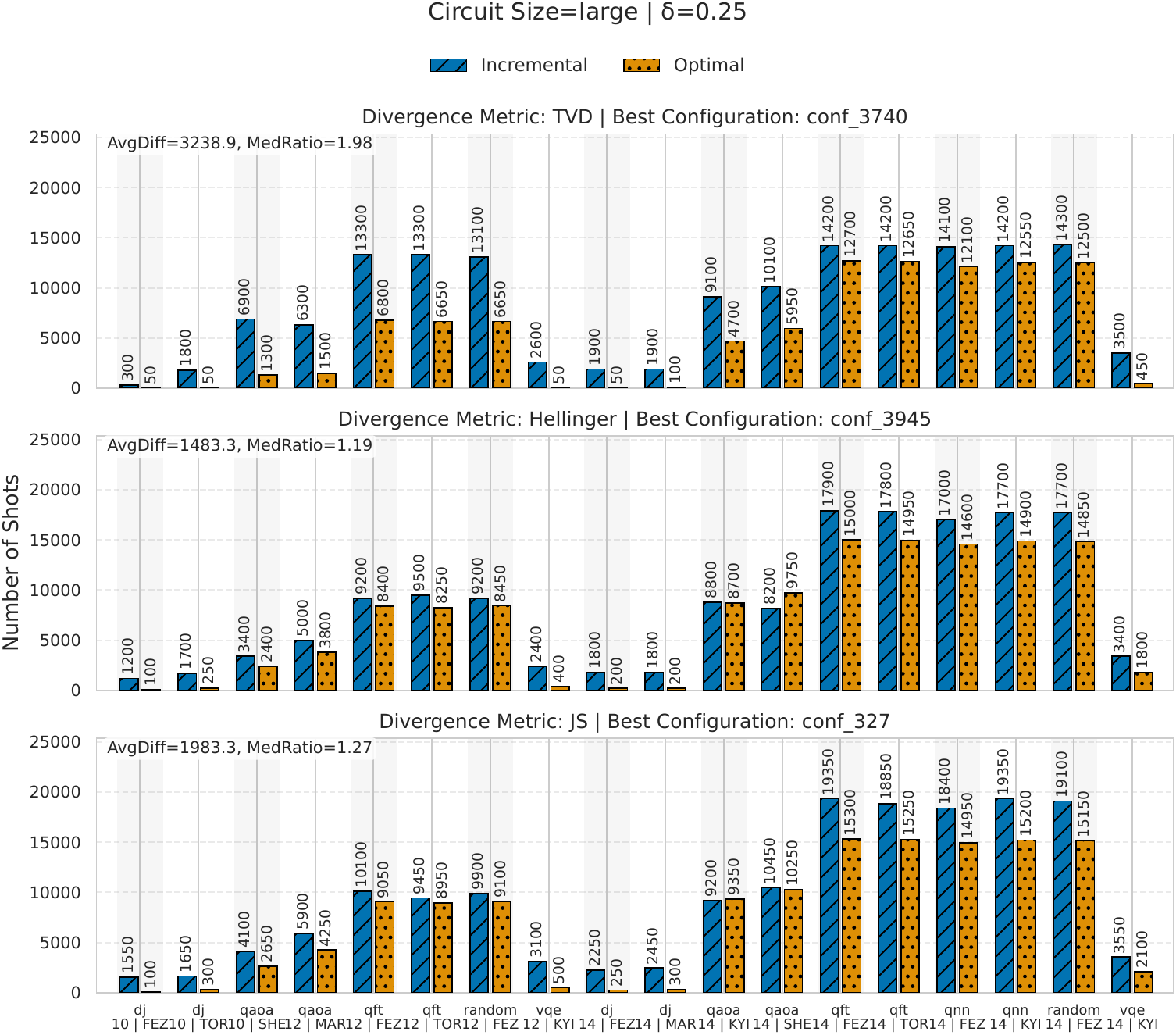}
  \caption{Large circuits and $\delta=0.25$: incremental shots (best configuration) vs.\ optimal shots on the test set.}
  \label{fig:best_large_d025}
\end{figure*}

\new{We conclude this section by inspecting, on the held-out test traces, how closely the \emph{best} configuration selected for each $(\textit{size},\delta,\textit{distance})$ setting matches the optimal number of shots. Figures~\ref{fig:best_small_d005}--\ref{fig:best_large_d025} are formatted in the same way. For a fixed \textit{size} and $\delta$, each plot contains three panels (TVD, Hellinger, and JS). In each panel, the blue bars represent the number of shots produced by \textsc{IncrementalExecution} using the selected best configuration, while the orange bars represent the optimal number of shots for the same trace. For each panel, we also report the two evaluation metrics on the test set: the average absolute deviation from the optimal shots (\textit{AvgDiff}, in shots) and the median ratio between incremental and optimal shots (\textit{MedRatio}). For readability, we abbreviate the backend names in the trace identifiers as follows:
\texttt{fake\_fez}$\rightarrow$\texttt{FEZ},
\texttt{fake\_kyiv}$\rightarrow$\texttt{KYI},
\texttt{fake\_marrakesh}$\rightarrow$\texttt{MAR},
\texttt{fake\_sherbrooke}$\rightarrow$\texttt{SHE},
\texttt{fake\_torino}$\rightarrow$\texttt{TOR}.}

\new{\paragraph{Small circuits, $\delta = 0.05$.}
Figure~\ref{fig:best_small_d005} shows that, for the strictest $\delta$ considered, the selected best configurations remain valid but tend to use a shot amount distant from the optimal value.
TVD (conf\_7) is the closest in absolute terms (AvgDiff $=1138.9$) with MedRatio $=1.67$, while Hellinger (conf\_3567) and JS (conf\_3650) exhibit larger deviations (AvgDiff $=3258.3$ and $3627.8$) with MedRatio around $1.75$ and $1.71$. In this regime, the configuration using the TVD is the one approximating best the optimal value.}

\new{\paragraph{Small circuits, $\delta = 0.10$.}
In Figure~\ref{fig:best_small_d01}, the three metrics still overshoot, with median ratios close to $2$.
TVD (conf\_358) achieves the smallest absolute deviation (AvgDiff $=450.0$, MedRatio $=1.99$), followed by Hellinger (conf\_465; AvgDiff $=847.2$, MedRatio $=2.06$) and JS (conf\_3682; AvgDiff $=1075.0$, MedRatio $=2.12$). This suggests that, for small circuits, relaxing $\delta$ improves absolute accuracy but does not necessarily reduce the typical multiplicative overhead, which remains close to a factor of two.}

\new{\paragraph{Small circuits, $\delta = 0.25$.}
Figure~\ref{fig:best_small_d025} corresponds to the most permissive delta and shows that the selected configurations are closer to the optimal in absolute terms.
TVD (conf\_525) attains AvgDiff $=166.7$ with MedRatio $=2.00$, while Hellinger (conf\_561) has a slightly larger AvgDiff $=216.7$ but the best MedRatio among the three ($1.90$). JS (conf\_1320) has AvgDiff $=208.3$ but a notably higher MedRatio ($2.80$), indicating that it more frequently overshoots the optimal budget even when the absolute deviation remains limited.}

\new{\paragraph{Large circuits, $\delta = 0.05$.}
Moving to large circuits, Figure~\ref{fig:best_large_d005} shows a different behaviour: the selected configurations track the optimal values more closely in \emph{relative} terms.
TVD (conf\_105) achieves MedRatio $=1.09$ (AvgDiff $=1508.3$), Hellinger (conf\_3559) is the closest multiplicatively with MedRatio $=1.03$ (AvgDiff $=1338.9$), and JS (conf\_275) has MedRatio $=1.06$ (AvgDiff $=2008.3$). While the absolute deviations are larger than in the small-circuit case (as expected from the larger budgets), the ratios indicate that the incremental budgets are generally near-optimal under all three metrics.}

\new{\paragraph{Large circuits, $\delta = 0.10$.}
Figure~\ref{fig:best_large_d01} confirms this trend: all three metrics yield very similar median ratios (around $1.13$--$1.14$), suggesting stable near-optimal behaviour in relative terms.
TVD (conf\_84) shows the smallest absolute deviation (AvgDiff $=1280.6$, MedRatio $=1.13$), while Hellinger (conf\_303) and JS (conf\_3921) are slightly more higher in absolute terms (AvgDiff $=1630.6$ and $1819.4$) with MedRatio $=1.14$ for both.}

\new{\paragraph{Large circuits, $\delta = 0.25$.}
At $\delta=0.25$ (Figure~\ref{fig:best_large_d025}) the metrics separate more clearly.
TVD (conf\_3740) deviates the most from the optimal values, with MedRatio $=1.98$ (AvgDiff $=3238.9$). In contrast, Hellinger (conf\_3945) and JS (conf\_327) remain closer to the optimum, achieving MedRatio $=1.19$ (AvgDiff $=1483.3$) and MedRatio $=1.27$ (AvgDiff $=1983.3$), respectively.
Thus, for large circuits under permissive $\delta$, Hellinger and JS better exploit the relaxed constraint to reduce shots, while TVD maintains a larger safety margin.}

\new{\paragraph{Overall discussion.}
Two high-level takeaways emerge.
First, circuit size strongly impacts the results: for \emph{small} circuits, the best configurations frequently overshoot the optimal values (MedRatio often close to $2$ or larger), even when AvgDiff becomes small at relaxed $\delta$; for \emph{large} circuits, instead, the best configurations track the optimal much more closely in relative terms for $\delta\in\{0.05,0.10\}$ (MedRatio $\approx 1.03$--$1.14$).
Second, the divergence metric matters most with higher values of $\delta$. For $\delta=0.25$ on large circuits, Hellinger and JS yield clearly better (lower) median ratios than TVD, indicating that they allow \textsc{IncrementalExecution} to save shots more aggressively while remaining valid. Conversely, under stricter deltas, the three metrics behave more similarly, and the main differences are expressed in absolute deviations rather than in typical multiplicative overhead.}

\new{\paragraph{Recap of Best Configuration Parameters}
Table~\ref{tab:best-configs} summarises the parameter settings of the best configurations for each circuit size, $\delta$, and convergence distance shown in this Section.}

\rev{\paragraph{Scalability of the framework} The \textsc{IncrementalExecution} framework is not tied to a specific number of qubits. Its stopping, stability, and allocation policies operate solely on empirical distributions collected during execution and therefore remain well defined for larger circuits. What changes with qubit count, depth, and circuit structure is not the applicability of the framework, but the configuration that yields the best reliability–cost trade-off. Larger circuits may induce broader or more complex output distributions, which can require stricter stopping thresholds, larger stability windows, larger shot budgets, or different divergence metrics. For this reason, the configurations reported here should be interpreted as best fixed configurations for the evaluated regimes, whereas the parameter-selection methodology introduced in this paper can be reused to tune \textsc{IncrementalExecution} for larger and deeper circuits.}

\begin{table}[t]
\centering
\caption{Best configuration parameters across circuit sizes and $\delta$ thresholds.}
\label{tab:best-configs}
\begin{tabular}{l c c c l l c c c}
\toprule
Config ID & Size & $\delta$ & Metric & Stop. crit. & Main parameters & $k$ & Next Shots & Default \\
\midrule
conf\_7     & Small & 0.05 & TVD &
Delta &
$\tau = 2$, $\varepsilon = 0.01$ &
3 & 100 & 50 \\

conf\_3567  & Small & 0.05 & Hellinger &
DMA &
$w = 5$, $\tau = 2$, $\varepsilon = 0.01$ &
5 & 100 & 100 \\

conf\_3650  & Small & 0.05 & JS &
DMA &
$w = 5$, $\tau = 2$, $\varepsilon = 0.01$ &
5 & 100 & 100 \\
\midrule
conf\_358   & Small & 0.10 & TVD &
DMA &
$w = 3$, $\tau = 3$, $\alpha = 0.5$, $\varepsilon = 0.025$ &
3 & 50 & 50 \\

conf\_465   & Small & 0.10 & Hellinger &
DMA &
$w = 3$, $\tau = 3$, $\alpha = 0.5$, $\varepsilon = 0.025$ &
5 & 50 & 50 \\

conf\_3682  & Small & 0.10 & JS &
Delta &
$\tau = 3$, $\varepsilon = 0.025$ &
1 & 100 & 100 \\
\midrule
conf\_525   & Small & 0.25 & TVD &
Delta &
$\tau = 1$, $\varepsilon = 0.05$ &
1 & 50 & 50 \\

conf\_561   & Small & 0.25 & Hellinger &
Delta &
$\tau = 1$, $\varepsilon = 0.05$ &
1 & 50 & 50 \\

conf\_1320  & Small & 0.25 & JS &
DMA &
$w = 5$, $\tau = 1$, $\varepsilon = 0.25$ &
3 & 100 & 50 \\
\midrule
conf\_105   & Large & 0.05 & TVD &
DMA &
$w = 3$, $\tau = 3$, $\alpha = 0.5$, $\varepsilon = 0.01$ &
3 & 100 & 50 \\

conf\_3559  & Large & 0.05 & Hellinger &
DMA &
$w = 3$, $\tau = 3$, $\alpha = 0.5$, $\varepsilon = 0.01$ &
5 & 100 & 100 \\

conf\_275   & Large & 0.05 & JS &
DMA &
$w = 3$, $\tau = 3$, $\varepsilon = 0.01$ &
5 & 100 & 50 \\
\midrule
conf\_84    & Large & 0.10 & TVD &
DMA &
$w = 3$, $\tau = 3$, $\alpha = 0.5$, $\varepsilon = 0.01$ &
3 & 50 & 50 \\

conf\_303   & Large & 0.10 & Hellinger &
Delta &
$\tau = 3$, $\varepsilon = 0.025$ &
5 & 100 & 50 \\

conf\_3921  & Large & 0.10 & JS &
DMA &
$w = 3$, $\tau = 3$, $\alpha = 0.5$, $\varepsilon = 0.025$ &
5 & 100 & 100 \\
\midrule
conf\_3740  & Large & 0.25 & TVD &
DMA &
$w = 3$, $\tau = 3$, $\alpha = 0.5$, $\varepsilon = 0.025$ &
3 & 100 & 100 \\

conf\_3945  & Large & 0.25 & Hellinger &
Delta &
$\tau = 2$, $\varepsilon = 0.05$ &
3 & 100 & 100 \\

conf\_327   & Large & 0.25 & JS &
Delta &
$\tau = 1$, $\varepsilon = 0.025$ &
5 & 50 & 50 \\
\bottomrule
\end{tabular}
\end{table}

\subsection{Best Configurations Compared to Hoeffding and Weissman Inequalities}
\label{sec:best-configs-comparison}

\new{We finally compare the best configurations identified for \textsc{IncrementalExecution} against two concentration-inequality baselines that provide \emph{a priori} sample-complexity requirements: the Hoeffding inequality~\cite{hoeffding1963probability} and the Weissman inequality~\cite{weissman2003inequalities}. In our setting, these bounds can be interpreted as worst-case upper bounds on the number of shots needed to guarantee that an empirical distribution is within a target tolerance $\delta$ under a fixed confidence level. We set this confidence level to $95\%$, aligning it with the empirical reliability observed for our selected configurations: as shown in Table~\ref{tab:best_configs_validity}, the best configurations achieve validity rates that are typically close to (and often equal to) $100\%$ on the test set, with the lowest observed values still around $95\%$ (e.g., $94.44\%$ in several settings).}

\new{Let $p$ denote the (unknown) true outcome distribution over an alphabet of size $K$ (in our experiments, we take $K=2^{n}$ for an $n$-qubit circuit), and let $\hat p_n$ be the empirical distribution obtained from $n$ i.i.d.\ shots. We write $\delta\in(0,1)$ for the target tolerance and set the failure probability to $\Delta := 1-\text{confidence\_level}$ (thus $\Delta=0.05$ for $95\%$ confidence).}

\paragraph{Hoeffding baseline (union bound).}
\new{Using Hoeffding's inequality for each outcome and a union bound over the $K$ outcomes, we obtain the standard $\ell_\infty$ control}
\begin{equation}
\Pr\!\left(\|\hat p_n - p\|_\infty \ge \varepsilon_\infty\right)
\;\le\; 2K\,\exp\!\left(-2n\varepsilon_\infty^2\right),
\label{eq:hoeffding_union_linf}
\end{equation}
\new{so that requiring the r.h.s.\ to be at most $\Delta$ yields the shot budget}
\begin{equation}
n \;\ge\; \frac{1}{2\varepsilon_\infty^2}\,\log\!\left(\frac{2K}{\Delta}\right).
\label{eq:hoeffding_union_linf_n}
\end{equation}

\new{Since our comparisons are stated in terms of total variation distance (TVD), we relate $\ell_\infty$ to TVD via
$\mathrm{TVD}(\hat p_n,p)=\tfrac12\|\hat p_n-p\|_1 \le \tfrac{K}{2}\|\hat p_n-p\|_\infty$.
Therefore, a sufficient condition for $\mathrm{TVD}(\hat p_n,p)\le \delta$ is
$\|\hat p_n-p\|_\infty \le \tfrac{2\delta}{K}$. Plugging $\varepsilon_\infty=\tfrac{2\delta}{K}$ into~\eqref{eq:hoeffding_union_linf_n} gives the TVD-calibrated Hoeffding baseline}
\begin{equation}
n \;\ge\; \frac{K^2}{8\delta^2}\,\log\!\left(\frac{2K}{\Delta}\right).
\label{eq:hoeffding_union_tvd_n}
\end{equation}

\paragraph{Weissman baseline (TVD / $\ell_1$ bound).}
\new{As an alternative that directly controls $\ell_1$, we use the inequality of Weissman \emph{et al.}~\cite{weissman2003inequalities}, which (in an appendix-friendly form) states that}
\begin{equation}
\Pr\!\left(\|\hat p_n - p\|_1 \ge \varepsilon_1\right)
\;\le\; (2^{K}-2)\,\exp\!\left(-\frac{n\varepsilon_1^2}{2}\right).
\label{eq:weissman_l1}
\end{equation}
\new{Since $\mathrm{TVD}(\hat p_n,p)=\tfrac12\|\hat p_n-p\|_1$, setting $\varepsilon_1=2\delta$ yields}
\begin{equation}
\Pr\!\left(\mathrm{TVD}(\hat p_n,p) \ge \delta\right)
\;\le\; (2^{K}-2)\,\exp\!\left(-2n\delta^2\right),
\label{eq:weissman_tvd}
\end{equation}
\new{and enforcing this probability to be at most $\Delta$ gives the corresponding shot requirement}
\begin{equation}
n \;\ge\; \frac{1}{2\delta^2}\,\log\!\left(\frac{2^{K}-2}{\Delta}\right).
\label{eq:weissman_tvd_n}
\end{equation}

\new{In the experimental tables that follow (Tables~\ref{tab:sota_d005}--\ref{tab:sota_d025}), we instantiate these expressions with confidence level $0.95$ (i.e., $\Delta=0.05$) and $K=2^{n}$ unless otherwise noted, and compare the resulting worst-case shot budgets against the empirical shot usage of the best \textsc{IncrementalExecution} configurations.}

\new{We report these baselines alongside (i) the optimal shot budget and (ii) the number of shots selected by \textsc{IncrementalExecution} when using the best configuration for each divergence metric (columns {Inc-TVD}, {Inc-Hell}, and {Inc-JS}) on the held-out test traces.}

\new{Tables~\ref{tab:sota_d005}--\ref{tab:sota_d025} present the comparison for $\delta\in\{0.05,0.10,0.25\}$. Each row corresponds to a test trace and reports the optimal and incremental number of shots. The \textit{Weissman} and \textit{Hoeffding} baselines, instead, are determined solely by the target tolerance $\delta$, the circuit size, and the chosen confidence level; hence, within each size group they take a single value, and we display them once using multirow entries.\footnote{\new{For readability, we show an excerpt with two representative test traces per circuit size. Within each row, we bold the smallest shot count among the three \textsc{IncrementalExecution} variants (Inc-TVD, Inc-Hell, Inc-JS). Full tables are in Appendix~\ref{app:full-sample-complexity}.}}}

\paragraph{Incremental approach (TVD, Hellinger and JS)}
\new{Consider the conditions assumed in our context: the outcome space
$\mathcal X$ is finite and $\hat P_n$ denotes the empirical distribution
constructed from measurement outcomes that are i.i.d.\ draws from a
target distribution $P$. Classical second-order limit results for
empirical distributions, in particular consequences of the Central
Limit Theorem (CLT), imply that the typical fluctuations of $\hat P_n$
around $P$ scale as $n^{-1/2}$. As a consequence, for any tolerance
$\delta>0$ and confidence level $1-\zeta\in(0,1)$, there exists
$n_0=n_0(\delta,\zeta,P)$ such that
\[
\Pr\!\left(\|\hat P_n-P\|_1 \le \delta\right) \ge 1-\zeta,
\qquad \forall n \ge n_0,
\]
with $n_0$ typically scaling on the order of $\delta^{-2}$.}

\new{As we are considering $\mathcal{D}$ which is
continuous and locally Lipschitz around $P$ with respect to the
$\ell_1$ norm. Then there exists a constant $L>0$ such that, for $n$
sufficiently large,
\[
\mathcal{D}(\hat P_n,P) \le L\,\|\hat P_n-P\|_1.
\]
Consequently, for any $\delta>0$ and $\zeta\in(0,1)$, there exists
$n_0'=n_0'(\delta,\zeta,P)$ such that
\[
\Pr\!\left(\mathcal{D}(\hat P_n,P) \le \delta\right) \ge 1-\zeta,
\qquad \forall n \ge n_0'.
\]
This explains why, in practice, the number of shots
needed in our incremental approach can be substantially smaller than distribution-free worst-case
bounds (like those expressed in Eq. \ref{eq:hoeffding_union_tvd_n} and \ref{eq:weissman_tvd_n}, which ignore the $n^{-1/2}$ second-order fluctuation behavior of
$\hat P_n$ around $P$ and instead control uniform deviations over the
entire probability simplex.. The above reasoning applies, in particular, to total variation
distance ($\mathrm{TVD}(P,Q)=\tfrac12\|P-Q\|_1$), Hellinger distance
($H^2(P,Q)\le \mathrm{TVD}(P,Q)$), and Jensen--Shannon divergence, which
is continuous and satisfies $\mathrm{JS}(P,Q)=\Theta(\mathrm{TVD}(P,Q)^2)$.
in the local regime.
}

\new{\paragraph{Reading the tables.}
Across all values of $\delta$, both inequality-based baselines are orders of magnitude larger than the optimal value and the number of shots selected by \textsc{IncrementalExecution}. This is expected: Hoeffding and Weissman provide worst-case, distribution-agnostic guarantees, whereas the optimal and \textsc{IncrementalExecution} exploit distributions obtained by executing observed traces.}

\new{\paragraph{Results for $\delta=0.05$.}
Table~\ref{tab:sota_d005} reports the strictest tolerance. Here, the two analytic baselines are extremely conservative: Hoeffding already reaches $8.27\times 10^{4}$ shots for 4-qubit circuits and grows to $1.80\times 10^{11}$ for 14 qubits, while Weissman ranges from $2.82\times 10^3$ (4 qubits) to $2.27\times 10^6$ (14 qubits). In contrast, \textsc{IncrementalExecution} typically remains in the $10^3$--$10^4$ range and is often within a small factor of the optimal value.
The gap becomes especially evident on large instances: for size 12 and 14, Inc-Hell and Inc-JS frequently saturate the $20{,}000$-shot budget in our runs (e.g., \texttt{qaoa\_12\_MAR} and \texttt{vqe\_14\_KYI}), whereas Inc-TVD still provides finite, often substantially smaller values (e.g., $15{,}650$ shots on \texttt{qaoa\_12\_MAR} and $7{,}750$ on \texttt{vqe\_14\_KYI}). Overall, at $\delta=0.05$, the TVD-based configuration has the lowest number of shots consistently among the three variants of \textsc{IncrementalExecution} in the shown traces.}

\begin{table}[t]
\centering
\setlength{\tabcolsep}{3pt}
\renewcommand{\arraystretch}{1.05}
\caption{Sample complexity comparison for $\delta = 0.05$ (excerpt; two test traces per circuit size).}
\label{tab:sota_d005}
\begin{tabular}{l c r r r r r r}
\toprule
Trace & Size & Optimal & Inc-TVD & Inc-Hell & Inc-JS & Weissman & Hoeffding \\
\midrule
dj\_4\_FEZ    & 4  & 100   & \textbf{550}   & 2000  & 2000  & \multirow{2}{*}{2818}      & \multirow{2}{*}{$8.27e+04$} \\
qaoa\_4\_SHE  & 4  & 300   & \textbf{1150}  & 3700  & 3800  &                              &                               \\
\midrule
qaoa\_6\_MAR  & 6  & 1800  & \textbf{3850}  & 6200  & 7300  & \multirow{2}{*}{9472}      & \multirow{2}{*}{$1.61e+06$} \\
vqe\_6\_KYI   & 6  & 150   & \textbf{1350}  & 4800  & 5300  &                              &                               \\
\midrule
dj\_8\_FEZ    & 8  & 250   & \textbf{1550}  & 5100  & 6200  & \multirow{2}{*}{$3.61e+04$} & \multirow{2}{*}{$3.03e+07$} \\
qaoa\_8\_KYI  & 8  & 5850  & \textbf{6550}  & 11200 & 14700 &                              &                               \\
\midrule
dj\_10\_FEZ   & 10 & 500   & \textbf{2550}  & 5600  & 9350  & \multirow{2}{*}{$1.43e+05$} & \multirow{2}{*}{$5.57e+08$} \\
qaoa\_10\_SHE & 10 & 13800 & \textbf{16650} & 20000 & 20000 &                              &                               \\
\midrule
qaoa\_12\_MAR & 12 & 14200 & \textbf{15650} & 20000 & 20000 & \multirow{2}{*}{$5.68e+05$} & \multirow{2}{*}{$1.01e+10$} \\
vqe\_12\_KYI  & 12 & 3150  & \textbf{5350}  & 20000 & 20000 &                              &                               \\
\midrule
dj\_14\_FEZ   & 14 & 2200  & \textbf{3950}  & 12400 & 19550 & \multirow{2}{*}{$2.27e+06$} & \multirow{2}{*}{$1.80e+11$} \\
vqe\_14\_KYI  & 14 & 8500  & \textbf{7750}  & 20000 & 20000 &                              &                               \\
\bottomrule
\end{tabular}
\end{table}

\new{\paragraph{Results for $\delta=0.10$.}
Moving to a looser tolerance (Table~\ref{tab:sota_d01}) reduces the needed number of shots for all methods, but the qualitative picture remains unchanged. The analytic baselines are still far above our online approaches: Hoeffding ranges from $2.07\times 10^{4}$ shots (4 qubits) to $4.49\times 10^{10}$ (14 qubits), and Weissman from $705$ to $5.68\times 10^{5}$. The incremental policies, instead, typically require a few thousand shots even at 12--14 qubits.
At this tolerance, we also start observing more ``metric-dependent'' crossovers: Inc-Hell can match or slightly improve over Inc-TVD on some traces (e.g., \texttt{qft\_6\_TOR} and \texttt{qft\_8\_FEZ}), suggesting that for moderate tolerances the tighter stopping behaviour induced by Hellinger can be advantageous for certain distributions. Inc-JS remains generally less sample-efficient in the excerpt, and it is often the highest among the three incremental variants.}

\begin{table}[t]
\centering
\setlength{\tabcolsep}{3pt}
\renewcommand{\arraystretch}{1.05}
\caption{Sample complexity comparison for $\delta = 0.10$ (excerpt; two test traces per circuit size).}
\label{tab:sota_d01}
\begin{tabular}{l c r r r r r r}
\toprule
Trace & Size & Optimal & Inc-TVD & Inc-Hell & Inc-JS & Weissman & Hoeffding \\
\midrule
dj\_4\_FEZ    & 4  & 50   & \textbf{150}  & 700  & 700  & \multirow{2}{*}{705}       & \multirow{2}{*}{$2.07e+04$} \\
qaoa\_4\_SHE  & 4  & 150  & \textbf{150}  & 1300 & 900  &                             &                               \\
\midrule
qaoa\_6\_MAR  & 6  & 650  & \textbf{1450} & 1950 & 2400 & \multirow{2}{*}{2368}      & \multirow{2}{*}{$4.02e+05$} \\
qft\_6\_TOR   & 6  & 850  & 2050          & \textbf{1950} & 2600 &                     &                               \\
\midrule
dj\_8\_FEZ    & 8  & 100  & \textbf{550}  & 2300 & 1900 & \multirow{2}{*}{9023}      & \multirow{2}{*}{$7.56e+06$} \\
qft\_8\_FEZ   & 8  & 3300 & 3650          & \textbf{3300} & 4900 &                     &                               \\
\midrule
dj\_10\_FEZ   & 10 & 100  & \textbf{1600} & 3450 & 4200 & \multirow{2}{*}{$3.56e+04$} & \multirow{2}{*}{$1.39e+08$} \\
qaoa\_10\_SHE & 10 & 6550 & 9950          & \textbf{9750} & 11000 &                    &                               \\
\midrule
qaoa\_12\_MAR & 12 & 7200 & \textbf{9000} & 14950 & 15900 & \multirow{2}{*}{$1.42e+05$} & \multirow{2}{*}{$2.52e+09$} \\
vqe\_12\_KYI  & 12 & 550  & \textbf{3400} & 6150  & 7500  &                     &                               \\
\midrule
dj\_14\_FEZ   & 14 & 500  & \textbf{2450} & 4850 & 5200 & \multirow{2}{*}{$5.68e+05$} & \multirow{2}{*}{$4.49e+10$} \\
vqe\_14\_KYI  & 14 & 2600 & \textbf{4500} & 11150 & 14000 &                    &                               \\
\bottomrule
\end{tabular}
\end{table}

\new{\paragraph{Results for $\delta=0.25$.}
Table~\ref{tab:sota_d025} shows the loosest tolerance, where all methods stop much earlier. Here, the number of shots for \textsc{IncrementalExecution} are often in the hundreds of shots for small circuits and in the few thousands for 12--14 qubits. While the analytic baselines also decrease substantially (e.g., Weissman is $113$ shots at size 4 and $9.09\times 10^{4}$ at size 14), Hoeffding remains extremely large, reaching $7.19\times 10^{9}$ at 14 qubits even under this relaxed tolerance.
This regime also makes the differences between divergence metrics more pronounced. In particular, Inc-Hell becomes the lowest variant on multiple traces (e.g., \texttt{qft\_6\_FEZ}, \texttt{qaoa\_12\_MAR}, and \texttt{vqe\_14\_KYI}), and Inc-JS can occasionally be competitive on specific traces (e.g., \texttt{dj\_10\_TOR}). These crossovers suggest that, once the target tolerance is sufficiently loose, the choice of the divergence metrics affects the stopping behaviour more than the global ``difficulty'' of meeting the constraint.}

\begin{table}[t]
\centering
\setlength{\tabcolsep}{3pt}
\renewcommand{\arraystretch}{1.05}
\caption{Sample complexity comparison for $\delta = 0.25$ (excerpt; two test traces per circuit size).}
\label{tab:sota_d025}
\begin{tabular}{l c r r r r r r}
\toprule
Trace & Size & Optimal & Inc-TVD & Inc-Hell & Inc-JS & Weissman & Hoeffding \\
\midrule
dj\_4\_FEZ    & 4  & 50  & \textbf{100} & 150 & 250 & \multirow{2}{*}{113}       & \multirow{2}{*}{3309} \\
qaoa\_4\_SHE  & 4  & 50  & \textbf{250} & 400 & \textbf{250} &              &              \\
\midrule
qaoa\_6\_MAR  & 6  & 100 & \textbf{300} & 350 & 450 & \multirow{2}{*}{379}       & \multirow{2}{*}{$6.43e+04$} \\
qft\_6\_FEZ   & 6  & 250 & 500          & \textbf{450} & 750 &                  &              \\
\midrule
dj\_8\_MAR    & 8  & 50  & \textbf{100} & 300 & 250 & \multirow{2}{*}{1444}      & \multirow{2}{*}{$1.21e+06$} \\
qft\_8\_FEZ   & 8  & 650 & \textbf{850} & \textbf{850} & 1050 &              &              \\
\midrule
dj\_10\_FEZ   & 10 & 50  & \textbf{300} & 1200 & 1550 & \multirow{2}{*}{5703}      & \multirow{2}{*}{$2.23e+07$} \\
dj\_10\_TOR   & 10 & 50  & 1800         & 1700 & \textbf{1650} &              &              \\
\midrule
qaoa\_12\_MAR & 12 & 1500 & 6300         & \textbf{5000} & 5900 & \multirow{2}{*}{$2.27e+04$} & \multirow{2}{*}{$4.03e+08$} \\
vqe\_12\_KYI  & 12 & 50   & 2600         & \textbf{2400} & 3100 &              &              \\
\midrule
dj\_14\_FEZ   & 14 & 50   & 1900         & \textbf{1800} & 2250 & \multirow{2}{*}{$9.09e+04$} & \multirow{2}{*}{$7.19e+09$} \\
vqe\_14\_KYI  & 14 & 450  & 3500         & \textbf{3400} & 3550 &              &              \\
\bottomrule
\end{tabular}
\end{table}

\new{\paragraph{Overall discussion.}
Taken together, Tables~\ref{tab:sota_d005}--\ref{tab:sota_d025} show that the two concentration-inequality baselines are systematically overconservative in our setting, by several orders of magnitude compared to both the optimal and the incremental shot values selected by \textsc{IncrementalExecution}. This overconservatism is not an artefact of tuning: it is inherent to worst-case, distribution-agnostic guarantees. Empirically, \textsc{IncrementalExecution} achieves shot values that are close to the optimal while maintaining high validity rates (Table~\ref{tab:best_configs_validity}) and scales far more gracefully with circuit size.}

\new{Across the three incremental variants, the best divergence depends on the tolerance. TVD is the most reliable choice under strict accuracy requirements (here, $\delta=0.05$), whereas Hellinger becomes increasingly competitive and can outperform TVD at looser thresholds (especially at $\delta=0.25$). JS is generally most costly in terms of shots, although it can occasionally be best on specific traces. Overall, these results motivate using TVD for stringent regimes and considering Hellinger as a strong alternative when the target tolerance is moderate to loose.}

\rev{\subsection{Comparison with fixed shot allocation strategies}}
\label{sec:aggregated-distance-profiles}

\rev{The previous comparison focuses on the number of shots selected by the best configurations and on their relation to concentration-inequality baselines. We enhance our analysis by comparing the residual distance between the distribution obtained by each strategy and the reference distribution obtained using the full budget of 20,000 shots. For each target tolerance $\delta$ and circuit-size group, we aggregate this distance over all matching executions using the median, and report the result separately for TVD, Hellinger distance, and JS divergence. The dashed horizontal line marks the target tolerance $\delta$; bars below this line therefore satisfy the target distance threshold after aggregation. The annotation box in each panel reports the median number of shots used by the optimal and incremental strategies.}

\rev{Importantly, these figures should not be interpreted as a simple ranking in which the lowest bar is always preferable. Fixed-shot baselines with larger budgets can naturally achieve smaller distances to the 20,000-shot reference distribution because they use substantially more shots. Our objective is to reach the required target distance while reducing the number of shots needed. Therefore, once a strategy falls below the $\delta$ line, the relevant question is how many shots were required to get there, rather than whether a substantially more expensive fixed-shot baseline achieves an even smaller distance.}

\begin{figure*}[t]
  \centering
  \includegraphics[width=\textwidth]{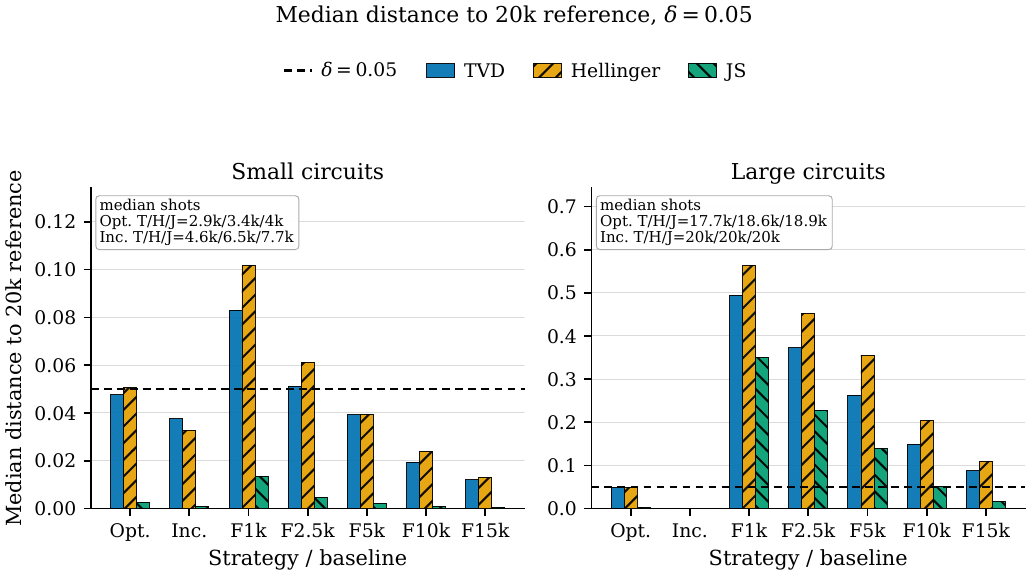}
  \caption{Median distance to the 20,000-shot reference distribution for
  $\delta=0.05$. The left panel reports small circuits, while the right panel
  reports large circuits. The dashed line marks the target tolerance. Strategy labels are abbreviated as follows: Opt. denotes the a posteriori optimal values, Inc. denotes \textsc{IncrementalExecution}, and F$x$k denotes a fixed-shot baseline using $x{,}000$ shots. The Inc. bars have zero height because \textsc{IncrementalExecution} reaches the full 20,000-shot budget in the median case; its distribution therefore coincides with the 20,000-shot reference distribution.}
  \label{fig:aggregated-distance-profile-median-delta-005}
\end{figure*}

\begin{figure*}[t]
  \centering
  \includegraphics[width=\textwidth]{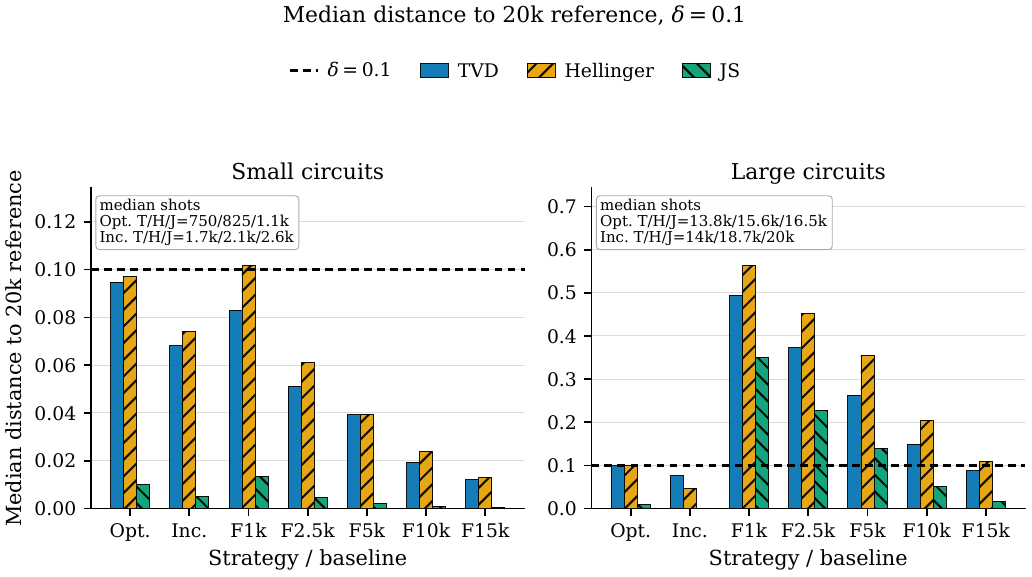}
  \caption{Median distance to the 20,000-shot reference distribution for
  $\delta=0.10$. The left panel reports small circuits, while the right panel
  reports large circuits. The dashed line marks the target tolerance. Strategy labels are abbreviated as follows: Opt. denotes the a posteriori
optimal values, Inc. denotes \textsc{IncrementalExecution}, and F$x$k denotes
a fixed-shot baseline using $x{,}000$ shots.}
  \label{fig:aggregated-distance-profile-median-delta-010}
\end{figure*}

\begin{figure*}[t]
  \centering
  \includegraphics[width=\textwidth]{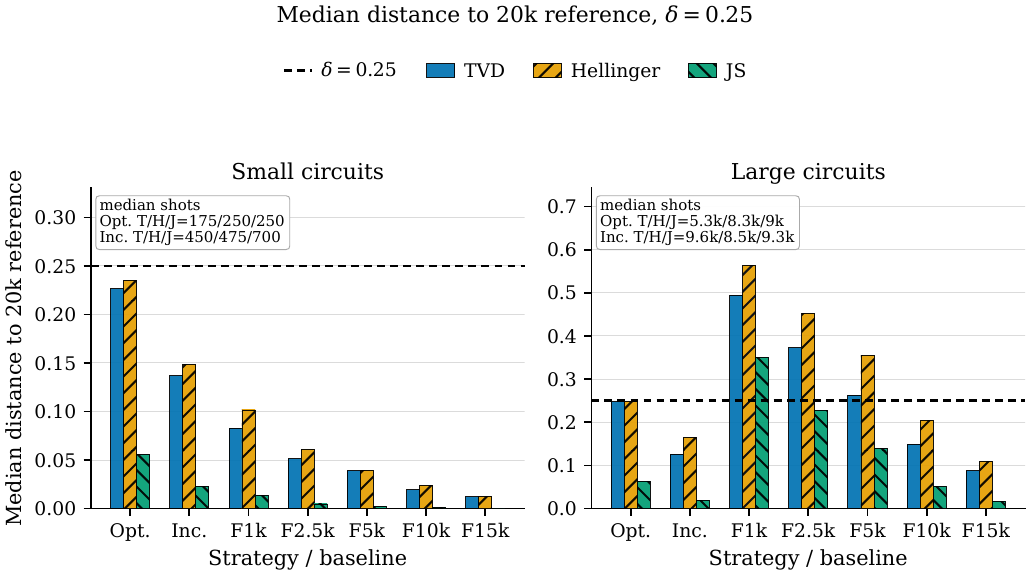}
  \caption{Median distance to the 20,000-shot reference distribution for
  $\delta=0.25$. The left panel reports small circuits, while the right panel
  reports large circuits. The dashed line marks the target tolerance. Strategy labels are abbreviated as follows: Opt. denotes the a posteriori
optimal values, Inc. denotes \textsc{IncrementalExecution}, and F$x$k denotes
a fixed-shot baseline using $x{,}000$ shots.}
  \label{fig:aggregated-distance-profile-median-delta-025}
\end{figure*}

\rev{Figures~\ref{fig:aggregated-distance-profile-median-delta-005}--\ref{fig:aggregated-distance-profile-median-delta-025} report the median distance profile across circuit-size groups and target tolerances. Median aggregation is used to focus on the typical behaviour of each strategy and to reduce the impact of extreme traces. For small circuits, \textsc{IncrementalExecution} remains below the target threshold across the considered values of $\delta$. At the strictest threshold, $\delta=0.05$, smaller fixed-shot baselines may remain above the target for at least one divergence metric, while larger fixed budgets approach or satisfy the required distance threshold. As $\delta$ becomes more permissive, especially for $\delta=0.10$ and $\delta=0.25$, more baselines fall below the threshold. However, smaller distances for high fixed-shot baselines should be interpreted together with their larger shot budgets: they are closer to the 20,000-shot reference because they continue executing even after the target distance may already have been reached.}

\rev{For large circuits, the same pattern becomes more pronounced. At $\delta=0.05$, the target is harder to satisfy, and several fixed-shot baselines remain above the threshold for one or more divergence metrics. In contrast, the optimal and incremental strategies lie close to the target boundary, which is the desired behaviour for a shot-optimisation method: stopping near the point where the required tolerance is met, rather than spending additional shots to further reduce the distance beyond what is required. At $\delta=0.10$ and $\delta=0.25$, \textsc{IncrementalExecution} remains below the target line, while high-budget fixed baselines can sometimes show smaller median distances. This does not contradict the benefit of \textsc{IncrementalExecution}, because those baselines use a larger, fixed number of shots independently of whether the target distance has already been achieved.}

\rev{Overall, the median profiles reinforce the main message of the evaluation: \textsc{IncrementalExecution} is not designed to dominate every fixed-shot baseline in absolute distance to the 20,000-shot reference distribution, since additional samples can always move the empirical distribution closer to that reference. Rather, it is designed to stop once the empirical distribution is close enough for the target tolerance. The relevant success condition is therefore whether the bars fall below the $\delta$ line. When that condition is met, the advantage of the incremental strategy lies in reaching the required distance threshold without committing in advance to a conservative fixed-shot budget.}

\rev{\subsection{Algorithm-wise Interpretation of the Results}}
\label{subsec:algorithm-wise-results}

\begin{figure}[t]
    \centering
    \includegraphics[width=\linewidth]{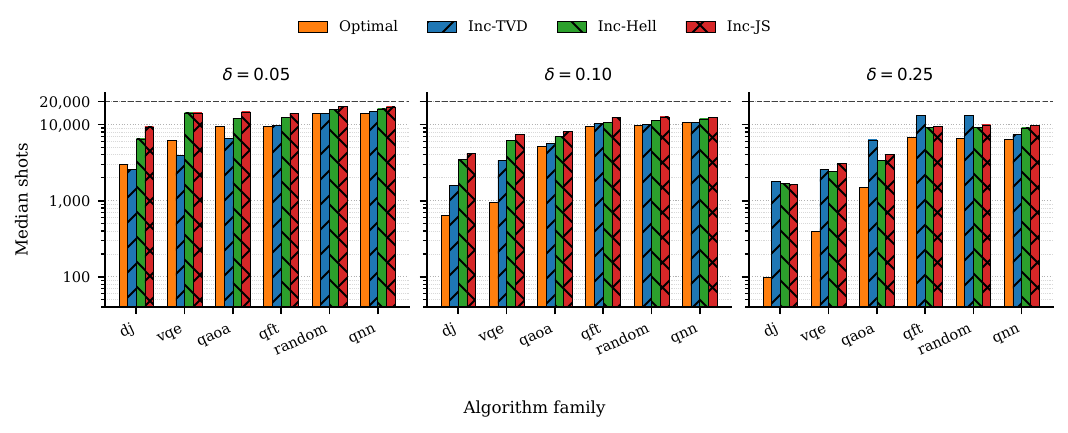}
    \caption{\rev{Algorithm-wise median sample-complexity profiles for the oracle and for the best \textsc{IncrementalExecution} configurations under TVD, Hellinger, and JS. Values are grouped by target tolerance $\delta$ and algorithm family. The dashed line marks the maximum budget $B=20{,}000$.}}
    \label{fig:algorithm-wise-sample-complexity}
\end{figure}

\rev{The previous analyses considered the behaviour of \textsc{IncrementalExecution} mainly with respect to its policy parameters, circuit size, backend noise model, and divergence metric. We now complement this discussion by interpreting the behaviour of the \textsc{IncrementalExecution} framework and its configurations from the perspective of the quantum algorithms included in the benchmark. Importantly, the algorithm label is not used by \textsc{IncrementalExecution}: the framework remains fully black-box and bases its decisions only on the observed measurement outcomes. Nevertheless, grouping the results of the best configurations by algorithm provides useful insight into the empirical difficulty of stabilising different classes of output distributions. Figure~\ref{fig:algorithm-wise-sample-complexity} summarises this perspective by reporting, for each algorithm family and target tolerance, the median oracle shot count and the median shot counts selected by the best \textsc{IncrementalExecution} configurations under the three divergence metrics.}

\rev{A first observation is that the benchmark algorithms exhibit markedly different stabilisation profiles. The \texttt{dj} instances are among the easiest cases for the framework. They often reach the a posteriori optimum with relatively few shots, especially for relaxed tolerances, and the best \textsc{IncrementalExecution} configurations rarely approach the full budget even at larger sizes. This behaviour is consistent with the intuition that these circuits tend to induce comparatively structured output distributions, for which the empirical distribution stabilises quickly. The static VQE instances show a similar, although more variable, behaviour: for loose tolerances they often require only a small fraction of the 20,000-shot budget, while under stricter tolerances the Hellinger- and JS-based configurations can become conservative. In these cases, TVD tends to provide the most economical stopping behaviour, as also visible in Figure~\ref{fig:algorithm-wise-sample-complexity}.}

\rev{The \texttt{qaoa} instances occupy an intermediate position. For small circuits, \textsc{IncrementalExecution} is generally able to stop well before the maximum budget, while for larger circuits and stricter tolerances the optimal and incremental shot counts grow substantially. This indicates that the difficulty of these instances is not solely algorithmic, but emerges from the interaction between algorithm family, circuit size, and target tolerance. In particular, when $\delta$ is relaxed, Hellinger-based configurations become competitive and sometimes require fewer shots than TVD-based configurations, suggesting that smoother divergences can better exploit approximate stabilisation in these cases.}

\rev{The hardest cases are represented by \texttt{qft}, \texttt{qnn}, and \texttt{random} circuits. For these algorithms, especially at 12 and 14 qubits, many traces have the a posteriori optimal shot count already close to the full budget under $\delta \in \{0.05,0.10\}$. Correspondingly, several \textsc{IncrementalExecution} configurations also stop near, or at, the 20,000-shot cap. This behaviour suggests that these circuits induce output distributions with larger effective support or slower empirical stabilisation, so that the point of diminishing returns is reached later. Even when $\delta=0.25$, these algorithm families typically require more shots than \texttt{dj} and static VQE, confirming that circuit size alone does not fully explain the observed difficulty.}

\rev{The algorithm-wise view also clarifies the role of the divergence metric. TVD is the most reliable choice in stringent regimes. This explains why TVD-based configurations are often preferable for \texttt{dj}, static VQE, and strict-$\delta$ settings. Conversely, Hellinger becomes increasingly attractive at moderate and loose tolerances, particularly for larger \texttt{qaoa}, \texttt{qft}, and static VQE instances, where its smoother response to small-probability fluctuations can reduce unnecessary sampling. JS generally behaves more conservatively in terms of shot usage, although it can still be competitive on specific traces.}

\rev{Overall, these results indicate that \textsc{IncrementalExecution} adapts not only to circuit size and backend noise, but also to the empirical distributional structure induced by the algorithm being executed. From a practical standpoint, this suggests that algorithm-aware policy selection can be beneficial when such information is available: TVD-based configurations are a robust default for strict accuracy requirements and more concentrated output distributions, while Hellinger-based configurations are a strong alternative for moderate-to-loose tolerances and algorithms whose output distributions stabilise more gradually. At the same time, these observations should be interpreted as tuning guidelines rather than assumptions of the framework: \textsc{IncrementalExecution} does not require algorithm-specific knowledge, and the algorithm-wise trends simply explain why different circuit families may benefit from different policy choices. A more detailed view of the per-trace sample-complexity results is reported in Appendix~\ref{app:full-sample-complexity}.}

\subsection{Concluding Discussion}
\label{sec:exp-conclusion}

\new{We now summarise the main findings of our experimental evaluation by explicitly linking them to the research questions introduced at the beginning of Section~\ref{sec:exp}.}

\new{\paragraph{RQ1: How the parameters of \textsc{IncrementalExecution} impact on finding the target accuracy?}
Our results show that reaching the target tolerance is highly sensitive to the stopping and stability parameters. In particular, validity concentrates in regimes where the stopping threshold is conservative (most notably $\varepsilon=0.01$), while larger thresholds quickly collapse the space of valid configurations. Stability enforcement plays a complementary role: increasing the stability window $k$ substantially raises the likelihood of finding valid configurations (especially under Hellinger and JS), with gains that largely saturate around $k\in\{3,5\}$. Finally, the stopping policy introduces a clear trade-off: \textsc{DMA} is generally more robust in stricter regimes (and is especially beneficial for the most demanding metric, TVD), whereas \textsc{Delta} tends to yield higher validity rates when $\delta$ is sufficiently permissive. Overall, these findings indicate that (i) $\varepsilon$ must typically be smaller than $\delta$ to obtain validity, and (ii) explicit stability enforcement is essential to avoid premature stopping driven by sampling noise.}

\new{\paragraph{RQ2: How the configuration of \textsc{IncrementalExecution} is affected by noise models and circuits?}
Circuit characteristics have a strong impact on feasibility, while the specific noise model plays a comparatively minor role within the considered backends. The partition into \emph{small} and \emph{large} circuits is reflected in consistently different validity regimes: small circuits admit many more valid configurations and exhibit near-perfect robustness for the selected best policies, whereas large circuits systematically reduce validity rates and make generalisation slightly less stable (e.g., $83.33\%$--$94.44\%$ in the worst cases). Conversely, the backend-level analysis shows that, for $\delta\in\{0.1,0.25\}$, validity percentages cluster tightly across the five simulated devices, suggesting that the main determinants of configuration feasibility are the target tolerance and circuit size rather than backend-specific noise idiosyncrasies. \rev{The algorithm-wise analysis further refines this observation: \texttt{dj} and static VQE instances are generally easier to stabilise, \texttt{qaoa} exhibits intermediate behaviour, whereas \texttt{qft}, \texttt{qnn}, and \texttt{random} circuits tend to require more shots, especially at larger sizes and under stricter tolerances. This suggests that algorithm family can provide a useful tip for policy tuning, even though \textsc{IncrementalExecution} itself remains fully black-box and does not rely on algorithm-specific information.} Backend-specific differences become visible mainly in stricter regimes (e.g., $\delta=0.05$), where validity is rare and small variations can be amplified.}

\new{\paragraph{RQ3: Which are the best configurations of \textsc{IncrementalExecution}?}
The best configurations selected with the train/test protocol confirm two consistent design patterns. First, robust best configurations typically combine conservative stopping (small $\varepsilon$ and moderate $\tau$) with stability enforcement ($k\ge 3$ for most settings). Second, the preferred stopping policy depends on the regime: \textsc{DMA} is frequently selected in stricter settings and for TVD-heavy regimes, while \textsc{Delta} appears more often among the best choices at looser tolerances and under Hellinger/JS, where it can exploit permissive constraints to stop earlier. Importantly, the selected configurations generalise well: for small circuits they achieve $100\%$ validity in all but one case (TVD at $\delta=0.10$), while for large circuits they still remain close to full validity, albeit with a few noticeable drops in the tightest settings. Hence, ``best'' configurations are not universal: they are conditional on circuit size and target tolerance, which motivates reporting best policies per $(\textit{size},\delta,\textit{metric})$ setting.}

\new{\paragraph{RQ4: How the best configurations of \textsc{IncrementalExecution} behave in comparison with the state of the art?}
Compared to concentration-inequality baselines (Hoeffding~\cite{hoeffding1963probability} and Weissman~\cite{weissman2003inequalities}), the best configurations of \textsc{IncrementalExecution} are orders of magnitude more sample-efficient in our setting. This gap is inherent: analytic inequalities provide worst-case, distribution-agnostic guarantees, whereas \textsc{IncrementalExecution} leverages online evidence from the observed output distributions to stop early. Empirically, the incremental budgets are consistently close to the \emph{a posteriori} optimal budgets, while maintaining high validity rates on the held-out test set, and they scale far more gracefully with circuit size than the inequality-based prescriptions.}

\new{At the same time, our results also highlight an important boundary case: in several traces (especially for larger circuits under stricter tolerances and for some metric/policy combinations), \textsc{IncrementalExecution} reaches the maximum budget of $20{,}000$ shots. This behaviour should not be interpreted only as ``policy inefficiency'': it can also signal that the fixed budget is itself too low to obtain a statistically meaningful estimate for those circuit/noise regimes, i.e., the experiment is still in a high-variance region where convergence has not yet occurred. In such cases, the number of shots required to reach a given tolerance may move closer to the (otherwise overconservative) analytic prescriptions, because the task effectively becomes harder and more ``worst-case-like''. Crucially, this is not a limitation of the incremental methodology per se. If the user is willing (or able) to allocate more shots than the initial budget, \textsc{IncrementalExecution} can simply continue beyond $20{,}000$ and still provide value: it will keep tracking convergence online and can identify the \emph{point of diminishing returns}, i.e., the regime where additional shots yield negligible improvements in the selected divergence. In other words, increasing the budget does not invalidate the approach; rather, it extends the observable convergence trajectory and allows the same policies to determine whether (and when) further execution becomes unproductive.}

\new{Finally, the metric comparison clarifies when each divergence is preferable: TVD is the most reliable choice under strict accuracy requirements (here, $\delta=0.05$), while Hellinger becomes increasingly competitive as $\delta$ loosens and can yield lower incremental budgets on large circuits at $\delta=0.25$. JS typically requires the most shots overall, but on isolated traces it can be the lowest.}

\new{\paragraph{Overall implications and limitations.}
Taken together, RQ1--RQ4 indicate that \textsc{IncrementalExecution} is a practical mechanism to reduce shot budgets while preserving a target distributional accuracy, provided that the given tolerance ($\delta$) is not excessively stringent and that stability is enforced to mitigate sampling fluctuations.
\rev{Circuit size, together with the algorithm-induced structure of the output distribution, is the primary factor shaping both feasibility and achievable savings, whereas backend noise variations are secondary in the regimes where validity is attainable.}
Finally, while our evaluation provides strong evidence that incremental policies outperform worst-case analytic baselines in realistic NISQ-like conditions, these conclusions are conditioned on simulator-based noise models and a finite reference budget ($20{,}000$ shots). Future work should validate these trends on real hardware and investigate adaptive batch sizing and policy selection objectives that explicitly optimise the reliability--savings frontier rather than a single combined score.}

%% file: src/applicability.tex
\section{Practical Applicability}
\label{sec:applicability}

A natural question arises regarding the feasibility of the proposed \textsc{IncrementalExecution} approach on real quantum hardware: \textit{can repeated, adaptive submissions be performed efficiently in practice, or do overheads such as queue times hinder their benefits?} While this concern is well-justified, recent advances in quantum cloud infrastructure mitigate these limitations and render incremental execution a viable strategy on current quantum platforms.

\subsection{Quantum Session Support}

Several quantum service providers, most notably \textit{IBM Quantum}, have introduced \emph{session-based execution models}~\cite{ibmsessions}. Within a session, users are allocated a continuous quantum processing window during which multiple circuit submissions can be issued with minimal latency. This capability enables rapid, successive dispatches of small shot batches, perfectly aligning with the needs of our incremental framework. By preserving contextual and hardware state across iterations, sessions reduce the per-batch overhead to negligible levels and significantly improve the responsiveness of adaptive execution schemes.

\subsection{Reservation-Based Execution Models}

Other providers, such as \textit{IonQ}~\cite{ionqreservation}, \textit{Rigetti}~\cite{rigettireservation}, and \textit{AWS Braket}~\cite{braketreservation}, offer alternative mechanisms that similarly support low-latency, iterative quantum-classical workloads. In particular, \emph{reservation-based models} allow users to pre-book dedicated time slots on quantum hardware, during which they can dispatch multiple jobs without re-entering the global job queue. This model provides users with predictable access and substantially reduced wait times, enabling the tight quantum-classical feedback loops required by \textsc{IncrementalExecution}. Some platforms further support job batching, in which a sequence of parameterised circuits can be submitted and evaluated efficiently within a single reservation window. These capabilities make it feasible to execute adaptive algorithms with controlled latency and sufficient throughput to maintain convergence guarantees, even in the presence of constrained hardware availability.

\subsection{Hybrid Execution Models in NISQ Workloads}

It is worth noting that the \textsc{IncrementalExecution} model proposed in this work closely parallels the hybrid quantum-classical loop that underpins many of the most prominent NISQ-era algorithms, including VQA, such as VQE, and QML approaches. These hybrid approaches involve parameterised quantum circuits that are iteratively updated by a classical optimiser based on quantum measurement results. Critically, each optimisation step requires fresh samples from the circuit with updated parameters — a process that is inherently iterative, non-deterministic, and highly sensitive to sampling noise.

The infrastructure, scheduling mechanisms, and interface layers developed for VQA-like applications are directly applicable to the \textsc{IncrementalExecution} paradigm we propose. Just as hybrid algorithms require timely feedback to guide optimisation, our framework requires rapid evaluation of intermediate measurement distributions to detect convergence and determine whether further execution is warranted.

This alignment is particularly significant given the central role that VQA and QML workloads play in current quantum computing research and application development. By leveraging the same architectural patterns and infrastructure already validated for these use cases, our framework not only becomes immediately implementable but also benefits from ongoing improvements in hybrid execution support.

\subsection{Using \textsc{IncrementalExecution} in VQA and QML Workloads}

\rev{This section discusses a possible integration scenario rather than an extension of our theoretical guarantees to fully non-stationary adaptive workloads. In a VQA or QML optimisation loop, the circuit parameters are updated between iterations; however, within a single iteration, the parameter values are fixed while the circuit is repeatedly sampled to estimate an objective, gradient, or output distribution. \textsc{IncrementalExecution} can be applied to this per-iteration static circuit evaluation. The optimisation trajectory as a whole remains non-stationary and requires additional mechanisms beyond the scope of this paper.}

\rev{A natural integration point for IncrementalExecution is the repeated sampling of a fixed parameterised circuit within one iteration of a hybrid workflow.} Instead of relying on fixed shot budgets for each circuit evaluation, it is possible to invoke \textsc{IncrementalExecution} at each step of the hybrid loop, dynamically determining the number of shots required to obtain a stable and accurate estimate of the circuit’s output distribution for the current parameter setting. This not only reduces shot waste in over-sampled evaluations but also prevents under-sampling in high-variance regions of the optimisation landscape.

Moreover, the framework’s support for the \texttt{initial\_guess} option allows it to exploit prior knowledge from the previous step’s output distribution. This is especially useful in hybrid algorithms, where subsequent parameter updates typically induce smooth, incremental changes in the quantum circuit’s output distribution. By bootstrapping, at the beginning of every iteration, the initial empirical distribution $\hat{P}_0$ with the empirical output of the previous iteration, convergence checks can begin from an already-informed prior, accelerating convergence and further reducing execution cost.

Recent work~\cite{kim2024distribution} supports this idea by empirically demonstrating that the output distributions of parametrised quantum circuits often evolve smoothly in parameter space. The authors propose using a form of distributional warm start, where knowledge of previously observed distributions informs the sample allocation strategy of subsequent iterations. Our framework can readily incorporate such strategies: for instance, the divergence between the current empirical distribution and the warm-start prior can guide the batch size allocation or convergence threshold, adapting more rapidly when parameter shifts are small.

\subsection{Summary}

In summary, while traditional execution paradigms on cloud-based quantum hardware may introduce latency barriers, emerging infrastructure developments, such as quantum sessions, reservation-based access, batched execution, and priority queues, actively support the kinds of adaptive, interactive workloads required by our framework. Moreover, the structural similarities between our incremental approach and widely adopted hybrid quantum-classical algorithms further reinforce its practical viability.

Importantly, VQA and QML circuits are particularly well-suited for integration with \textsc{IncrementalExecution}: not only does the framework align with their feedback-driven nature, but it can also exploit inter-step distributional smoothness to accelerate convergence and reduce shot usage. By leveraging distributional priors, either directly from prior steps or via more sophisticated mechanisms such as those proposed in~\cite{kim2024distribution}, the framework supports intelligent shot allocation across optimisation steps, paving the way for more efficient, noise-resilient hybrid quantum workflows.

The proposed execution strategy is thus well-aligned with the current operational realities and architectural trajectories of NISQ-era quantum computing.

%% file: src/threats.tex
\section{Threats to Validity}
\label{sec:threats}

We discuss potential threats to the validity of our findings, organised according to commonly accepted categories: internal, external, construct, and conclusion validity~\cite{wohlin2012experimentation}.

\subsection{Internal Validity}

Internal validity concerns whether the experimental results are attributable to the variables being manipulated rather than to confounding factors. Our experiments rely on synthetic execution traces generated by IBM’s noisy \texttt{FakeBackend} simulators. While these simulators emulate real device noise based on recent calibration data, they may not fully capture temporal noise drift, scheduling effects, or transient failures encountered in real hardware. Nevertheless, by standardising the simulation environment across all tested configurations, we ensure that relative comparisons between policy variants remain valid.

Additionally, care was taken to avoid implementation bias: all stopping criteria, distance functions, and framework internals were independently tested and validated. Outlier removal and convergence filtering were performed systematically and without favouring specific configurations.

\subsection{External Validity}

External validity refers to the generalisability of results beyond the specific circuits, backends, and parameter settings used in the study. Although our benchmark includes a diverse set of \new{180} circuit-backend combinations spanning multiple quantum algorithms (QPE, Grover, QFT, random circuits) and realistic noise models, it is still limited in scope. Results may not generalise to quantum algorithms with fundamentally different output characteristics or to future quantum processors with radically different noise behaviours. \new{In particular, results may not generalise to \emph{variational} or other \emph{adaptive} workloads, where the output distribution changes across optimisation iterations and the stationarity assumption underlying our stopping criterion no longer holds without additional mechanisms.}

\rev{Our evaluation is also limited to circuits up to 14 qubits. This limitation concerns the empirical validation of specific configurations, not the applicability of \textsc{IncrementalExecution} itself. The framework and policies are qubit-agnostic: they require only online access to batches of measurement outcomes and do not assume a fixed circuit size, topology, or noise model. However, evaluating larger and deeper circuits requires substantially more resources because the a posteriori reference distribution used for validation becomes more expensive to estimate reliably as the effective output support grows. Consequently, we leave a systematic evaluation beyond 14 qubits and for wider/deeper circuit families to future work.}

Furthermore, we did not evaluate execution on real quantum hardware. While Section~\ref{sec:applicability} argues that the framework is compatible with modern hybrid and session-based runtimes, empirical confirmation on actual devices remains future work.

\subsection{Construct Validity}

Construct validity addresses whether the evaluation measures faithfully capture the intended properties—in our case, convergence and accuracy. The use of Total Variation Distance as the sole divergence metric, while standard and interpretable, may not capture all relevant aspects of output quality, especially for structured or task-specific distributions. However, our convergence criteria are generic and metric-agnostic; alternative divergence measures (e.g., Jensen–Shannon, Hellinger) could easily be substituted.


\subsection{Conclusion Validity}

Conclusion validity concerns the statistical soundness and reproducibility of the results. Our findings are based on over $7.3M$ experiments across a carefully designed parameter grid. Outlier filtering, validation/test splitting, and relative performance metrics were used to ensure robustness. However, we did not apply formal statistical hypothesis testing or confidence interval estimation, which could be considered in future work to further strengthen conclusions.

Additionally, while we focused on relative performance comparisons, absolute performance could vary depending on backend availability, calibration profiles, and real-time load conditions. These factors should be taken into account when interpreting results in operational settings.

\subsection{Summary}

\new{In summary, although our results are subject to the standard limitations of simulation-based evaluation and to the deliberate scope restriction to static, non-variational circuit executions, we have taken multiple steps to ensure fairness, reproducibility, and relevance of our conclusions. Future work will include validation on real quantum hardware, extensions to adaptive/variational workloads via mechanisms to handle non-stationary output distributions, and broader evaluation across additional circuit families and divergence measures. We also plan to study system-level considerations for integrating online stopping into cloud-based quantum runtimes, including the impact of batching, session semantics, and backend scheduling variability.}

%% file: src/conclusions.tex
\section{Conclusions and Future Work}
\label{sec:conclusions}

\new{In this paper, we investigated the following main research question: \emph{\textbf{MRQ:} Given a static, non-variational quantum circuit and a noisy quantum processing unit (QPU), is it possible to determine how many shots are sufficient to reach a required target accuracy?} Our results provide a positive answer within the static execution scope considered here. Specifically, we showed that it is possible to determine a sufficient shot count \emph{online}---without assumptions on circuit structure or backend noise models---by monitoring distributional stabilisation and stopping at the point of diminishing returns, i.e., when additional shots no longer produce statistically meaningful changes in the empirical output distribution.}

\new{Central to our approach is the concept of the point of diminishing returns, which provides an observable stabilisation signal without requiring access to the unknown target distribution. To ground evaluation in a principled baseline, we further introduced the notion of a posteriori optimality, which retrospectively identifies the minimal prefix of an execution trace whose empirical distribution matches the best available estimate under the given budget.}

\new{The primary objective of this paper was to establish and validate \textsc{IncrementalExecution} as a general-purpose solution within the scope of static circuit executions. Through an extensive empirical evaluation comprising over 7.3M experiments across 33,750 policy configurations and 180 circuit--backend pairs, our results indicate that:
\begin{itemize}
    \item \textbf{Answer to the main research question:} \new{for static circuits, one can determine \emph{online} a shot count that is sufficient to meet a user-defined stability/accuracy target by detecting distributional stabilisation and stopping at the observed point of diminishing returns, without requiring circuit- or noise-model assumptions.}
   \item \textbf{Policy-driven flexibility for static circuits:} \textsc{IncrementalExecution} can be configured to reach user-specified stability/accuracy targets across the evaluated execution traces, without relying on circuit-specific or backend-specific prior knowledge. This demonstrates that effective shot optimisation can be achieved in a black-box manner when the circuit remains fixed during execution.
   \item \textbf{Near-optimal online stopping:} Compared to the a posteriori optimal baseline (which assumes access to the full execution history), the overhead incurred by online decision making is typically small, while achieving comparable distributional fidelity under the adopted divergence criterion.
\end{itemize}}

\new{Our evaluation also highlights that there is no universally best configuration: optimal performance depends on properties of the executed circuit and backend (e.g., circuit size and output complexity). This suggests that systematic configuration and meta-optimisation can yield additional gains. As an initial step, we showed that circuit size provides a meaningful signal for guiding such tuning, and we anticipate that richer circuit descriptors could further improve configuration selection.}

\new{Finally, we emphasise that the framework is not only conceptually principled but also practically deployable. Emerging support for session-based execution, reservation models, and hybrid runtimes across quantum cloud platforms makes online stopping strategies increasingly viable in operational settings, where reducing unnecessary shots can translate into tangible savings in time and cost.}

\new{While the IncrementalExecution framework is, in principle, applicable to a broad class of stochastic sampling processes, quantum computing constitutes a domain in which its advantages are particularly pronounced. In this setting, shot budgets translate directly into monetary cost, wall-clock latency, and hardware contention, while oversampling cannot overcome accuracy plateaus induced by noise. As a result, the ability to detect the point of diminishing returns online is uniquely valuable for quantum workloads. In summary, although our methodology is formulated in general statistical terms, it is motivated by, and specifically tailored to, the practical constraints of quantum computing. By abstracting away circuit- and noise-specific assumptions, we offer a broadly applicable solution to a problem widely recognised as timely and important in the quantum computing literature: determining how many shots are sufficient.}

\new{Looking ahead, several directions could further extend the scope and impact of IncrementalExecution:}
\begin{description}
    \item[\new{Beyond fixed-budget baselines.}] \new{In this work, optimality is defined relative to a fixed budget $\mathcal{B}$ via the reference distribution $\hat{P}_\mathcal{B}$. A natural extension is to study a \emph{budget-less} formulation, where the goal is to determine a stopping point without assuming an a priori maximum number of shots. Addressing this setting requires principled criteria for termination under uncertainty (e.g., confidence guarantees or asymptotic stabilisation tests), and we plan to investigate both theoretical and empirical solutions.}

    \item[\new{Integration with complementary execution techniques.}] \new{We currently consider executing a circuit on a single backend. Future work will embed \textsc{IncrementalExecution} into broader quantum workflow optimisations~\cite{bisicchia2024quantum}, including \textit{shot-wise distribution}~\cite{bisicchia2024distributingQCE,bisicchia2023distributing,bisicchia2024distributing,bisicchia2023dispatching}, \textit{circuit composition}~\cite{ohkura2022simultaneous,das2019case,bisicchia2025maximizing}, and \textit{circuit cutting}~\cite{tang2021cutqc,bisicchia2024cut,bisicchia2026cut}. These integrations raise interesting questions about how stabilisation signals compose across distributed and decomposed executions.} \rev{A particularly promising direction is the integration of \textsc{IncrementalExecution} with \textit{circuit cutting}. Large circuits could be decomposed into smaller fragments whose output distributions are sampled independently. Since our current experiments already identify effective configurations for smaller static circuits, these configurations could serve as initial policies for fragment-level execution. \textsc{IncrementalExecution} would then determine, for each fragment, when additional shots no longer significantly improve the fragment distribution. The main open question is how fragment-level stabilisation errors compose during the reconstruction of the full-circuit observable or distribution. Studying the interaction between circuit cutting, fragment-wise shot optimisation, and reconstruction error therefore constitutes an interesting direction for scaling the approach to larger circuits.}

    \item[\new{Learning-based policy design.}] \new{We plan to develop learning-based strategies (e.g., reinforcement learning) to synthesise shot-allocation and stopping policies from data. Such approaches could move beyond static heuristics by conditioning decisions on circuit descriptors, observed stabilisation dynamics, and backend characteristics, learning policies that generalise across circuit families and execution environments.}

    \item[\new{Validation on real quantum hardware.}] \new{While our experiments use realistic noisy simulators, validating the framework on real devices is essential. Real hardware introduces additional factors, including queueing delays, calibration drift, transient failures, and runtime scheduling variability, that may affect both the stopping signal and the realised savings. Empirical evaluation under these conditions is a key next step.}

    \item[\new{Automated hyperparameter tuning and configuration selection.}] \new{Our results show that no single policy configuration is uniformly optimal across all circuit--backend pairs. Future work will therefore explore automated tuning and selection mechanisms, including Bayesian optimisation, evolutionary strategies, and AutoML techniques. By linking circuit properties (e.g., qubit count, depth, output sparsity) to effective hyperparameter settings, we aim to provide practical guidelines and tooling for deploying \textsc{IncrementalExecution} in real-world quantum software workflows.}

    \item[\new{Extension to variational and adaptive workloads.}] \new{A particularly important direction is extending \textsc{IncrementalExecution} to \emph{variational} and other \emph{adaptive} algorithms. Unlike the static setting studied here, these workloads induce non-stationary output distributions because the circuit changes across optimisation iterations. Supporting such settings will require additional mechanisms (e.g., phase-aware stopping rules, change-point detection, or iteration-coupled policies) to disentangle stabilisation from distributional shifts.}
\end{description}

%% file: src/thAnalysis.tex
\appendix

\section{Complete data comparison of incremental shots with Weissman and Hoeffding}
\label{app:full-sample-complexity}

This appendix section reports the complete per-trace tables underlying the best-configuration comparison discussed in Section~\ref{sec:best-configs-comparison}. For each tolerance $\delta\in\{0.05,0.10,0.25\}$, we list (i) the oracle shot budget, (ii) the shot budgets selected by \textsc{IncrementalExecution} using the best configuration for each divergence metric (Inc-TVD, Inc-Hell, Inc-JS), and (iii) the two \emph{a priori} inequality baselines (Weissman and Hoeffding) at $95\%$ confidence. As in the main text, the Weissman and Hoeffding values depend only on $\delta$ and the circuit size, and therefore remain constant within each ($\delta$, size) group. For the sake of presentation, Tables~\ref{tab:appendix-sota-d005-s}--\ref{tab:appendix-sota-d025-l} divide the test set traces by circuit sizes, small and large.

\begin{table}[ht!]
\centering
\footnotesize
\setlength{\tabcolsep}{3pt}
\renewcommand{\arraystretch}{1.05}
\caption{Full sample comparison for $\delta = 0.05$, all small circuits test traces.}
\label{tab:appendix-sota-d005-s}
\begin{tabular}{l c r r r r r r}
\toprule
Trace & Size & Oracle & Inc-TVD & Inc-Hell & Inc-JS & Weissman & Hoeffding \\
\midrule
dj\_4\_FEZ & 4 & 100 & \textbf{550} & 2000 & 2000 & 2818 & $8.27e+04$ \\
dj\_4\_FEZ & 4 & 150 & \textbf{550} & 2000 & 2000 & 2818 & $8.27e+04$ \\
dj\_4\_TOR & 4 & 50 & \textbf{550} & 2300 & 3000 & 2818 & $8.27e+04$ \\
dj\_4\_TOR & 4 & 300 & \textbf{550} & 2300 & 3000 & 2818 & $8.27e+04$ \\
dj\_4\_TOR & 4 & 550 & \textbf{550} & 2300 & 3000 & 2818 & $8.27e+04$ \\
qaoa\_4\_SHE & 4 & 300 & \textbf{1150} & 3700 & 3800 & 2818 & $8.27e+04$ \\
qaoa\_4\_SHE & 4 & 1150 & \textbf{1150} & 3700 & 3800 & 2818 & $8.27e+04$ \\
qaoa\_4\_SHE & 4 & 1600 & \textbf{1150} & 3700 & 3800 & 2818 & $8.27e+04$ \\
qaoa\_6\_MAR & 6 & 1800 & \textbf{3850} & 6200 & 7300 & 9472 & $1.61e+06$ \\
qaoa\_6\_MAR & 6 & 3000 & \textbf{3850} & 6200 & 7300 & 9472 & $1.61e+06$ \\
qaoa\_6\_MAR & 6 & 3350 & \textbf{3850} & 6200 & 7300 & 9472 & $1.61e+06$ \\
qft\_6\_FEZ & 6 & 3450 & \textbf{4850} & 6400 & 7300 & 9472 & $1.61e+06$ \\
qft\_6\_FEZ & 6 & 3000 & \textbf{4850} & 6400 & 7300 & 9472 & $1.61e+06$ \\
qft\_6\_FEZ & 6 & 3600 & \textbf{4850} & 6400 & 7300 & 9472 & $1.61e+06$ \\
qft\_6\_TOR & 6 & 4000 & \textbf{5350} & 6600 & 7900 & 9472 & $1.61e+06$ \\
qft\_6\_TOR & 6 & 3500 & \textbf{5350} & 6600 & 7900 & 9472 & $1.61e+06$ \\
qft\_6\_TOR & 6 & 4250 & \textbf{5350} & 6600 & 7900 & 9472 & $1.61e+06$ \\
random\_6\_FEZ & 6 & 2250 & \textbf{4350} & 6100 & 7400 & 9472 & $1.61e+06$ \\
random\_6\_FEZ & 6 & 2700 & \textbf{4350} & 6100 & 7400 & 9472 & $1.61e+06$ \\
random\_6\_FEZ & 6 & 3600 & \textbf{4350} & 6100 & 7400 & 9472 & $1.61e+06$ \\
vqe\_6\_KYI & 6 & 150 & \textbf{1350} & 4800 & 5300 & 9472 & $1.61e+06$ \\
vqe\_6\_KYI & 6 & 1500 & \textbf{1350} & 4800 & 5300 & 9472 & $1.61e+06$ \\
vqe\_6\_KYI & 6 & 1550 & \textbf{1350} & 4800 & 5300 & 9472 & $1.61e+06$ \\
dj\_8\_FEZ & 8 & 250 & \textbf{1550} & 5100 & 6200 & $3.61e+04$ & $3.03e+07$ \\
dj\_8\_FEZ & 8 & 3200 & \textbf{1550} & 5100 & 6200 & $3.61e+04$ & $3.03e+07$ \\
dj\_8\_FEZ & 8 & 3850 & \textbf{1550} & 5100 & 6200 & $3.61e+04$ & $3.03e+07$ \\
dj\_8\_MAR & 8 & 350 & \textbf{1250} & 6500 & 6500 & $3.61e+04$ & $3.03e+07$ \\
dj\_8\_MAR & 8 & 3050 & \textbf{1250} & 6500 & 6500 & $3.61e+04$ & $3.03e+07$ \\
dj\_8\_MAR & 8 & 3600 & \textbf{1250} & 6500 & 6500 & $3.61e+04$ & $3.03e+07$ \\
qaoa\_8\_KYI & 8 & 5850 & \textbf{6550} & 11200 & 14700 & $3.61e+04$ & $3.03e+07$ \\
qaoa\_8\_KYI & 8 & 8100 & \textbf{6550} & 11200 & 14700 & $3.61e+04$ & $3.03e+07$ \\
qaoa\_8\_KYI & 8 & 9350 & \textbf{6550} & 11200 & 14700 & $3.61e+04$ & $3.03e+07$ \\
qaoa\_8\_SHE & 8 & 5350 & \textbf{6350} & 12000 & 13900 & $3.61e+04$ & $3.03e+07$ \\
qaoa\_8\_SHE & 8 & 7850 & \textbf{6350} & 12000 & 13900 & $3.61e+04$ & $3.03e+07$ \\
qaoa\_8\_SHE & 8 & 9400 & \textbf{6350} & 12000 & 13900 & $3.61e+04$ & $3.03e+07$ \\
qft\_8\_FEZ & 8 & 8700 & \textbf{9650} & 12500 & 13700 & $3.61e+04$ & $3.03e+07$ \\
qft\_8\_FEZ & 8 & 7800 & \textbf{9650} & 12500 & 13700 & $3.61e+04$ & $3.03e+07$ \\
qft\_8\_FEZ & 8 & 9450 & \textbf{9650} & 12500 & 13700 & $3.61e+04$ & $3.03e+07$ \\
qft\_8\_TOR & 8 & 8450 & \textbf{9850} & 11700 & 13900 & $3.61e+04$ & $3.03e+07$ \\
qft\_8\_TOR & 8 & 7850 & \textbf{9850} & 11700 & 13900 & $3.61e+04$ & $3.03e+07$ \\
qft\_8\_TOR & 8 & 9450 & \textbf{9850} & 11700 & 13900 & $3.61e+04$ & $3.03e+07$ \\
qnn\_8\_FEZ & 8 & 7250 & \textbf{8150} & 12100 & 13600 & $3.61e+04$ & $3.03e+07$ \\
qnn\_8\_FEZ & 8 & 7450 & \textbf{8150} & 12100 & 13600 & $3.61e+04$ & $3.03e+07$ \\
qnn\_8\_FEZ & 8 & 9250 & \textbf{8150} & 12100 & 13600 & $3.61e+04$ & $3.03e+07$ \\
qnn\_8\_KYI & 8 & 8200 & \textbf{9550} & 11500 & 14100 & $3.61e+04$ & $3.03e+07$ \\
qnn\_8\_KYI & 8 & 7550 & \textbf{9550} & 11500 & 14100 & $3.61e+04$ & $3.03e+07$ \\
qnn\_8\_KYI & 8 & 9200 & \textbf{9550} & 11500 & 14100 & $3.61e+04$ & $3.03e+07$ \\
random\_8\_FEZ & 8 & 7600 & \textbf{8250} & 11500 & 14700 & $3.61e+04$ & $3.03e+07$ \\
random\_8\_FEZ & 8 & 8100 & \textbf{8250} & 11500 & 14700 & $3.61e+04$ & $3.03e+07$ \\
random\_8\_FEZ & 8 & 9500 & \textbf{8250} & 11500 & 14700 & $3.61e+04$ & $3.03e+07$ \\
vqe\_8\_KYI & 8 & 1000 & \textbf{2450} & 8400 & 8400 & $3.61e+04$ & $3.03e+07$ \\
vqe\_8\_KYI & 8 & 5700 & \textbf{2450} & 8400 & 8400 & $3.61e+04$ & $3.03e+07$ \\
vqe\_8\_KYI & 8 & 6700 & \textbf{2450} & 8400 & 8400 & $3.61e+04$ & $3.03e+07$ \\
\bottomrule
\end{tabular}
\end{table}

\clearpage
\begin{table}[ht!]
\centering
\footnotesize
\setlength{\tabcolsep}{3pt}
\renewcommand{\arraystretch}{1.05}
\caption{Full sample comparison for $\delta = 0.05$, all large circuits test traces.}
\label{tab:appendix-sota-d005-l}
\begin{tabular}{l c r r r r r r}
\toprule
Trace & Size & Oracle & Inc-TVD & Inc-Hell & Inc-JS & Weissman & Hoeffding \\
\midrule
dj\_10\_FEZ & 10 & 500 & \textbf{2550} & 5600 & 9350 & $1.43e+05$ & $5.57e+08$ \\
dj\_10\_FEZ & 10 & 6750 & \textbf{2550} & 5600 & 9350 & $1.43e+05$ & $5.57e+08$ \\
dj\_10\_FEZ & 10 & 7750 & \textbf{2550} & 5600 & 9350 & $1.43e+05$ & $5.57e+08$ \\
dj\_10\_TOR & 10 & 1550 & \textbf{3950} & 9000 & 15250 & $1.43e+05$ & $5.57e+08$ \\
dj\_10\_TOR & 10 & 10050 & \textbf{3950} & 9000 & 15250 & $1.43e+05$ & $5.57e+08$ \\
dj\_10\_TOR & 10 & 10650 & \textbf{3950} & 9000 & 15250 & $1.43e+05$ & $5.57e+08$ \\
qaoa\_10\_SHE & 10 & 13800 & \textbf{16650} & 20000 & 20000 & $1.43e+05$ & $5.57e+08$ \\
qaoa\_10\_SHE & 10 & 15400 & \textbf{16650} & 20000 & 20000 & $1.43e+05$ & $5.57e+08$ \\
qaoa\_10\_SHE & 10 & 16350 & \textbf{16650} & 20000 & 20000 & $1.43e+05$ & $5.57e+08$ \\
qaoa\_12\_MAR & 12 & 14200 & \textbf{15650} & 20000 & 20000 & $5.68e+05$ & $1.01e+10$ \\
qaoa\_12\_MAR & 12 & 17550 & \textbf{15650} & 20000 & 20000 & $5.68e+05$ & $1.01e+10$ \\
qaoa\_12\_MAR & 12 & 17850 & \textbf{15650} & 20000 & 20000 & $5.68e+05$ & $1.01e+10$ \\
qft\_12\_FEZ & 12 & 18650 & \textbf{20000} & 20000 & 20000 & $5.68e+05$ & $1.01e+10$ \\
qft\_12\_FEZ & 12 & 18550 & \textbf{20000} & 20000 & 20000 & $5.68e+05$ & $1.01e+10$ \\
qft\_12\_FEZ & 12 & 18900 & \textbf{20000} & 20000 & 20000 & $5.68e+05$ & $1.01e+10$ \\
qft\_12\_TOR & 12 & 18600 & \textbf{20000} & 20000 & 20000 & $5.68e+05$ & $1.01e+10$ \\
qft\_12\_TOR & 12 & 18550 & \textbf{20000} & 20000 & 20000 & $5.68e+05$ & $1.01e+10$ \\
qft\_12\_TOR & 12 & 18900 & \textbf{20000} & 20000 & 20000 & $5.68e+05$ & $1.01e+10$ \\
random\_12\_FEZ & 12 & 18550 & \textbf{20000} & 20000 & 20000 & $5.68e+05$ & $1.01e+10$ \\
random\_12\_FEZ & 12 & 18650 & \textbf{20000} & 20000 & 20000 & $5.68e+05$ & $1.01e+10$ \\
random\_12\_FEZ & 12 & 18950 & \textbf{20000} & 20000 & 20000 & $5.68e+05$ & $1.01e+10$ \\
vqe\_12\_KYI & 12 & 3150 & \textbf{5350} & 20000 & 20000 & $5.68e+05$ & $1.01e+10$ \\
vqe\_12\_KYI & 12 & 13400 & \textbf{5350} & 20000 & 20000 & $5.68e+05$ & $1.01e+10$ \\
vqe\_12\_KYI & 12 & 14300 & \textbf{5350} & 20000 & 20000 & $5.68e+05$ & $1.01e+10$ \\
dj\_14\_FEZ & 14 & 2200 & \textbf{3950} & 12400 & 19550 & $2.27e+06$ & $1.80e+11$ \\
dj\_14\_FEZ & 14 & 12450 & \textbf{3950} & 12400 & 19550 & $2.27e+06$ & $1.80e+11$ \\
dj\_14\_FEZ & 14 & 12850 & \textbf{3950} & 12400 & 19550 & $2.27e+06$ & $1.80e+11$ \\
dj\_14\_MAR & 14 & 2350 & \textbf{3550} & 13200 & 14350 & $2.27e+06$ & $1.80e+11$ \\
dj\_14\_MAR & 14 & 10050 & \textbf{3550} & 13200 & 14350 & $2.27e+06$ & $1.80e+11$ \\
dj\_14\_MAR & 14 & 10650 & \textbf{3550} & 13200 & 14350 & $2.27e+06$ & $1.80e+11$ \\
qaoa\_14\_KYI & 14 & 17550 & \textbf{20000} & 20000 & 20000 & $2.27e+06$ & $1.80e+11$ \\
qaoa\_14\_KYI & 14 & 19350 & \textbf{20000} & 20000 & 20000 & $2.27e+06$ & $1.80e+11$ \\
qaoa\_14\_KYI & 14 & 19400 & \textbf{20000} & 20000 & 20000 & $2.27e+06$ & $1.80e+11$ \\
qaoa\_14\_SHE & 14 & 17900 & \textbf{20000} & 20000 & 20000 & $2.27e+06$ & $1.80e+11$ \\
qaoa\_14\_SHE & 14 & 19500 & \textbf{20000} & 20000 & 20000 & $2.27e+06$ & $1.80e+11$ \\
qft\_14\_FEZ & 14 & 18950 & \textbf{20000} & 20000 & 20000 & $2.27e+06$ & $1.80e+11$ \\
qft\_14\_FEZ & 14 & 19750 & \textbf{20000} & 20000 & 20000 & $2.27e+06$ & $1.80e+11$ \\
qft\_14\_FEZ & 14 & 19800 & \textbf{20000} & 20000 & 20000 & $2.27e+06$ & $1.80e+11$ \\
qft\_14\_TOR & 14 & 18950 & \textbf{20000} & 20000 & 20000 & $2.27e+06$ & $1.80e+11$ \\
qft\_14\_TOR & 14 & 19800 & \textbf{20000} & 20000 & 20000 & $2.27e+06$ & $1.80e+11$ \\
qnn\_14\_FEZ & 14 & 18950 & \textbf{20000} & 20000 & 20000 & $2.27e+06$ & $1.80e+11$ \\
qnn\_14\_FEZ & 14 & 19750 & \textbf{20000} & 20000 & 20000 & $2.27e+06$ & $1.80e+11$ \\
qnn\_14\_FEZ & 14 & 19800 & \textbf{20000} & 20000 & 20000 & $2.27e+06$ & $1.80e+11$ \\
qnn\_14\_KYI & 14 & 18950 & \textbf{20000} & 20000 & 20000 & $2.27e+06$ & $1.80e+11$ \\
qnn\_14\_KYI & 14 & 19750 & \textbf{20000} & 20000 & 20000 & $2.27e+06$ & $1.80e+11$ \\
qnn\_14\_KYI & 14 & 19800 & \textbf{20000} & 20000 & 20000 & $2.27e+06$ & $1.80e+11$ \\
random\_14\_FEZ & 14 & 18950 & \textbf{20000} & 20000 & 20000 & $2.27e+06$ & $1.80e+11$ \\
random\_14\_FEZ & 14 & 19750 & \textbf{20000} & 20000 & 20000 & $2.27e+06$ & $1.80e+11$ \\
random\_14\_FEZ & 14 & 19800 & \textbf{20000} & 20000 & 20000 & $2.27e+06$ & $1.80e+11$ \\
vqe\_14\_KYI & 14 & 8500 & \textbf{7750} & 20000 & 20000 & $2.27e+06$ & $1.80e+11$ \\
vqe\_14\_KYI & 14 & 17050 & \textbf{7750} & 20000 & 20000 & $2.27e+06$ & $1.80e+11$ \\
vqe\_14\_KYI & 14 & 17300 & \textbf{7750} & 20000 & 20000 & $2.27e+06$ & $1.80e+11$ \\
\bottomrule
\end{tabular}
\end{table}
 \clearpage

\begin{table}[ht!]
\centering
\footnotesize
\setlength{\tabcolsep}{3pt}
\renewcommand{\arraystretch}{1.05}
\caption{Full sample comparison for $\delta = 0.10$, all small circuits test traces.}
\label{tab:appendix-sota-d01-s}
\begin{tabular}{l c r r r r r r}
\toprule
Trace & Size & Oracle & Inc-TVD & Inc-Hell & Inc-JS & Weissman & Hoeffding \\
\midrule
dj\_4\_FEZ & 4 & 50 & \textbf{150} & 700 & 700 & 705 & $2.07e+04$ \\
dj\_4\_FEZ & 4 & 100 & \textbf{150} & 700 & 700 & 705 & $2.07e+04$ \\
dj\_4\_TOR & 4 & 50 & \textbf{150} & 550 & 500 & 705 & $2.07e+04$ \\
qaoa\_4\_SHE & 4 & 150 & \textbf{150} & 1300 & 900 & 705 & $2.07e+04$ \\
qaoa\_4\_SHE & 4 & 350 & \textbf{150} & 1300 & 900 & 705 & $2.07e+04$ \\
qaoa\_4\_SHE & 4 & 400 & \textbf{150} & 1300 & 900 & 705 & $2.07e+04$ \\
qaoa\_6\_MAR & 6 & 650 & \textbf{1450} & 1950 & 2400 & 2368 & $4.02e+05$ \\
qaoa\_6\_MAR & 6 & 1350 & \textbf{1450} & 1950 & 2400 & 2368 & $4.02e+05$ \\
qaoa\_6\_MAR & 6 & 1500 & \textbf{1450} & 1950 & 2400 & 2368 & $4.02e+05$ \\
qft\_6\_FEZ & 6 & 1000 & \textbf{1750} & 1800 & 2500 & 2368 & $4.02e+05$ \\
qft\_6\_FEZ & 6 & 850 & \textbf{1750} & 1800 & 2500 & 2368 & $4.02e+05$ \\
qft\_6\_FEZ & 6 & 1050 & \textbf{1750} & 1800 & 2500 & 2368 & $4.02e+05$ \\
qft\_6\_TOR & 6 & 850 & 2050 & \textbf{1950} & 2600 & 2368 & $4.02e+05$ \\
qft\_6\_TOR & 6 & 800 & 2050 & \textbf{1950} & 2600 & 2368 & $4.02e+05$ \\
qft\_6\_TOR & 6 & 1100 & 2050 & \textbf{1950} & 2600 & 2368 & $4.02e+05$ \\
random\_6\_FEZ & 6 & 600 & \textbf{1700} & 1800 & 2600 & 2368 & $4.02e+05$ \\
random\_6\_FEZ & 6 & 700 & \textbf{1700} & 1800 & 2600 & 2368 & $4.02e+05$ \\
random\_6\_FEZ & 6 & 850 & \textbf{1700} & 1800 & 2600 & 2368 & $4.02e+05$ \\
vqe\_6\_KYI & 6 & 50 & \textbf{150} & 1450 & 1500 & 2368 & $4.02e+05$ \\
vqe\_6\_KYI & 6 & 300 & \textbf{150} & 1450 & 1500 & 2368 & $4.02e+05$ \\
dj\_8\_FEZ & 8 & 100 & \textbf{550} & 2300 & 1900 & 9023 & $7.56e+06$ \\
dj\_8\_FEZ & 8 & 600 & \textbf{550} & 2300 & 1900 & 9023 & $7.56e+06$ \\
dj\_8\_FEZ & 8 & 950 & \textbf{550} & 2300 & 1900 & 9023 & $7.56e+06$ \\
dj\_8\_MAR & 8 & 50 & \textbf{150} & 1400 & 2000 & 9023 & $7.56e+06$ \\
dj\_8\_MAR & 8 & 700 & \textbf{150} & 1400 & 2000 & 9023 & $7.56e+06$ \\
qaoa\_8\_KYI & 8 & 2000 & \textbf{2350} & 4150 & 5400 & 9023 & $7.56e+06$ \\
qaoa\_8\_KYI & 8 & 3350 & \textbf{2350} & 4150 & 5400 & 9023 & $7.56e+06$ \\
qaoa\_8\_KYI & 8 & 3950 & \textbf{2350} & 4150 & 5400 & 9023 & $7.56e+06$ \\
qaoa\_8\_SHE & 8 & 1700 & \textbf{2300} & 4150 & 4600 & 9023 & $7.56e+06$ \\
qaoa\_8\_SHE & 8 & 3350 & \textbf{2300} & 4150 & 4600 & 9023 & $7.56e+06$ \\
qaoa\_8\_SHE & 8 & 3950 & \textbf{2300} & 4150 & 4600 & 9023 & $7.56e+06$ \\
qft\_8\_FEZ & 8 & 3300 & 3650 & \textbf{3300} & 4900 & 9023 & $7.56e+06$ \\
qft\_8\_FEZ & 8 & 2800 & 3650 & \textbf{3300} & 4900 & 9023 & $7.56e+06$ \\
qft\_8\_FEZ & 8 & 3850 & 3650 & \textbf{3300} & 4900 & 9023 & $7.56e+06$ \\
qft\_8\_TOR & 8 & 3600 & 3500 & \textbf{3350} & 4800 & 9023 & $7.56e+06$ \\
qft\_8\_TOR & 8 & 3000 & 3500 & \textbf{3350} & 4800 & 9023 & $7.56e+06$ \\
qft\_8\_TOR & 8 & 3950 & 3500 & \textbf{3350} & 4800 & 9023 & $7.56e+06$ \\
qnn\_8\_FEZ & 8 & 2550 & \textbf{3100} & 3700 & 5000 & 9023 & $7.56e+06$ \\
qnn\_8\_FEZ & 8 & 2950 & \textbf{3100} & 3700 & 5000 & 9023 & $7.56e+06$ \\
qnn\_8\_FEZ & 8 & 3750 & \textbf{3100} & 3700 & 5000 & 9023 & $7.56e+06$ \\
qnn\_8\_KYI & 8 & 3150 & 3550 & \textbf{3500} & 4500 & 9023 & $7.56e+06$ \\
qnn\_8\_KYI & 8 & 3100 & 3550 & \textbf{3500} & 4500 & 9023 & $7.56e+06$ \\
qnn\_8\_KYI & 8 & 3900 & 3550 & \textbf{3500} & 4500 & 9023 & $7.56e+06$ \\
random\_8\_FEZ & 8 & 2750 & \textbf{3100} & 3600 & 5300 & 9023 & $7.56e+06$ \\
random\_8\_FEZ & 8 & 3350 & \textbf{3100} & 3600 & 5300 & 9023 & $7.56e+06$ \\
random\_8\_FEZ & 8 & 4150 & \textbf{3100} & 3600 & 5300 & 9023 & $7.56e+06$ \\
vqe\_8\_KYI & 8 & 200 & \textbf{1100} & 2700 & 2700 & 9023 & $7.56e+06$ \\
vqe\_8\_KYI & 8 & 750 & \textbf{1100} & 2700 & 2700 & 9023 & $7.56e+06$ \\
vqe\_8\_KYI & 8 & 950 & \textbf{1100} & 2700 & 2700 & 9023 & $7.56e+06$ \\
\bottomrule
\end{tabular}
\end{table}

\clearpage

\begin{table}[ht!]
\centering
\footnotesize
\setlength{\tabcolsep}{3pt}
\renewcommand{\arraystretch}{1.05}
\caption{Full sample comparison for $\delta = 0.10$, all large circuits test traces.}
\label{tab:appendix-sota-d01-l}
\begin{tabular}{l c r r r r r r}
\toprule
Trace & Size & Oracle & Inc-TVD & Inc-Hell & Inc-JS & Weissman & Hoeffding \\
\midrule
dj\_10\_FEZ & 10 & 100 & \textbf{1600} & 3450 & 4200 & $3.56e+04$ & $1.39e+08$ \\
dj\_10\_FEZ & 10 & 1600 & \textbf{1600} & 3450 & 4200 & $3.56e+04$ & $1.39e+08$ \\
dj\_10\_FEZ & 10 & 1750 & \textbf{1600} & 3450 & 4200 & $3.56e+04$ & $1.39e+08$ \\
dj\_10\_TOR & 10 & 500 & \textbf{2200} & 4450 & 4700 & $3.56e+04$ & $1.39e+08$ \\
dj\_10\_TOR & 10 & 2400 & \textbf{2200} & 4450 & 4700 & $3.56e+04$ & $1.39e+08$ \\
dj\_10\_TOR & 10 & 2800 & \textbf{2200} & 4450 & 4700 & $3.56e+04$ & $1.39e+08$ \\
qaoa\_10\_SHE & 10 & 6550 & 9950 & \textbf{9750} & 11000 & $3.56e+04$ & $1.39e+08$ \\
qaoa\_10\_SHE & 10 & 9150 & 9950 & \textbf{9750} & 11000 & $3.56e+04$ & $1.39e+08$ \\
qaoa\_10\_SHE & 10 & 10300 & 9950 & \textbf{9750} & 11000 & $3.56e+04$ & $1.39e+08$ \\
qaoa\_12\_MAR & 12 & 7200 & \textbf{9000} & 14950 & 15900 & $1.42e+05$ & $2.52e+09$ \\
qaoa\_12\_MAR & 12 & 13250 & \textbf{9000} & 14950 & 15900 & $1.42e+05$ & $2.52e+09$ \\
qaoa\_12\_MAR & 12 & 13750 & \textbf{9000} & 14950 & 15900 & $1.42e+05$ & $2.52e+09$ \\
qft\_12\_FEZ & 12 & 15400 & \textbf{17150} & 18150 & 20000 & $1.42e+05$ & $2.52e+09$ \\
qft\_12\_FEZ & 12 & 15600 & \textbf{17150} & 18150 & 20000 & $1.42e+05$ & $2.52e+09$ \\
qft\_12\_FEZ & 12 & 16450 & \textbf{17150} & 18150 & 20000 & $1.42e+05$ & $2.52e+09$ \\
qft\_12\_TOR & 12 & 15200 & \textbf{17200} & 19350 & 20000 & $1.42e+05$ & $2.52e+09$ \\
qft\_12\_TOR & 12 & 15550 & \textbf{17200} & 19350 & 20000 & $1.42e+05$ & $2.52e+09$ \\
qft\_12\_TOR & 12 & 16550 & \textbf{17200} & 19350 & 20000 & $1.42e+05$ & $2.52e+09$ \\
random\_12\_FEZ & 12 & 15200 & \textbf{17100} & 19250 & 20000 & $1.42e+05$ & $2.52e+09$ \\
random\_12\_FEZ & 12 & 15700 & \textbf{17100} & 19250 & 20000 & $1.42e+05$ & $2.52e+09$ \\
random\_12\_FEZ & 12 & 16600 & \textbf{17100} & 19250 & 20000 & $1.42e+05$ & $2.52e+09$ \\
vqe\_12\_KYI & 12 & 550 & \textbf{3400} & 6150 & 7500 & $1.42e+05$ & $2.52e+09$ \\
vqe\_12\_KYI & 12 & 5250 & \textbf{3400} & 6150 & 7500 & $1.42e+05$ & $2.52e+09$ \\
vqe\_12\_KYI & 12 & 6200 & \textbf{3400} & 6150 & 7500 & $1.42e+05$ & $2.52e+09$ \\
dj\_14\_FEZ & 14 & 500 & \textbf{2450} & 4850 & 5200 & $5.68e+05$ & $4.49e+10$ \\
dj\_14\_FEZ & 14 & 4150 & \textbf{2450} & 4850 & 5200 & $5.68e+05$ & $4.49e+10$ \\
dj\_14\_FEZ & 14 & 4700 & \textbf{2450} & 4850 & 5200 & $5.68e+05$ & $4.49e+10$ \\
dj\_14\_MAR & 14 & 600 & \textbf{2300} & 5150 & 6200 & $5.68e+05$ & $4.49e+10$ \\
dj\_14\_MAR & 14 & 3100 & \textbf{2300} & 5150 & 6200 & $5.68e+05$ & $4.49e+10$ \\
dj\_14\_MAR & 14 & 3750 & \textbf{2300} & 5150 & 6200 & $5.68e+05$ & $4.49e+10$ \\
qaoa\_14\_KYI & 14 & 13350 & \textbf{13400} & 20000 & 20000 & $5.68e+05$ & $4.49e+10$ \\
qaoa\_14\_KYI & 14 & 17400 & \textbf{13400} & 20000 & 20000 & $5.68e+05$ & $4.49e+10$ \\
qaoa\_14\_KYI & 14 & 17650 & \textbf{13400} & 20000 & 20000 & $5.68e+05$ & $4.49e+10$ \\
qaoa\_14\_SHE & 14 & 14350 & \textbf{14600} & 20000 & 20000 & $5.68e+05$ & $4.49e+10$ \\
qaoa\_14\_SHE & 14 & 17750 & \textbf{14600} & 20000 & 20000 & $5.68e+05$ & $4.49e+10$ \\
qaoa\_14\_SHE & 14 & 17950 & \textbf{14600} & 20000 & 20000 & $5.68e+05$ & $4.49e+10$ \\
qft\_14\_FEZ & 14 & 17750 & \textbf{17800} & 20000 & 20000 & $5.68e+05$ & $4.49e+10$ \\
qft\_14\_FEZ & 14 & 19050 & \textbf{17800} & 20000 & 20000 & $5.68e+05$ & $4.49e+10$ \\
qft\_14\_FEZ & 14 & 19150 & \textbf{17800} & 20000 & 20000 & $5.68e+05$ & $4.49e+10$ \\
qft\_14\_TOR & 14 & 17750 & \textbf{17750} & 20000 & 20000 & $5.68e+05$ & $4.49e+10$ \\
qft\_14\_TOR & 14 & 19000 & \textbf{17750} & 20000 & 20000 & $5.68e+05$ & $4.49e+10$ \\
qft\_14\_TOR & 14 & 19100 & \textbf{17750} & 20000 & 20000 & $5.68e+05$ & $4.49e+10$ \\
qnn\_14\_FEZ & 14 & 17600 & \textbf{17700} & 20000 & 20000 & $5.68e+05$ & $4.49e+10$ \\
qnn\_14\_FEZ & 14 & 19000 & \textbf{17700} & 20000 & 20000 & $5.68e+05$ & $4.49e+10$ \\
qnn\_14\_FEZ & 14 & 19050 & \textbf{17700} & 20000 & 20000 & $5.68e+05$ & $4.49e+10$ \\
qnn\_14\_KYI & 14 & 17700 & \textbf{17750} & 20000 & 20000 & $5.68e+05$ & $4.49e+10$ \\
qnn\_14\_KYI & 14 & 19000 & \textbf{17750} & 20000 & 20000 & $5.68e+05$ & $4.49e+10$ \\
qnn\_14\_KYI & 14 & 19100 & \textbf{17750} & 20000 & 20000 & $5.68e+05$ & $4.49e+10$ \\
random\_14\_FEZ & 14 & 17700 & \textbf{17800} & 20000 & 20000 & $5.68e+05$ & $4.49e+10$ \\
random\_14\_FEZ & 14 & 18950 & \textbf{17800} & 20000 & 20000 & $5.68e+05$ & $4.49e+10$ \\
random\_14\_FEZ & 14 & 19050 & \textbf{17800} & 20000 & 20000 & $5.68e+05$ & $4.49e+10$ \\
vqe\_14\_KYI & 14 & 2600 & \textbf{4500} & 11150 & 14000 & $5.68e+05$ & $4.49e+10$ \\
vqe\_14\_KYI & 14 & 11400 & \textbf{4500} & 11150 & 14000 & $5.68e+05$ & $4.49e+10$ \\
vqe\_14\_KYI & 14 & 12050 & \textbf{4500} & 11150 & 14000 & $5.68e+05$ & $4.49e+10$ \\
\bottomrule
\end{tabular}
\end{table}

\clearpage

\begin{table}[ht!]
\centering
\footnotesize
\setlength{\tabcolsep}{3pt}
\renewcommand{\arraystretch}{1.05}
\caption{Full sample comparison for $\delta = 0.25$, all small circuits test traces.}
\label{tab:appendix-sota-d025-s}
\begin{tabular}{l c r r r r r r}
\toprule
Trace & Size & Oracle & Inc-TVD & Inc-Hell & Inc-JS & Weissman & Hoeffding \\
\midrule
dj\_4\_FEZ & 4 & 50 & \textbf{100} & 150 & 250 & 113 & 3309 \\
dj\_4\_TOR & 4 & 50 & \textbf{100} & 200 & 250 & 113 & 3309 \\
qaoa\_4\_SHE & 4 & 50 & \textbf{250} & 400 & 250 & 113 & 3309 \\
qaoa\_6\_MAR & 6 & 100 & \textbf{300} & 350 & 450 & 379 & $6.43e+04$ \\
qaoa\_6\_MAR & 6 & 150 & \textbf{300} & 350 & 450 & 379 & $6.43e+04$ \\
qft\_6\_FEZ & 6 & 250 & 500 & \textbf{450} & 750 & 379 & $6.43e+04$ \\
qft\_6\_FEZ & 6 & 300 & 500 & \textbf{450} & 750 & 379 & $6.43e+04$ \\
qft\_6\_TOR & 6 & 200 & \textbf{500} & 500 & 750 & 379 & $6.43e+04$ \\
qft\_6\_TOR & 6 & 250 & \textbf{500} & 500 & 750 & 379 & $6.43e+04$ \\
random\_6\_FEZ & 6 & 150 & \textbf{400} & 400 & 650 & 379 & $6.43e+04$ \\
random\_6\_FEZ & 6 & 250 & \textbf{400} & 400 & 650 & 379 & $6.43e+04$ \\
vqe\_6\_KYI & 6 & 50 & \textbf{150} & 250 & 250 & 379 & $6.43e+04$ \\
dj\_8\_FEZ & 8 & 50 & \textbf{150} & 250 & 250 & 1444 & $1.21e+06$ \\
dj\_8\_MAR & 8 & 50 & \textbf{100} & 300 & 250 & 1444 & $1.21e+06$ \\
qaoa\_8\_KYI & 8 & 300 & \textbf{550} & 750 & 850 & 1444 & $1.21e+06$ \\
qaoa\_8\_KYI & 8 & 700 & \textbf{550} & 750 & 850 & 1444 & $1.21e+06$ \\
qaoa\_8\_KYI & 8 & 850 & \textbf{550} & 750 & 850 & 1444 & $1.21e+06$ \\
qaoa\_8\_SHE & 8 & 350 & \textbf{500} & 850 & 850 & 1444 & $1.21e+06$ \\
qaoa\_8\_SHE & 8 & 600 & \textbf{500} & 850 & 850 & 1444 & $1.21e+06$ \\
qaoa\_8\_SHE & 8 & 750 & \textbf{500} & 850 & 850 & 1444 & $1.21e+06$ \\
qft\_8\_FEZ & 8 & 650 & \textbf{850} & 850 & 1050 & 1444 & $1.21e+06$ \\
qft\_8\_FEZ & 8 & 800 & \textbf{850} & 850 & 1050 & 1444 & $1.21e+06$ \\
qft\_8\_FEZ & 8 & 900 & \textbf{850} & 850 & 1050 & 1444 & $1.21e+06$ \\
qft\_8\_TOR & 8 & 650 & \textbf{850} & 1100 & 1050 & 1444 & $1.21e+06$ \\
qft\_8\_TOR & 8 & 750 & \textbf{850} & 1100 & 1050 & 1444 & $1.21e+06$ \\
qft\_8\_TOR & 8 & 900 & \textbf{850} & 1100 & 1050 & 1444 & $1.21e+06$ \\
qnn\_8\_FEZ & 8 & 500 & \textbf{700} & 1050 & 950 & 1444 & $1.21e+06$ \\
qnn\_8\_FEZ & 8 & 750 & \textbf{700} & 1050 & 950 & 1444 & $1.21e+06$ \\
qnn\_8\_FEZ & 8 & 900 & \textbf{700} & 1050 & 950 & 1444 & $1.21e+06$ \\
qnn\_8\_KYI & 8 & 650 & \textbf{800} & 1050 & 1050 & 1444 & $1.21e+06$ \\
qnn\_8\_KYI & 8 & 750 & \textbf{800} & 1050 & 1050 & 1444 & $1.21e+06$ \\
qnn\_8\_KYI & 8 & 900 & \textbf{800} & 1050 & 1050 & 1444 & $1.21e+06$ \\
random\_8\_FEZ & 8 & 550 & \textbf{700} & 1000 & 950 & 1444 & $1.21e+06$ \\
random\_8\_FEZ & 8 & 800 & \textbf{700} & 1000 & 950 & 1444 & $1.21e+06$ \\
random\_8\_FEZ & 8 & 950 & \textbf{700} & 1000 & 950 & 1444 & $1.21e+06$ \\
vqe\_8\_KYI & 8 & 50 & \textbf{200} & 500 & 450 & 1444 & $1.21e+06$ \\
vqe\_8\_KYI & 8 & 150 & \textbf{200} & 500 & 450 & 1444 & $1.21e+06$ \\
\bottomrule
\end{tabular}
\end{table}

\clearpage
\begin{table}[ht!]
\centering
\footnotesize
\setlength{\tabcolsep}{3pt}
\renewcommand{\arraystretch}{1.05}
\caption{Full sample comparison for $\delta = 0.25$, all large circuits test traces.}
\label{tab:appendix-sota-d025-l}
\begin{tabular}{l c r r r r r r}
\toprule
Trace & Size & Oracle & Inc-TVD & Inc-Hell & Inc-JS & Weissman & Hoeffding \\
\midrule
dj\_10\_FEZ & 10 & 50 & \textbf{300} & 1200 & 1550 & 5703 & $2.23e+07$ \\
dj\_10\_FEZ & 10 & 100 & \textbf{300} & 1200 & 1550 & 5703 & $2.23e+07$ \\
dj\_10\_TOR & 10 & 50 & 1800 & 1700 & \textbf{1650} & 5703 & $2.23e+07$ \\
dj\_10\_TOR & 10 & 250 & 1800 & 1700 & \textbf{1650} & 5703 & $2.23e+07$ \\
dj\_10\_TOR & 10 & 300 & 1800 & 1700 & \textbf{1650} & 5703 & $2.23e+07$ \\
qaoa\_10\_SHE & 10 & 1300 & 6900 & \textbf{3400} & 4100 & 5703 & $2.23e+07$ \\
qaoa\_10\_SHE & 10 & 2400 & 6900 & \textbf{3400} & 4100 & 5703 & $2.23e+07$ \\
qaoa\_10\_SHE & 10 & 2650 & 6900 & \textbf{3400} & 4100 & 5703 & $2.23e+07$ \\
qaoa\_12\_MAR & 12 & 1500 & 6300 & \textbf{5000} & 5900 & $2.27e+04$ & $4.03e+08$ \\
qaoa\_12\_MAR & 12 & 3800 & 6300 & \textbf{5000} & 5900 & $2.27e+04$ & $4.03e+08$ \\
qaoa\_12\_MAR & 12 & 4250 & 6300 & \textbf{5000} & 5900 & $2.27e+04$ & $4.03e+08$ \\
qft\_12\_FEZ & 12 & 6800 & 13300 & \textbf{9200} & 10100 & $2.27e+04$ & $4.03e+08$ \\
qft\_12\_FEZ & 12 & 8400 & 13300 & \textbf{9200} & 10100 & $2.27e+04$ & $4.03e+08$ \\
qft\_12\_FEZ & 12 & 9050 & 13300 & \textbf{9200} & 10100 & $2.27e+04$ & $4.03e+08$ \\
qft\_12\_TOR & 12 & 6650 & 13300 & 9500 & \textbf{9450} & $2.27e+04$ & $4.03e+08$ \\
qft\_12\_TOR & 12 & 8250 & 13300 & 9500 & \textbf{9450} & $2.27e+04$ & $4.03e+08$ \\
qft\_12\_TOR & 12 & 8950 & 13300 & 9500 & \textbf{9450} & $2.27e+04$ & $4.03e+08$ \\
random\_12\_FEZ & 12 & 6650 & 13100 & \textbf{9200} & 9900 & $2.27e+04$ & $4.03e+08$ \\
random\_12\_FEZ & 12 & 8450 & 13100 & \textbf{9200} & 9900 & $2.27e+04$ & $4.03e+08$ \\
random\_12\_FEZ & 12 & 9100 & 13100 & \textbf{9200} & 9900 & $2.27e+04$ & $4.03e+08$ \\
vqe\_12\_KYI & 12 & 50 & 2600 & \textbf{2400} & 3100 & $2.27e+04$ & $4.03e+08$ \\
vqe\_12\_KYI & 12 & 400 & 2600 & \textbf{2400} & 3100 & $2.27e+04$ & $4.03e+08$ \\
vqe\_12\_KYI & 12 & 500 & 2600 & \textbf{2400} & 3100 & $2.27e+04$ & $4.03e+08$ \\
dj\_14\_FEZ & 14 & 50 & 1900 & \textbf{1800} & 2250 & $9.09e+04$ & $7.19e+09$ \\
dj\_14\_FEZ & 14 & 200 & 1900 & \textbf{1800} & 2250 & $9.09e+04$ & $7.19e+09$ \\
dj\_14\_FEZ & 14 & 250 & 1900 & \textbf{1800} & 2250 & $9.09e+04$ & $7.19e+09$ \\
dj\_14\_MAR & 14 & 100 & 1900 & \textbf{1800} & 2450 & $9.09e+04$ & $7.19e+09$ \\
dj\_14\_MAR & 14 & 200 & 1900 & \textbf{1800} & 2450 & $9.09e+04$ & $7.19e+09$ \\
dj\_14\_MAR & 14 & 300 & 1900 & \textbf{1800} & 2450 & $9.09e+04$ & $7.19e+09$ \\
qaoa\_14\_KYI & 14 & 4700 & 9100 & \textbf{8800} & 9200 & $9.09e+04$ & $7.19e+09$ \\
qaoa\_14\_KYI & 14 & 8700 & 9100 & \textbf{8800} & 9200 & $9.09e+04$ & $7.19e+09$ \\
qaoa\_14\_KYI & 14 & 9350 & 9100 & \textbf{8800} & 9200 & $9.09e+04$ & $7.19e+09$ \\
qaoa\_14\_SHE & 14 & 5950 & 10100 & \textbf{8200} & 10450 & $9.09e+04$ & $7.19e+09$ \\
qaoa\_14\_SHE & 14 & 9750 & 10100 & \textbf{8200} & 10450 & $9.09e+04$ & $7.19e+09$ \\
qaoa\_14\_SHE & 14 & 10250 & 10100 & \textbf{8200} & 10450 & $9.09e+04$ & $7.19e+09$ \\
qft\_14\_FEZ & 14 & 12700 & \textbf{14200} & 17900 & 19350 & $9.09e+04$ & $7.19e+09$ \\
qft\_14\_FEZ & 14 & 15000 & \textbf{14200} & 17900 & 19350 & $9.09e+04$ & $7.19e+09$ \\
qft\_14\_FEZ & 14 & 15300 & \textbf{14200} & 17900 & 19350 & $9.09e+04$ & $7.19e+09$ \\
qft\_14\_TOR & 14 & 12650 & \textbf{14200} & 17800 & 18850 & $9.09e+04$ & $7.19e+09$ \\
qft\_14\_TOR & 14 & 14950 & \textbf{14200} & 17800 & 18850 & $9.09e+04$ & $7.19e+09$ \\
qft\_14\_TOR & 14 & 15250 & \textbf{14200} & 17800 & 18850 & $9.09e+04$ & $7.19e+09$ \\
qnn\_14\_FEZ & 14 & 12100 & \textbf{14100} & 17000 & 18400 & $9.09e+04$ & $7.19e+09$ \\
qnn\_14\_FEZ & 14 & 14600 & \textbf{14100} & 17000 & 18400 & $9.09e+04$ & $7.19e+09$ \\
qnn\_14\_FEZ & 14 & 14950 & \textbf{14100} & 17000 & 18400 & $9.09e+04$ & $7.19e+09$ \\
qnn\_14\_KYI & 14 & 12550 & \textbf{14200} & 17700 & 19350 & $9.09e+04$ & $7.19e+09$ \\
qnn\_14\_KYI & 14 & 14900 & \textbf{14200} & 17700 & 19350 & $9.09e+04$ & $7.19e+09$ \\
qnn\_14\_KYI & 14 & 15200 & \textbf{14200} & 17700 & 19350 & $9.09e+04$ & $7.19e+09$ \\
random\_14\_FEZ & 14 & 12500 & \textbf{14300} & 17700 & 19100 & $9.09e+04$ & $7.19e+09$ \\
random\_14\_FEZ & 14 & 14850 & \textbf{14300} & 17700 & 19100 & $9.09e+04$ & $7.19e+09$ \\
random\_14\_FEZ & 14 & 15150 & \textbf{14300} & 17700 & 19100 & $9.09e+04$ & $7.19e+09$ \\
vqe\_14\_KYI & 14 & 450 & 3500 & \textbf{3400} & 3550 & $9.09e+04$ & $7.19e+09$ \\
vqe\_14\_KYI & 14 & 1800 & 3500 & \textbf{3400} & 3550 & $9.09e+04$ & $7.19e+09$ \\
vqe\_14\_KYI & 14 & 2100 & 3500 & \textbf{3400} & 3550 & $9.09e+04$ & $7.19e+09$ \\
\bottomrule
\end{tabular}
\end{table}